\documentclass[11pt]{article}
\usepackage[letterpaper,includefoot,top=1in,bottom=1in,left=1in,right=1in]{geometry}
\usepackage[centertags,sumlimits]{amsmath}
\usepackage{amsfonts}
\usepackage{amssymb,graphics,psfrag}
\usepackage{array,epsfig,multirow,stmaryrd,graphicx}
\usepackage[all,cmtip]{xy}
\usepackage{mathrsfs}
\usepackage[bookmarks,breaklinks]{hyperref}
\usepackage{multirow}
\usepackage[bookmarks,breaklinks]{hyperref}
\usepackage{amsthm}
\usepackage{color}
\usepackage{slashed}

\usepackage[numbers,square,comma, compress]{natbib} 

\numberwithin{equation}{section}
\numberwithin{table}{section}

\renewcommand{\Im}{\operatorname{Im}}
\renewcommand{\Re}{\operatorname{Re}}

\newcommand{\beq}{\begin{equation}}
\newcommand{\eeq}{\end{equation}}
\newcommand{\be}{\begin{equation}}
\newcommand{\ee}{\end{equation}}
\newcommand{\bea}{\begin{eqnarray}}
\newcommand{\eea}{\end{eqnarray}}   
\newcommand{\ben}{\begin{eqnarray*}}
\newcommand{\een}{\end{eqnarray*}}                  
\newcommand{\ba}{\begin{aligned}}
\newcommand{\ea}{\end{aligned}}
\newcommand{\bt}{\begin{tabular}}
\newcommand{\et}{\end{tabular}}
\newcommand{\bc}{\begin{center}}
\newcommand{\ec}{\end{center}}
\newcommand{\cT}{\mathcal{T}}

\newcommand{\cD}{\mathcal{D}}
\newcommand{\cL}{\mathcal{L}}

\newcommand{\cK}{\mathcal{K}}
\newcommand{\cN}{\mathcal{N}}

\newcommand{\cH}{\mathcal{H}}

\newcommand{\cI}{\mathcal{I}}
\newcommand{\cJ}{\mathcal{J}}

\newcommand{\cV}{\mathcal{V}}

\newcommand{\bbZ}{\mathbb{Z}}

\newcommand{\bbC}{\mathbb{C}}

\newcommand{\nn}{\nonumber}

\newcommand{\cref}{{\bf [check ref]}}

\usepackage{comment}

\newcommand{\sixcT}{{\boldsymbol \cT}}
\newcommand{\sixB}{{\bf B}}
\newcommand{\sixsigma}{{\boldsymbol \sigma}}
\newcommand{\sixlambda}{{\boldsymbol \lambda}}
\newcommand{\sixcH}{{\boldsymbol \cH}}

\newcommand{\da}{\dot{a}}
\newcommand{\dbi}{\dot{b}}
\newcommand{\dc}{\dot{c}}
\newcommand{\ddi}{\dot{d}}

\newcommand{\KKcharge}{{n}}

\entrymodifiers={+!!<0pt,\fontdimen22\textfont2>}

\begin{document}

\baselineskip=16pt
\setlength{\parskip}{6pt}

\begin{titlepage}
\begin{flushright}
\parbox[t]{1.4in}{
\flushright MPP-2012-127}
\end{flushright}

\begin{center}

\vspace*{1.5cm}

{\LARGE \bf Non-Abelian Tensor Towers and \\[.2cm]
  (2,0) Superconformal Theories
}

\vskip 1.5cm

\renewcommand{\thefootnote}{}

\begin{center}
    \normalsize \bf{Federico Bonetti}, \normalsize \bf{Thomas W.~Grimm} and \normalsize \bf{Stefan Hohenegger}\footnote{\texttt{bonetti, grimm, shoheneg @mpp.mpg.de}} 
\end{center}
\vskip 0.5cm

 \emph{ Max Planck Institute for Physics, \\ 
        F\"ohringer Ring 6, 80805 Munich, Germany} 
%\\[0.25cm]
%\emph{$^{a,c,d}$ Department of Physics and Astronomy,\\
%University of Pennsylvania, Philadelphia, PA 19104-6396, USA}
%\\[0.15cm]
% \vspace*{0.5cm}

\end{center}

\vskip 1.5cm
\addtocounter{footnote}{-1}
\renewcommand{\thefootnote}{\arabic{footnote}}

\begin{center} {\bf ABSTRACT } \end{center}

With the aim to study six-dimensional $(2,0)$ superconformal theories with non-Abelian tensor multiplets 
we propose a five-dimensional superconformal action with eight supersymmetries for an infinite tower of 
non-Abelian vector, tensor and hypermultiplets. It describes the dynamics  
of the complete spectrum of the $(2,0)$ theories compactified on a circle coupled to an 
additional vector multiplet containing the circle radius and the Kaluza-Klein vector arising from 
the six-dimensional metric. All couplings are only given in terms of group theoretical constants 
and the Kaluza-Klein levels. After superconformal symmetry is reduced to Poincar\'e supersymmetry
we find a Kaluza-Klein inspired action coupling super-Yang-Mills theory to an infinite tower of massive non-Abelian tensors. 
We explore the possibility to restore sixteen supersymmetries by using techniques 
known from harmonic superspace. Namely, additional bosonic coordinates on a four-sphere are introduced to 
enhance the R-symmetry group. Maximally supersymmetric Yang-Mills 
theories and the Abelian $(2,0)$ tensor theories are recovered as special cases of our construction. 
Finally, we comment on the generation of an anomaly balancing Wess-Zumino term for the R-symmetry vector at one loop.

%%%%%%%%%%%%%%%%%%%%%%%%%%%%%%%%%%%%%%%%%%%%%%%%%%%%%%%%%%%%%%%%%%%%%%%%%%%%%%%%%%%%%%%%%%%%%%%%%%%%%%%%%%%%%%%%%%%%%%%%%%%%%%%%%%%%%%%%%%%%%%%%%%%
%%%%%%%%%%%%%%%%%%%%%%%%%

\vskip 0.5cm
\hfill {September, 2012}
\end{titlepage}

%\preprint{}
%\tableofcontents

\section{Introduction}

Six-dimensional interacting theories with $(2,0)$ supersymmetry have resisted a 
complete understanding since their discovery \cite{Witten:1995zh}. 
These theories propagate tensor multiplets which transform under a non-Abelian group $G$ and are supersymmetric with sixteen 
supercharges. They can arise as the low energy limit of multiple M5-branes or in the compactification of Type IIB string theory on singular geometries. 
So far no six-dimensional action for these theories is known and it remains unclear if such an action actually exists. 
Furthermore, given the fact that they have no dimensionless parameter a perturbative 
expansion seems not possible even given an action. Much research has been directed to nevertheless analyze the properties 
of these theories by using string dualities, matrix theory or a partial knowledge of required 
couplings \cite{Indirect20,Ganor:1998ve,Harvey:1998bx,Intriligator:2000eq}. Reviews on the $(2,0)$ theories 
can be found in \cite{Seiberg:1997ax,Witten:2009at,Gaiotto:2010be}.

There are various complications that have to be addressed in order to find a formulation of six-dimensional $(2,0)$ theories.
Two separate problems are particularly prominent. Firstly, 
the tensor multiplets have to contain a two-form with self-dual field strength. There is no canonical six-dimensional action 
for such chiral two-forms \cite{Marcus:1982yu}. Different solutions to this problem 
have been proposed, based on breaking of manifest Lorentz invariance, introduction of auxiliary fields, or
a holographic approach \cite{Siegel:1983es,Townsend:1983xs,Bonetti:2012fn}. Secondly, finding an action 
with interacting tensors transforming under a non-Abelian group is complicated due to the fact that  
$(2,0)$ theories admit no vectors in their spectrum. Even forming a covariant derivative for the 
tensors seems challenging. Recent discussions about the various complications 
in formulating $(2,0)$ theories can be found in \cite{Lambert:2010wm,Ho:2011ni,Lambert:2011gb,Linander:2011jy,Chu:2012um,Palmer:2012ya,Ho:2012nt}.

In this work we approach six-dimensional $(2,0)$ theories by studying 
a five-dimensional action for an infinite tower of modes that 
can be interpreted as Kaluza-Klein states. 
We propose that using this perspective
one can address both the self-duality as well as the non-Abelian gaugings at the level of 
an action. Indeed, we will write a five-dimensional superconformal action with $\cN=2$ supersymmetry, 
i.e.~with eight supercharges, whose spectrum, however, is chosen to consist of the degrees of freedom of the six-dimensional $(2,0)$ tensor multiplets compactified 
on a circle together with an additional 
$\cN=2$ vector multiplet containing the circle radius and the Kaluza-Klein vector of the 
six-dimensional metric. The infinite tower of $\cN=2$ non-Abelian tensor and hypermultiplets 
are interpreted as Kaluza-Klein excitations since they are gauged by the Kaluza-Klein vector. 
In accord with the six-dimensional self-duality the tensors turn out to have only 
Chern-Simons type kinetic terms. 
The zero modes play a distinctive role both in imposing the 
self-duality constraint and in the implementation of the non-Abelian 
gaugings with gauge group $G$. Our action fits in the general $\cN=2$ superconformal framework 
of \cite{Bergshoeff:2001hc,Bergshoeff:2002qk} that extends and applies
\cite{Cremmer:1980gs, Gunaydin:1983bi, Gunaydin:1999zx,Ceresole:2000jd,Bergshoeff:2004kh}.\footnote{Recent progress on the
construction of $(1,0)$ superconformal theories in six dimensions can be found in \cite{Samtleben:2011fj,Akyol:2012cq}.} It is 
crucial, however, that all couplings in our theory are only given in terms of group 
theoretical constants associated to $G$ and the Kaluza-Klein levels.

The five-dimensional superconformal invariance of the $\cN=2$ action is implemented in a way 
compatible with a subgroup of the six-dimensional superconformal group.
This implies that the additional vector multiplet,  containing the circle radius and the Kaluza-Klein vector,
has to transform in accord with the six-dimensional line element. However, in order to more directly 
interpret the $\cN=2$ superconformal action as a Kaluza-Klein theory, one has to fix superconformal 
invariance. We consider a restriction of the action that 
preserves $\cN=2$ Poincar\'e supersymmetry by giving a vacuum expectation 
value to the entire multiplet containing the circle radius and the Kaluza-Klein vector. 
After this gauge-fixing the infinite tower of tensor multiplets and hypermultiplets will gain a mass proportional
to the Kaluza-Klein scale set by the circle radius. The non-Abelian
gaugings and the realization of only half the maximal supersymmetry, however, 
prevent us from lifting the five-dimensional theory directly to six dimensions. Analyzing 
two special cases is instructive.
 
Considering zero modes alone the restricted $\cN=2$ action reduces to only maximally supersymmetric
Yang-Mills theory with gauge group $G$. The zero mode sector is automatically invariant under 
sixteen supercharges and is thus $\cN=4$ supersymmetric. This theory was recently discussed 
in connection with $(2,0)$ theories in refs.~\cite{Douglas:2010iu,Lambert}. 
It was proposed that the full dynamics of $(2,0)$ theories is encoded 
in maximally supersymmetric five-dimensional Yang-Mills theories at the non-perturbative level. 
In particular, the full tower of massive Kaluza-Klein modes was not included in the perturbative 
super Yang-Mills action but rather recovered as a subset of  
non-perturbative excitations of the action. This point of view should be 
contrasted with the approach taken in this work. Here we include the Kaluza-Klein modes directly 
in the five-dimensional superconformal action and propose a set of consistent interactions. 
We believe that this gives a promising starting point to give a lower-dimensional action 
formulation of $(2,0)$ theories.  

A precise identification of the five-dimensional theory as Kaluza-Klein action of 
a $(2,0)$ theory can be given in an Abelian setting. In this 
case a six-dimensional pseudoaction for a number of non-interacting $(2,0)$ self-dual 
tensor multiplets can be given. We show that the circle compactification allows to derive 
a proper five-dimensional $\cN=4$ action keeping all Kaluza-Klein modes. The resulting theory 
provides a supersymmetric completion of the five-dimensional bosonic
tensor action considered in \cite{Townsend:1983xs,Bonetti:2012fn}. We find that this action 
matches precisely with our restricted $\cN=2$ superconformal theory in an Abelian setting.
As for the zero mode action alone, supersymmetry is automatically enhanced to $\cN=4$ in 
this Abelian case.

It is an interesting question whether a maximally supersymmetric theory with 
a non-Abelian tensor tower exists in five dimensions. 
At first, following \cite{Awada:1985ep, Gunaydin:1985cu, Dall'Agata:2001vb, Schon:2006kz}, it appears not to be possible to formulate such 
an action in an obvious manner, since the constraints imposed by having sixteen supercharges 
are not compatible with the non-Abelian gaugings. We discuss a possibility
how one might circumvent these problems by studying an enhancement of the $\cN=2$ R-symmetry 
group $SU(2)_R$ to the R-symmetry group $USp(4)_R$ of an $\cN=4$ theory.
This is achieved through introducing an additional four-sphere $S^4$
and promoting all fields to functions on the latter. Already in the $\cN=2$ setting  
the field content can be grouped into proper $\cN=4$ multiplets. R-symmetry invariance of the action is obtained 
by integrating over $S^4$ which realizes $USp\,(4)$ transformations as coordinate 
reparametrizations. More abstractly,  
we patch together infinitely many copies of 
the $\cN=2$ supersymmetric action discussed before such that the resulting action displays $USp\,(4)_R$ 
R-symmetry. This procedure follows closely in spirit ideas of harmonic 
superspace \cite{Galperin:1984av, Ivanov:1984ut, Galperin:1984bu,Hartwell:1994rp,Howe:1995md,HSS} and indeed 
we devote some time to discuss the group-theoretic underpinning in more detail.

In the final section of this work we also make first comments on the quantum properties of 
the five-dimensional action. This is motivated by \cite{Bonetti:2012fn,BonettiGrimmHohenegger2}, where it
was shown that a five-dimensional action containing all Kaluza-Klein modes can be used to 
study six-dimensional gravitational anomalies. In five dimensions the latter are encoded in a
one-loop contribution to a Chern-Simons term for the background Kaluza-Klein vector 
with all excited Kaluza-Klein modes running in the loop. 
In the context of our action including infinite tensor towers, we investigate here how Kaluza-Klein modes can contribute to 
a five-dimensional Wess-Zumino term involving the background R-symmetry vector. 
Such a term is already expected in six dimensions to be induced by the breaking of the gauge 
group $G\to H\times U(1)$. It compensates for the shifted anomaly counting after massive modes arising in this 
breaking are integrated out \cite{tHooft,Intriligator:2000eq}.   
We consider the case $G=SU(N)$ so that the corresponding $(2,0)$ theory 
is the world-volume theory of a stack of M5-branes. Considering the theory already in the broken 
phase $SU(N-1)\times U(1)$ we investigate how integrating out Kaluza-Klein modes in an effective five-dimensional 
treatment potentially modifies the large $N$ scaling.  We find that the 
six-dimensional scaling behavior seems not to be modified through effects intrinsic to our effective 
five-dimensional approach. This hints to the fact that the five-dimensional theory can indeed be used to study 
properties of the $(2,0)$ theory even at the quantum level. We therefore see the presented one-loop test 
as a first step towards extracting the conformal anomaly using our 
five-dimensional action. The study of this anomaly has 
attracted much 
attention \cite{Henningson:1998gx,Harvey:1998bx,Intriligator:2000eq,Bastianelli:2000hi,Yi:2001bz,Maxfield:2012aw}. 
It also seems to be related to the number of degrees of freedom of $(2,0)$ theories 
which have been counted using various different 
methods \cite{Klebanov:1996un,Bolognesi:2011rq,Bolognesi:2011nh,Kim:2011mv,Kim:2012av,Kallen:2012zn}.

The paper is organized as follows. Sections \ref{Gaugings_and_bosonic_action}--\ref{sec:susyactions} 
contain our discussion of the $\cN=2$ superconformal non-Abelian theory. In section~\ref{Gaugings_and_bosonic_action} 
we present the spectrum of the theory including the extension that allows to implement five-dimensional 
superconformal invariance in section~\ref{N=2spectrum}. Section~\ref{sec:susyactions} discusses the $\cN=2$ 
action with superconformal symmetry and in the gauge fixed supersymmetric phase. 
Two special cases with $\cN=4$ supersymmetry are presented. Section~\ref{sec:N=4completion} 
contains the extension of the R-symmetry group and discusses the necessary group-theoretic foundations. 
Section~\ref{Sect:LoopCheck} contains a test of the quantum properties of the 
proposed action. Finally, we end with a summary of our results and a brief discussion of further directions. 
The main body of this paper is accompanied by several appendices. %, which contain our notation and present 
% further technical details. 
Our conventions and useful identifies are summarized in appendix \ref{appendix_conventions}, 
with index conventions relegated to appendix \ref{index-appendix}. An Abelian $(2,0)$ pseudo-action for 
a set of free tensor multiplets can be found in appendix \ref{ab_tensors_app}. The last 
appendix \ref{Sect:Superconformal5} discusses the realization of Weyl scaling symmetry in an 
$USp(4)_R$ covariant framework.

%%%%%%%%%%%%%%%%%%%%%%%%%%%%%%%%%%%%%%%%%%%%%%%%%%%%%%%%%%%%%%%%%%%%%%%%%%%%%%%%%%%%%%
\section{Supersymmetric spectrum and non-Abelian gauging} \label{Gaugings_and_bosonic_action}
%%%%%%%%%%%%%%%%%%%%%%%%%%%%%%%%%%%%%%%%%%%%%%%%%%%%%%%%%%%%%%%%%%%%%%%%%%%%%%%%%%%%%%

This section is devoted to the discussion of the supersymmetric
spectrum of the five-dimensional theories of non-Abelian tensors
which will be constructed in the following sections.
Our starting point consists of a number of tensor multiplets of six-dimensional rigid $(2,0)$ superconformal symmetry.
This spectrum is dimensionally reduced on a circle and the resulting $\cN = 4$
supermultiplets are described. Moreover, a mechanism 
for a non-Abelian gauging of tensors is implemented.
The decomposition of the $\cN = 4$ spectrum into $\cN=2$ multiplets and 
the discussion of conformal invariance is relegated to section \ref{N=2spectrum}.

\subsection{$(2,0)$ tensor multiplets on a circle and five-dimensional $\cN =4$ spectrum} \label{6d_and_N4}

Let $\sixcT^I$ be a collection of $(2,0)$ tensor multiplets in six dimensions. The index $I$ plays
here the role of a degeneracy index, but will be identified with an adjoint index of 
a non-Abelian gauge group in subsection \ref{non-Abelian_gauging}.
Boldface symbols 
will be used throughout to denote six-dimensional quantities.
The field content of $\sixcT^I$ is given by
\begin{equation} \label{sixcT_definition}
 \sixcT^I = (\sixB_{\boldsymbol{ \mu\nu}}^I ,\sixsigma^{I\, ij} , \sixlambda^{I\, i}) \ ,
\end{equation}
where $\sixB^I_{\boldsymbol{\mu\nu}}$ is a tensor (2-form), $\sixsigma^{I\, ij}$ are scalars, $\sixlambda^{I\, i }$ are spin-$1/2$ fermions. 
In our conventions, the supersymmetry parameter is a left-handed Weyl spinor, 
the tensors have negative chirality, i.e.~their field strength $\sixcH^I = d \sixB^I $ obey 
the anti-self-duality constraint $* \sixcH^I = - \sixcH^I$, and the fermions $\sixlambda^{I\,i}$ are right-handed Weyl spinors.
Indices $i,j = 1, \dots 4$ are indices of the  $\bf{4}$ representation
of $USp(4)_R$, the $R$-symmetry group of the $(2,0)$ supersymmetry algebra. 
The tensors ${\bf B}^I_{\boldsymbol{\mu\nu}}$ are singlets of $USp(4)_R$, the fermions $\sixlambda^{I\, i}$ transform in the
$\bf 4$ representation, while the scalars $\sixsigma^{I\, ij}$ belong to the $\bf 5$ representation, i.e.~they are anti-symmetric and traceless 
\begin{equation}
\sixsigma^{I\, ij}  = - \sixsigma^{I\, ji}  \ ,  \qquad
\Omega_{ij} \sixsigma^{I\, ij}  = 0 \ .
\end{equation}
In the last equation $\Omega_{ij}$ is the primitive antisymmetric invariant of $USp(4)_R$. We refer
the reader to Appendix \ref{appendix_conventions} for our conventions.
Tensor multiplets are pseudo-real, i.e.~they satisfy 
\begin{equation} \label{6d_reality}
  (\sixcT^I)^* = \sixcT^I \ : \quad
\begin{cases}
 \ \ \bar \sixB^I_{\boldsymbol{\mu\nu}} \equiv (\sixB^I_{\boldsymbol{\mu\nu}})^* = \sixB^I_{\boldsymbol{\mu\nu}} \ , \\[.1cm]
 \ \ \bar\sixsigma^I_{ ij} \equiv (\sixsigma^{I\, ij})^* = \Omega_{ik} \Omega_{jl} \sixsigma^{I\, kl} \ , \\[.1cm]
 \ \ \bar \sixlambda^{I\, i} \equiv (\sixlambda^I_{ i})^\dagger {\boldsymbol \gamma}^0  = \Omega^{ij} (\sixlambda_{j}^I)^{\sf T} \mathbf{ C} \ .
\end{cases}
\end{equation}
The last line encodes the usual symplectic-Majorana condition. The quantities  
${\boldsymbol \gamma}^0, \mathbf{C}$ are the timelike gamma matrix and the charge conjugation matrix in six dimensions, respectively. 

The $(2,0)$ Poincar\'e superalgebra can be enlarged to 
the superconformal algebra $OSp(8^*|4)$ \cite{Kac:1977em, Nahm:1977tg}. This requires
the introduction of new generators for dilatations, conformal boosts,   
special supersymmetry transformations, and R-symmetry transformations.
The action of these generators on physical fields can be found in \cite{Claus:1997cq, Bergshoeff:1999db}.
A more detailed discussion the rigid superconformal theory 
will be given in sections \ref{N=2spectrum} and \ref{N=2superconf_action} in the context of $\cN =2 $ supersymmetry
in five dimensions. In this section we just focus on the Weyl weights, which
are the charges under dilatations. For the fields in the tensor multiplets
$\sixcT^I$ they are collected in Table \ref{6d_table}.

\begin{table} 
\centering 
 \begin{tabular}{|c |c |c |c |c |c|} \hline
 \rule[-.2cm]{0cm}{0.7cm}multiplet & fields & type & comments & $USp(4)_R$ rep & Weyl weight  \\ \hline \hline
   &\rule[-.2cm]{0cm}{0.7cm} $\sixB^I_{\boldsymbol {\mu\nu}}$ & anti-self-dual tensor & pseudo-real &  {\bf 1} & 0 \\
 \parbox{3cm}{\vspace{-0.7cm}\begin{center}$\sixcT^I$ \\ massless tensor\\ multiplet\end{center}\vspace{-0.8cm}} & 
 \rule[-.2cm]{0cm}{0.7cm}  $\sixsigma^{I\, ij}$ & scalar & pseudo-real & {\bf 5} & 2 \\
 & \rule[-.2cm]{0cm}{0.7cm}  $\sixlambda^{I\, i}$ & right-handed spinor & pseudo-real &{\bf 4} & 5/2 \\ \hline 
\end{tabular}
\caption{Field content of an on-shell tensor multiplet $\sixcT$ of rigid $(2,0)$ superconformal symmetry 
in six dimensions. The precise
formulation of the reality properties of the fields is found in \eqref{6d_reality}. 
}
\label{6d_table}
\end{table}

A supersymmetric, two-derivative pseudoaction for a collection of non-interacting 
tensor multiplets $\sixcT^I$ can be found in
Appendix \ref{ab_tensors_app}, along with the $(2,0)$ supersymmetry transformations. 
A proper action is in general not known, since there is no obvious way to impose 
the anti-self-duality constraint consistently with non-Abelian gauge invariance.
In particular, there are no vectors in the spectrum which could be used as gauge connections. 
Indeed, $(2,0)$ gauge theories of tensors are conjectured to be a non-Abelian generalization
of gerbes, with 2-form connections \cite{Breen:2001ie,Baez:2010ya,Ho:2012nt} (see also \cite{Saemann:2011nb}). Our strategy 
is to avoid these difficulties, by performing the gauging in the reduced five-dimensional theory. 

We compactify one spatial dimension on a circle 
using the standard Kaluza-Klein ansatz for the metric,
\begin{equation} \label{KK_metric}
 {\mathbf g}_{\boldsymbol{\mu\nu}} d \mathbf{x}^{\boldsymbol \mu} d \mathbf{x}^{\boldsymbol \nu} = 
g_{\mu\nu} dx^\mu dx^\nu + r^2(dy - A^0_\mu dx^\mu)^2 \ .
\end{equation}
On the right hand side $g_{\mu\nu}$ is the five-dimensional metric, $r$ is the radius
of the circle, $y \sim y + 2\pi$ is the compact coordinate along the circle,
 and $A^0_\mu$ is the Kaluza-Klein vector with Abelian field strength $F^0 = dA^0$. In the rigid limit, $g_{\mu\nu}$
is the flat Minkowski metric, $r$ is constant and $A^0$ vanishes. Later on, we will promote these 
quantities to fields, however, since they will play a crucial role in the 
superconformal theories of section \ref{sec:susyactions}. 

Upon compactification on a circle, the scalars $\sixsigma^{I\, ij}$ and the spinors $\sixlambda^{I\, i}$
give rise to a Kaluza-Klein tower of five-dimensional scalars $\sigma^{I\, ij}_n$ and spinors $\lambda^{I\, i}_n$, where $n \in \bbZ$.
More precisely we write
\begin{equation} \label{KK_sigma_lambda}
 \boldsymbol{\sigma}^{I\, ij} = r^{-1} \sum_{n \in \bbZ} e^{iny} \sigma^{I\, ij}_n \ , \qquad
 \boldsymbol{\lambda}^{I\, i} = r^{-1}  \sum_{n \in \bbZ} e^{iny} \lambda^{I\, i}_n \otimes \eta \ ,
\end{equation}
where $\eta$ is a constant two-component spinor. Note that we have included a factor of $r^{-1}$ 
in the Kaluza-Klein ansatz, in order to have five-dimensional fields $\sigma^{I\,ij}, \lambda^{I\,i}$
of canonical dimensions $1$ and $3/2$, respectively. These fields are also the natural variables compatible 
with the lower-dimensional supersymmetry.
As far as the tensors are concerned, 
reduction of $\sixB^I_{\boldsymbol{\mu \nu}}$ furnishes both a tower of tensors $B^I_{n\, \mu\nu}$
 and of vectors $A^I_{n\, \mu}$ in five dimensions. 
We can write
\begin{equation} \label{basic_KK}
 \sixB^I = \sum_{n \in \bbZ} e^{iny} \left[ B^I_n + A^I_n \wedge (dy - A^0) \right] \ .
\end{equation}
As a consequence of the six-dimensional anti-self-duality constraint, $B^I_{n\, \mu\nu}$ and $A^I_{n\, \mu}$
do not contain independent degrees of freedom. On the one hand, 
the anti-self-duality constraint can be used to eliminate the tensor zero modes $B^I_{0\, \mu\nu}$ from the spectrum of 
the five-dimensional theory, keeping the vector zero modes $A^I_\mu \equiv A^I_{0\, \mu}$ only. On the other hand, 
excited modes $B^I_{n\, \mu\nu}, A^I_{n\, \mu}$ are related by a St\"uckelberg-like symmetry in the invariant derivative 
$F_n^I = d A^I_n + i n B_n^I$ \cite{Townsend:1983xs,Bonetti:2012fn}. In this way $B^I_{n\, \mu\nu}$ can `eat' $A^I_{n\, \mu}$ 
and become a massive tensor field in five dimensions. 
In conclusion, reduction of ${\bf B}^I_{\boldsymbol{\mu\nu}}$
yields a massless vector $A^I_{\mu}$ and a tower of complex massive
tensors $B^I_{n\, \mu\nu}$.
A purely bosonic Lagrangian for $A^I_\mu, B_{n\,\mu\nu}^I$ coupled to the 
Kaluza-Klein vector $A^0_\mu $
has been discussed in \cite{Bonetti:2012fn} and takes the form
\bea \label{purely_bosonic_action}
 \cL_{\rm tens} &=& d_{IJ} \Big[ -\tfrac {1}{4} r^{-1} F^I_{\mu \nu} F^{J\, \mu \nu}  
                    - \tfrac18  \epsilon^{\mu \nu \lambda \rho \sigma} A^0_\mu\, F^I_{\nu \lambda}\, F^J_{\rho \sigma} \Big]\\
                  &&+  {\textstyle \sum\limits_{n=1}^\infty} d_{IJ} \Big[ - \tfrac12 r^{-1} \bar F^I_{n\, \mu \nu} F^{J\, \mu \nu}_{n} 
                    + \tfrac{i}{4 n}  \epsilon^{\mu \nu \lambda \rho \sigma} \bar F^I_{n\, \mu \nu} \, \cD^{\rm KK}_{\lambda} F^J_{n\, \rho \sigma}
\Big] \ . \nn 
\eea
On the right hand side we have introduced the Abelian field strength $F^I= dA^I$
and we have used the St\"uckelberg gauge-fixed expression for the tensors 
\beq \label{def-Fn}
    F^I_{n\, \mu\nu} = i n B^I_{n\, \mu\nu}\ .
\eeq 
It will be convenient to use this rescaled $F^I_{n\, \mu\nu}$ to represent the tensors in the remainder of this work.
Indices $I,J$ are contracted
with a constant metric $d_{IJ}$. In section~\ref{non-Abelian_gauging} it will be related to
group-theoretical invariants after the degeneracy index $I$ is promoted to a gauge index. 
We have also made use of the shorthand notation $\cD^{\rm KK}_\mu X_n = \partial_\mu X_n + i n A^0_\mu X_n $ 
for generic Kaluza-Klein modes $X_n$. More information about this covariant derivative will be given 
in section \ref{sec:mass_KKgauging}.

The main purpose of our work is to provide a supersymmetric non-Abelian
generalization of the action \eqref{purely_bosonic_action}. As a first step,
we discuss how five-dimensional fields are organized in $\cN = 4$ multiplets.
 The $R$-symmetry group is again $USp(4)_R$, and 
the transformation properties of the fields under $R$-symmetry are unaffected by dimensional reduction.
The vector zero mode $A^I$ combines with the zero modes $\sigma^{I\, ij} \equiv \sigma_0^{I\, ij}$ and
$\lambda^{I\, i} \equiv \lambda^{I\, i}_0$ into a single vector multiplet which we will denote as
\begin{equation} \label{cV_definition}
 \cV^I = (A^I_\mu , \sigma^{I\, ij} , \lambda^{I\, i}) \ .
\end{equation}
Each massive tensor $F^I_n$ combines with the corresponding excited modes $\sigma^{I\, ij}_n, \lambda^{I\, i}_n$
into a massive tensor multiplet
\begin{equation} \label{cT_definition}
 \cT_n^I = (F^I_{n\,\mu\nu} , \sigma^{I\, ij}_n , \lambda^{I\, i}_n) \ , \qquad n \in \bbZ^* \ .
\end{equation}
As a consequence of the reality conditions \eqref{6d_reality} in six dimensions, the vector multiplet is pseudo-real,
\begin{equation} \label{5d_reality}
 (\cV^I)^* = \cV^I \ : \quad
\begin{cases}
 \ \ \bar A^I_\mu \equiv (A^I_\mu)^* = A^I_\mu \ , \\[.1cm]
 \ \ \bar\sigma_{ ij}^I \equiv (\sigma^{I\, ij})^* = \Omega_{ik} \Omega_{jl} \sigma^{I\, kl} \\[.1cm]
 \ \ \bar\lambda^{I\, i} \equiv (\lambda_i^I)^\dagger \gamma^0  = \Omega^{ij} (\lambda_j^I)^{\sf T} C  \ ,
\end{cases}
\end{equation}
and the tensor multiplets satisfy 
\begin{equation} \label{5d_tensor_reality}
 (\cT^I_n)^* = \cT_{-n}^I \ : \quad
\begin{cases}
 \ \ \bar F_{n\, \mu\nu}^I \equiv (F_{n\, \mu\nu}^I)^* = F^I_{-n\, \mu\nu} \ , \\[.1cm]
 \ \ \bar\sigma_{n\, ij}^I \equiv (\sigma^{I ij}_n)^*   = \Omega_{ik} \Omega_{jl} \sigma^{I kl}_{-n} \ , \\[.1cm]
 \ \ \bar\lambda^{I\, i}_n \equiv (\lambda^I_{n\, i})^\dagger \gamma^0  = \Omega^{ij} (\lambda_{-n\,j}^I)^{\sf T} C  \ .
\end{cases}
\end{equation}
We can thus restrict our attention to positive $n$ only, to
avoid a redundant description of the same degrees of freedom.
Note that now $\gamma^0, C$ refer to spinors in five dimensions. Our conventions about five-dimensional spinors
are collected in Appendix \ref{appendix_conventions} along with some useful identities. It is interesting to 
contrast the reality condition for spinors on zero modes and on excited modes: the former is the usual symplectic-Majorana
condition, but the latter relates two different symplectic multiplets, $\lambda_n^i$ and $\lambda_{-n}^i$, and
imposes no constraint on either of them separately. In this respect $\lambda_n^i$ is referred to as `complex'.
As discussed in Appendix \ref{appendix_conventions}, every complex symplectic spinor as $\lambda^{I\, i}_n$
is equivalent to a doublet of symplectic-Majorana spinors.

Since there is no known extension of the five-dimensional $\cN=4$ Poincar\'e superalgebra
to a superconformal algebra \cite{Kac:1977em, Nahm:1977tg}, there is no well-defined notion of Weyl weight for 
$\cN=4$ supermultiplets. Six-dimensional superconformal $(2,0)$ symmetry, however,
implies a (classical) scaling symmetry of the five-dimensional $\cN=4$ theory.
From the metric ansatz \eqref{KK_metric} we infer that the compactification
radius $r$ has scaling weight $-1$, as will be further discussed in section \ref{sec:conformal_symmetry}. The scaling weights
of all fields in vector and tensor multiplets can be extracted by comparing the six-dimensional Weyl weights listed in Table \ref{6d_table}
with the Kaluza-Klein ans\"atze \eqref{KK_sigma_lambda}, \eqref{basic_KK}. 
They are found in Table \ref{N4_table}, together with
a summary of $USp(4)_R$ representations.

\begin{table} 
\centering 
 \begin{tabular}{|c |c |c |c |c| c |}
 \hline
 \rule[-.2cm]{0cm}{0.7cm} multiplet & fields  & type & comments & $USp(4)_R$ rep & Scaling weight \\ \hline \hline
  &  \rule[-.2cm]{0cm}{0.7cm}  $A^I_\mu \equiv A^I_{0\, \mu}$ & vector & pseudo-real &  {\bf 1} & $0$ \\
\parbox{3cm}{\begin{center}\vspace{-0.7cm}$\cV^I$\\massless vector\\multiplet\end{center}\vspace{-0.8cm}}    
     &  \rule[-.2cm]{0cm}{0.7cm} $\sigma^{I\, ij} \equiv \sigma_0^{I\, ij}$ & scalar & pseudo-real &  {\bf 5} & $1$ \\
       & \rule[-.2cm]{0cm}{0.7cm} $\lambda^{I\, i} \equiv \lambda_0^{I\, i}$ & spinor & pseudo-real &  {\bf 4} & $3/2$ \\ \hline
 & \rule[-.2cm]{0cm}{0.7cm}  $F^I_{n\, \mu\nu}$ & tensor & complex &  {\bf 1} & $0$ \\
\parbox{3cm}{\begin{center}\vspace{-0.7cm} $\cT^I_n$\\massive tensor\\multiplet\end{center}
\vspace{-0.8cm}}  & \rule[-.2cm]{0cm}{0.7cm} $\sigma^{I\, ij}_n$ & scalar & complex &  {\bf 5} & $1$  \\
 & \rule[-.2cm]{0cm}{0.7cm}  $\lambda^{I\, i}_n$ & spinor & complex &  {\bf 4} & $3/2$ \\
 \hline
\end{tabular}
\caption{Field content of $\cN=4$ vector multiplets $\cV^I$ and tensor multiplets $\cT^I_n$ in five dimensions. The precise
formulation of the pseudo-reality properties of the fields in $\cV^I$ is found in \eqref{5d_reality}.
The last column collects the weights with respect to the five-dimensional 
scaling symmetry inherited from full six-dimensional conformal invariance.}
\label{N4_table}
\end{table}

\subsection{Mass scale and Kaluza-Klein gauging} \label{sec:mass_KKgauging}

Let us analyze in more detail the role played by the compactification radius $r$ and the
Kaluza-Klein vector $A^0$.
The $(2,0)$ theory we started from has no mass scale.\footnote{This holds in the rigid limit
 $\kappa \rightarrow 0$, where $\kappa$ is the six-dimensional gravitational constant with $[\kappa]=-2$. }
In contrast, the dimensionally reduced theory has a mass scale set by the inverse of the compactification radius $r$.
In particular, the $n$th excited modes $F_{n\, \mu\nu}, \sigma_n^{ij} , \lambda_n^i$ have masses proportional to
\begin{equation} \label{KK_mass}
 m_n = n r^{-1} \ ,
\end{equation}
as can be seen by comparing the mass and kinetic terms for the respective fields
as given below.
In order to infer this, we recall that $B_{n\, \mu\nu}$ is related with $F_{n\,\mu\nu}$ by the 
rescaling \eqref{def-Fn}.
It is worth recalling the role of $r$ in the conjectured equivalence between
$(2,0)$ theories and five-dimensional super-Yang-Mills theories \cite{Lambert:2010wm,Douglas:2010iu}. 
Even if a complete formulation of $(2,0)$ theories in the non-Abelian
case is not available, upon compactification on a circle they have to yield super-Yang-Mills
in the massless sector, corresponding to the multiplets $\cV^I$ in our notation.
The Yang-Mills coupling constant in five dimensions has mass dimension $[g] = -1/2$, 
and is identified with the compactification radius,
\begin{equation}
 g^2 = r \ ,
\end{equation}
consistently with the fact that $(2,0)$ theories have no tunable parameter
and compactification is the only source of a mass scale.

The Kaluza-Klein field can be interpreted as a
five-dimensional gauge connection which is needed when a global $U(1)$
symmetry is promoted to a local symmetry. This $U(1)$ symmetry will
be denoted $U(1)_{\rm KK}$. Since it will play a key role in our formulation of the non-Abelian 
five-dimensional action, let us discuss this symmetry in more detail and introduce some 
useful notation. $U(1)_{KK}$ originates from constant shifts of the compact coordinate 
$y' = y - \Lambda$. These can be undone by redefining the $n$th Kaluza-Klein mode of a field $X$ 
as $X_n' =  e^{i n\Lambda} X_n$, as can be seen from \eqref{basic_KK}, Thus, the $n$th Kaluza-Klein mode of any field
has electric charge $n$ under $U(1)_{\rm KK}$. The associated infinitesimal transformation reads
\begin{equation} \label{KK_trans}
 \delta_{\rm KK}(\lambda) X_n =  i n \lambda X_n \ .
\end{equation}
If we demand  
\begin{equation} \label{A0_gauge}
 \delta_{\rm KK}(\lambda) A^0_\mu = - \partial_\mu \lambda  \ ,
\end{equation}
we can gauge $U(1)_{\rm KK}$ by introducing the covariant derivative
\begin{equation} \label{KK_minimal}
\cD^{\rm KK}_\mu X_n= \partial_\mu X_n + i n A^0_\mu X_n \ .
\end{equation}
From a six-dimensional perspective,
$A^0$ is identified with fluctuations of the off-diagonal components 
of the metric, as can be seen from \eqref{KK_metric}. Its gauge transformation \eqref{A0_gauge} is just a 
special case of a six-dimensional diffeomorphism along the circle, and the minimal coupling to $X_n$ \eqref{KK_minimal}
is required by six-dimensional covariance.

In section \ref{sec:susyactions} it will prove useful to rewrite the $U(1)_{\rm KK}$ gauging  in terms of
real fields. To this end, we exploit the isomorphism $U(1)_{\rm KK} \cong SO(2)_{\rm KK}$ and for
any complex field $X_\KKcharge$ of charge $\KKcharge$ we introduce the $SO(2)_{\rm KK}$ doublet $X_\KKcharge^{\alpha}$, $\alpha =1,2$
via
\begin{equation} \label{compl_splitting}
 X_\KKcharge = \tfrac{1}{\sqrt{2}} \big( X_n^{\alpha=1} + i X_n^{\alpha =2}\big)\ . 
\end{equation}
Since the action of $U(1)_{\rm KK}$ on $X_\KKcharge$ is given by $X'_\KKcharge = e^{i \KKcharge \Lambda} X_\KKcharge$, 
 the corresponding action of $SO(2)_{\rm KK}$ on $X_\KKcharge^\alpha$ reads
\begin{equation} \label{SO2_action}
 X_\KKcharge^{\prime \alpha} = M^\alpha_{\phantom{A}\beta} X_\KKcharge^\beta \ , \qquad
M^\alpha_{\phantom{A}\beta} = 
\begin{pmatrix}
 \cos \KKcharge\,\Lambda & - \sin \KKcharge\, \Lambda \\
 \sin \KKcharge\, \Lambda & \cos \KKcharge\, \Lambda
\end{pmatrix}=\delta^{\alpha\gamma}\left(\delta_{\gamma\beta}-n\Lambda\epsilon_{\gamma\beta}+\mathcal{O}(\Lambda^2)\right) \ .
\end{equation}
The Kaluza-Klein covariant derivative of the doublet $X_\KKcharge^\alpha$ is therefore
\begin{equation} \label{SO2_covariant_derivative}
 \cD^{\rm KK}_\mu X_\KKcharge^\alpha = \partial_\mu X_\KKcharge^\alpha + \KKcharge \epsilon_{\beta \gamma} \delta^{\gamma \alpha}\, A^0_\mu X_{\KKcharge}^\beta \ ,
\end{equation}
where we have chosen the representation  
\begin{equation} \label{def-epsilon_alphabeta}
 \epsilon_{\alpha\beta} = 
\begin{pmatrix}
 0 & 1 \\
 -1 & 0 
\end{pmatrix}
\end{equation}
for the antisymmetric invariant of $SO(2)_{\rm KK}$. In the last equations
we have implicitly assumed that $X_\KKcharge$ is a boson. As explained in Appendix \ref{appendix_conventions},
the same formalism can be applied to symplectic spinors.

As a first application we present the reformulation of the bosonic action \eqref{purely_bosonic_action} with $SO(2)_{\rm KK}$ doublets instead of 
complex fields. Inserting \eqref{compl_splitting} for the tensors $F^I_n$ into \eqref{purely_bosonic_action} we find 
\bea \label{purely_bosonic_action_alphabeta}
 \cL_{\rm tens} &=& d_{IJ} \Big[ -\tfrac {1}{4} r^{-1} F^I_{\mu \nu} F^{J\, \mu \nu}  - \tfrac18  \epsilon^{\mu \nu \lambda \rho \sigma} A^0_\mu\, F^I_{\nu \lambda}\, F^J_{\rho \sigma} \Big]\\
&&+  {\textstyle \sum\limits_{n=1}^\infty} d_{IJ} \Big[ - \tfrac14 r^{-1} \delta_{\alpha \beta} F^{I \alpha}_{n\, \mu \nu} F^{J\beta\, \mu \nu}_{n} 
                             - \tfrac{1}{8 n} \epsilon_{\alpha \beta} \epsilon^{\mu \nu \lambda \rho \sigma} F^{I\alpha }_{n\, \mu \nu} \, \cD^{\rm KK}_{\lambda} F^{J\beta}_{n\, \rho \sigma}
\Big] \ , \nn 
\eea
where we have used the identities \eqref{rewrite_alphabeta}. These terms together with 
the Kaluza-Klein gauging, and the non-Abelian gaugings that we introduce next, 
turn out to be sufficient to determine the key characteristic data of the complete 
supersymmetric theory discussed in section \ref{sec:susyactions}.

\subsection{Non-Abelian gauge transformation and covariant derivative} \label{non-Abelian_gauging}

In our discussion of the five-dimensional spectrum zero modes and excited modes 
are treated on a very different footing, at the expense of manifest six-dimensional Lorentz symmetry.
However, this enables us to implement a non-Abelian gauging, since we can use massless vectors in five dimensions 
as gauge connections, and treat all other fields as charged matter. This implementation is the only 
straightforward gauging compatible with the Kaluza-Klein charges under the assumption that the gauge parameter is 
neutral under $U(1)_{\rm KK}$. 
The same strategy has been proposed in the literature in a similar context,
see e.g.~\cite{Ho:2011ni,Ho:2012nt}. Identifying a possible six-dimensional interpretation for this non-democratic gauging is a non-trivial task and is 
left for future research.

To define a non-Abelian gauging we first identify the degeneracy index $I$ with the adjoint
index of some non-Abelian group $G$. More precisely,
we let $I$ enumerate the elements $t_I$ of a basis of anti-Hermitean generators of the associated Lie algebra,
so that $I = 1, \dots |G| \equiv {\rm dim}(G)$.
We introduce the 
structure constants and the Cartan-Killing metric by 
\beq \label{TT=fT}
   [t_I, t_J] = - f_{IJ}^{\phantom{IJ}K} t_K\ ,  \qquad d_{IJ} = \text{Tr}(t_I t_J)\ .
\eeq
Both $f_{IJ}^{\phantom{IJ}K}$ and $d_{IJ}$ are real. We assume $d_{IJ}$ is non-singular and positive definite, and we use it together with its inverse $d^{IJ}$
to raise and lower adjoint indices. For example, $f_{IJK}= d_{IL} {f_{JK}}^{L}$. Furthermore, we take $f_{IJK}$ to be completely 
antisymmetric. The groups under consideration are taken to be of A-D-E type.

In order to realize a non-Abelian gauging of the 
spectrum \eqref{cV_definition}, \eqref{cT_definition}, we interpret $A^I$ as a gauge connection, while all other fields will 
be seen as adjoint matter. More precisely, we postulate the following infinitesimal transformation rules 
under the action of the non-Abelian gauge group $G$,
\begin{equation} \label{gauge_trans}
 \delta_G(\alpha) A^I_\mu = \partial_\mu \alpha^I +   \,  f_{JK}^{\phantom{JK}I} A^J_\mu \alpha^K \ , \qquad
 \delta_G(\alpha) X^I = -  f_{JK}^{\phantom{JK}I} \alpha^J X^K  \ ,
\end{equation}
where $\alpha$ is the scalar gauge parameter
and $X^I$ is any field among $\sigma^{I\, ij}, \lambda^{I\, i}, \sigma^{I\, ij}_n, \lambda^{I\, i}_n, F_{n\, \mu\nu}^I$ ($n >0$).
Recalling \eqref{KK_minimal}, we see that the full $G \times U(1)_{\rm KK}$ covariant derivative of any adjoint field $X^I_\KKcharge$ with Kaluza-Klein charge
 $\KKcharge$ is given by \footnote{
Since we work in flat space, we do not have to introduce a spacetime connection and covariant derivative.}
\begin{equation} \label{covariant_derivative_gen}
 \cD_\mu X^I_\KKcharge = \partial_\mu X^I_\KKcharge + i \KKcharge \,A^0_\mu X^I_\KKcharge +   f_{JK}^{\phantom{JK}I} A^J_\mu X^K_\KKcharge \ .
\end{equation}
We note here that $\mathcal{D}_\mu X_n^I$ has the same charge under $U(1)_{KK}$ as $X_n^I$ itself. The non-Abelian field-strength of $A^I$ reads
\begin{equation}
 F^I_{\mu\nu} = 2 \partial_{[\mu} A_{\nu]}^I +   f_{JK}^{\phantom{JK}I} A^J_\mu A^K_\nu \ ,
\end{equation}
transforms in the adjoint representation, satisfies the  Bianchi identity $\cD_{[\mu} F^I_{\nu\rho]} = 0$, and
enters the commutator of covariant derivatives as specified by
\begin{equation}
 [\cD_\mu , \cD_\nu] X^I_\KKcharge =  i \KKcharge F^0_{\mu\nu} X^I_\KKcharge +  f_{JK}^{\phantom{JK}I} F^J_{\mu\nu} X^K_\KKcharge \ . 
\end{equation}
The algebra of gauge transformations closes on all fields, according to
\begin{equation}
 [\delta_G(\alpha_1) , \delta_G (\alpha_2)] = \delta_G(\alpha_3) \ , \qquad
\alpha_3^I =  f_{JK}^{\phantom{JK}I} \alpha^J_1 \alpha^K_2 \ .
\end{equation}

\section{Spectrum in terms of $\cN=2$ superconformal multiplets} \label{N=2spectrum}

Since non-Abelian gaugings of tensor multiplets are not consistent 
with standard $\cN=4$ actions determined in \cite{Awada:1985ep, Gunaydin:1985cu, Dall'Agata:2001vb, Schon:2006kz}, 
we first consider an $\cN=2$ formulation. Upon reduction, we get $\cN = 2$ vector, tensor 
and hypermultiplets, and we can exploit the $\cN=2$ rigid
superconformal formalism of refs. \cite{Bergshoeff:2001hc,Bergshoeff:2002qk}. 

\subsection{Splitting of $\cN = 4$ multiplets}

To rewrite the $\cN = 4$ spectrum in terms of $\cN = 2$
supermultiplets, we consider the splitting of the original $R$-symmetry group 
$USp(4)_R$ according to
\begin{equation} \label{R_breaking}
 USp(4)_R \rightarrow SU(2)_R \times SU(2) \ ,
\end{equation}
where the first factor is the $R$-symmetry group of the $\cN = 2$ algebra, 
and the second factor is an extra global symmetry
of the theory. We use indices $a,b = 1,2$ for the ${\bf 2}$ representation of $SU(2)_R$, 
while indices $\da, \dbi = 1, 2$ refer 
to the ${\bf 2}$ representation of $SU(2)$. Under \eqref{R_breaking}
the branching rules for the relevant representations of $USp(4)_R$ read
\begin{align} \label{rep_decomp}
\mathbf{5} \ \ \; &\rightarrow \ (\mathbf{1},\mathbf{1}) + (\mathbf{2},\mathbf{2})\ ,
& \mathbf{4} \ \ \, &\rightarrow \ (\mathbf{2},\mathbf{1}) + (\mathbf{1},\mathbf{2})\ , \\
 \sigma^{I\, ij}_n  \ &\rightarrow \ \ \   \phi^I_n\ \ \ ,\  \ q^{I\, a \dbi}_n \ \ , \qquad \quad  
& \lambda^{I\, i}_n    \ &\rightarrow \ \ \,    \chi^{I\, a}_n \ \ , \ \ \, \zeta^{I\, \dbi}_n \ \ \, ,  \nn
\end{align}
where the entries in the brackets correspond to the two $SU(2)$'s, 
and we have introduced the bosonic fields $\phi^I_n$, $q^{I\, a \dbi}_n$, and the fermionic fields 
$\chi^{I\, a}_n$, $\zeta^{I\, \dbi}_n$ which will be discussed in more detail in the following.

Let us summarize the complete multiplets of rigid $\cN = 2$
supersymmetry  originating
from the $\cN = 4$ spectrum of section \ref{6d_and_N4}. 
Firstly, we find the vector multiplets 
\beq \label{spec_hatcV}
\widehat \cV^I=(A^I_\mu,\phi^I, \chi^{I\,a}, Y_{ab}^I) \equiv (A^I_{0\, \mu}, \phi^I_0, \chi^{I\, a}_0 , Y^I_{0\, ab}) \ .
 \eeq 
The vector $A^I_\mu$ is still identified with the gauge connection. The real scalar $\phi^I$ is a singlet 
$(\mathbf 1 , \mathbf 1)$ under $SU(2)_R \times SU(2)$
and originates from $\sigma^{I\, ij}$. The spinor $\chi^{I\, a}$ belongs to the $( \mathbf 2 , \mathbf 1)$
representation and comes from the decomposition of $\lambda^{I\,i}$.
The scalars $Y^I_{ab}=Y^I_{ba}$ are auxiliary fields of the $\cN = 2$ superconformal
formalism and transform in the $(\mathbf 3, \mathbf 1)$ representation. They would arise from the 
decomposition of auxiliary fields in the (linearized) off-shell $\cN=4$ vector multiplet (see e.g.~\cite{Howe:1981ev}) that 
transform in higher irreducible representations of $USp\,(4)_R$. The multiplets $\widehat \cV^I$ are pseudo-real,
\begin{equation} \label{5d_N2_reality}
 (\widehat \cV^I)^* = \widehat \cV^I \ : \quad
\begin{cases}
 \ \ \bar A^I_\mu \equiv (A^I_\mu)^* = A^I_\mu \ , \\[.1cm]
 \ \ \bar\phi^I \equiv (\phi^I)^* = \phi^I \ , \\[.1cm]
 \ \ \bar\chi^{I\, a} \equiv (\chi_a^I)^\dagger \gamma^0  = \epsilon^{ab} (\chi_b^I)^{\sf T} C \ , \\[.1 cm]
 \ \ \bar Y^{I\, ab} \equiv (Y^I_{ab})^* = \epsilon^{ac} \epsilon^{bd} Y^I_{cd} \ , \hspace{2 cm} 
\end{cases}
\end{equation}
where 
 $\epsilon^{ab}$ is the primitive antisymmetric invariant of $SU(2)_R$.

Secondly, in a completely
analogous fashion we have the tensor multiplets
 \beq    \label{spec_hatcT}
    \widehat  \cT_n^I = (F_{n\, \mu\nu}^{I},\phi^{I}_n,\chi_n^{I\, a}, Y^I_{n\, ab})\ , \qquad n>0 \ ,
 \eeq 
with the scalars $\phi^I_n$ in the $(\mathbf 1, \mathbf 1)$ representation,
the spinors $\chi^{I\, a}$ in the $(\mathbf 2, \mathbf 1)$ representation,
and the auxiliary fields $Y^I_{n\, ab}$ in the $(\mathbf 3, \mathbf 1)$ representation.
 In contrast to their counterparts in $\widehat \cV^I$, all fields 
in $\widehat \cT_n$ are complex and will become massive
after breaking of conformal invariance, as discussed in more detail below.

 Finally, we find the hypermultiplets
\beq \label{spec_hatcH}
  \widehat \cH^I_0 \equiv \widehat \cH^I = (q^{I \,a\dbi},\zeta^{I\,\dbi}) \equiv 
(q^{I \,a\dbi}_0,\zeta^{I\,\dbi}_0)\ , \qquad 
 \widehat \cH^I_n = (q^{I \,a\dbi}_n,\zeta^{I\,\dbi}_n)\ ,\qquad n > 0\ . 
\eeq
They consist of scalars $q^{I \,a\dbi}_n$ that are the $(\mathbf 2, \mathbf 2)$ component
of $\sigma_n^{I\,ij}$ under the branching \eqref{rep_decomp}, and of spinors $\zeta^{I\,\dbi}_n$
that belong to the $(\mathbf 1, \mathbf 2)$ branch in the reduction of $\lambda^{I\, i}_n$. 
For $n>0$ the hypermultiplet is complex and massive
(in the broken phase of conformal symmetry). For $n=0$ it is massless and pseudo-real,
\begin{equation} \label{5d_N2_hyper_reality}
 (\widehat \cH^I)^* = \widehat \cH^I \ : \quad
\begin{cases}
 \ \ \bar q^I_{ a \da} \equiv (q^{I\, a \da})^* = \epsilon_{ab} \epsilon_{\da \dbi} q^{I\, b \dbi} \ , \\[.1 cm]
 \ \ \bar \zeta^{I\, \da} \equiv (\zeta^I_{\da})^\dagger \gamma^0 = \epsilon^{\da \dbi} \zeta^{I\,\sf T}_{\dbi} C \ , 
\end{cases}
\end{equation}
where we have made use of the primitive antisymmetric
invariants $\epsilon^{ab}, \epsilon^{\da \dbi}$ of $SU(2)_R$ and $SU(2)$.
The Weyl weights of all the fields introduced in this section
are collected in Table \ref{N2_table}, along with a summary of $SU(2)_R \times SU(2)$ 
representations. The matching of the Weyl weights of $\cN=4$ fields and $\cN=2$ fields will 
be discussed in the next subsection.

\begin{table} 
\centering 
 \begin{tabular}{|c |c |c |c |c | c|}
 \hline
\rule[-.2cm]{0cm}{0.7cm}  \! \! \! multiplet  \! \! \!  & fields  & type & comments & $SU(2)_R \times SU(2)$ rep & Weyl weight  \\ \hline \hline
 \multirow{4}{*}{\vspace{-.7cm} \parbox{3cm}{\begin{center}$\widehat \cV^I$\\massless vector\\multiplet\end{center}}} 
  \rule[-.2cm]{0cm}{0.7cm} & $A^I_\mu \equiv A^I_{0\, \mu}$ & vector & pseudo-real &  $(\mathbf 1 , \mathbf 1)$ & 0  \\
  \rule[-.2cm]{0cm}{0.7cm} & $\phi^I \equiv \phi^I_0$ & scalar & pseudo-real &  $(\mathbf 1 , \mathbf 1)$ & 1 \\
  \rule[-.2cm]{0cm}{0.7cm} & $\chi^{I\, a} \equiv \chi^{I\, a}_0$ & spinor & pseudo-real &  $(\mathbf 2, \mathbf 1)$ & 3/2 \\ 
  \rule[-.2cm]{0cm}{0.7cm} & $Y^I_{ab} \equiv Y^I_{0\, ab}$ & scalar & auxiliary &  $(\mathbf 3, \mathbf 1)$ & 2 \\ \hline
 \multirow{2}{*}{\vspace{-.1cm} \parbox{3cm}{\begin{center}\vspace{-0.3cm}$\widehat \cH^I \equiv \widehat \cH^I_0$\\massless hyperm.\vspace{0.3cm}\end{center}}}  
  \rule[-.2cm]{0cm}{0.7cm} & $q^{I\, a \dbi} \equiv q_0^{I\, a \dbi}$ & scalar & pseudo-real &  $(\mathbf 2 , \mathbf 2)$ & 3/2  \\
  \rule[-.2cm]{0cm}{0.7cm} & $\zeta^{I\, \da} \equiv \zeta^{I\, \da}_0$ & spinor & pseudo-real &  $(\mathbf 1, \mathbf 2)$ & 2 \\ \hline
 \multirow{4}{*}{\vspace{-.7cm} \parbox{3cm}{\begin{center}$\widehat \cT^I_n$\\massive tensor \\ multiplet\end{center}}} 
  \rule[-.2cm]{0cm}{0.7cm} & $F^I_{n\, \mu\nu}$ & tensor & complex &  $(\mathbf 1 , \mathbf 1)$  & 0\\
  \rule[-.2cm]{0cm}{0.7cm} & $\phi^I_n$ & scalar & complex &  $(\mathbf 1 , \mathbf 1)$ & 1 \\
  \rule[-.2cm]{0cm}{0.7cm} & $\chi^{I\, a}_n$ & spinor & complex &  $(\mathbf 2, \mathbf 1)$ & 3/2 \\ 
  \rule[-.2cm]{0cm}{0.7cm} & $Y^I_{n\, ab}$ & scalar & auxiliary &  $(\mathbf 3, \mathbf 1)$ & 2 \\ \hline
\multirow{2}{*}{\vspace{-.1cm} \parbox{3cm}{\begin{center}\vspace{-0.3cm}$\widehat \cH^I_n$\\ massive hyperm.\end{center}}}  
 \rule[-.2cm]{0cm}{0.7cm} & $q^{I\, a \dbi}_n$ & scalar & complex &  $(\mathbf 2 , \mathbf 2)$ & 3/2  \\
  \rule[-.2cm]{0cm}{0.7cm} & $\zeta^{I\, \da}_n$ & spinor & complex &  $(\mathbf 1, \mathbf 2)$ & 2 \\ \hline\hline
 \multirow{4}{*}{\vspace{-.6cm}
\parbox{3cm}{\begin{center}$\widehat \cV^0$\\massless vector
\\ multiplet \end{center}}
}   \rule[-.2cm]{0cm}{0.7cm} & $A^0_\mu$ & vector & pseudo-real &  $(\mathbf 1 , \mathbf 1)$ & 0  \\
  \rule[-.2cm]{0cm}{0.7cm} & $\phi^0$ & scalar & pseudo-real &  $(\mathbf 1 , \mathbf 1)$ & 1 \\
 \rule[-.2cm]{0cm}{0.7cm} &  $\chi^{0\, a}$ & spinor & pseudo-real &  $(\mathbf 2, \mathbf 1)$ & 3/2 \\ 
  \rule[-.2cm]{0cm}{0.7cm} & $Y^0_{ab}$ & scalar & auxiliary &  $(\mathbf 3, \mathbf 1)$ & 2 \\ \hline 
\end{tabular}
\caption{Field content of $\cN=2$ vector multiplet $\widehat \cV$, tensor multiplets $\widehat \cT_n$ and hypermultiplets $\widehat \cH, \widehat \cH_n$ in five dimensions. 
The additional multiplet $\widehat \cV^0$ is included. The precise 
formulation of the pseudo-reality properties of the fields in $\widehat \cV^I, \widehat \cH^I$
is found in \eqref{5d_N2_reality} and \eqref{5d_N2_hyper_reality}, respectively.
The specification `massless' or `massive' applies to the broken phase of conformal symmetry.
}
\label{N2_table}
\end{table}

\subsection{Restoration of five-dimensional conformal symmetry} \label{sec:conformal_symmetry}

It is important to clarify the role of conformal symmetry 
in our discussion. Our goal is a five-dimensional
action that is able to capture some crucial ingredients
of a non-Abelian $(2,0)$ model. This six-dimensional
theory is invariant under rigid conformal transformations \cite{Claus:1997cq,Bergshoeff:1999db}, 
i.e.~transformations that leave the six-dimensional line-element
invariant up to a factor. We refrain from a complete account on the transformation 
properties of the six-dimensional fields. In our discussion we restrict our
attention mostly to the Weyl weights of the fields as listed in Table \ref{6d_table}.

If we compactify the six-dimensional theory on a circle using \eqref{KK_metric}, we expect some generators of 
the six-dimensional conformal symmetry to be spontaneously broken. The remaining generators are those which act 
only on the five-dimensional line element. In particular, the Weyl invariance discussed above will be broken, 
unless we also allow for a rescaling of the compactification radius, i.e.~unless we would consider 
transformations of the form
\beq  \label{WeylRescaling}
 g_{\mu\nu}dx^\mu d x^\nu = ds^2 \quad \mapsto \quad \Omega^{-2}ds^2\,, 
\qquad \qquad \quad r \quad  \mapsto \quad  \Omega^{-1} r\,.
\eeq
Another way to see that Weyl invariance is compromised in the dimensionally reduced theory 
is to notice that the multiplets $\widehat \cT_n, \widehat \cH_n$ have become massive with 
masses $m_n$ given in \eqref{KK_mass}. Since Weyl invariance is incompatible with massive fields, the Kaluza-Klein 
masses $m_n$ break conformal invariance explicitly. This 
can be remedied by allowing them to transform as $m_n\mapsto \Omega\; m_n$ as 
can be inferred from \eqref{WeylRescaling}.

Note that the $\cN=2$ Poincar\'e supersymmetry algebra
does admit a superconformal extension, given by 
the exceptional superalgebra $F^2(4)$ \cite{Kac:1977em, Nahm:1977tg}.
This is in contrast with the $\cN=4$ case considered
before. 
In practice, five-dimensional rigid superconformal invariance is restored by introducing additional 
five-dimensional degrees of freedom. Indeed, we can promote the radius $r$ to the scalar component 
of a full $\cN=2$ vector multiplet
\beq \label{def-cV0}
  \widehat \cV^{0}= (A^0_\mu, \phi^0,\chi^{0\, a} , Y^0_{ab}) \ ,
\eeq
where $A^0_\mu$ can be identified with the Kaluza-Klein vector introduced in (\ref{KK_metric}). 
We can combine this vector multiplet with the physical vector multiplets
introduced in the last section and denote them collectively as
\beq
\widehat \cV^{\widehat I} = (\widehat \cV^0 , \widehat \cV^I) \ , \qquad \quad \widehat I = 0 , 1 \dots , |G| \ .
\eeq

Using the multiplet $\widehat \cV^0$ we can make the $\cN=4$ to $\cN=2$ split 
of the spectrum more explicit. 
We follow the split \eqref{rep_decomp} and we match 
the scaling weights of Table \ref{N4_table} 
with the Weyl weights of Table \ref{N2_table}
to infer that the proper map from $\cN=4$ to $\cN=2$ multiplets is of the form
\beq \label{schematic_sigma}
\sigma^{I\, ij}_n \mapsto 
\begin{pmatrix}
 \tfrac{1}{\sqrt 2} \epsilon^{ab} \, \phi^I_n &
 (\phi^0)^{-1/2} q^{I\, a \dbi}_n \\
 - (\phi^0)^{-1/2} q^{I\,  b \da}_n & 
 - \tfrac{1}{\sqrt 2} \epsilon^{\da \dbi} \, \phi^I_n
\end{pmatrix} \ , \qquad \quad
\lambda^{I\, i}_n \mapsto 
\begin{pmatrix}
 \chi^{I\, a}_n \\
 \sqrt 2 (\phi^0)^{-1/2} \zeta^{I\,\da}_n
\end{pmatrix} \ , \qquad
 n \ge 0 \ .
\eeq
Prefactors are chosen for later convenience. Note that the split \eqref{R_breaking} is not unique. 
The different allowed decompositions 
of the indices $i$ into $a,\dbi$  will play a crucial role in section \ref{sec:N=4completion}.

In the action of section \ref{N=2superconf_action} the additional multiplet $\widehat \cV^0$ will couple to 
all other multiplets making the action superconformally invariant.
To give a direct link with the Kaluza-Klein reduction it will be
convenient to return to the broken phase of the superconformal symmetry by 
setting the additional fields to a fixed value. This requires to set 
\beq \label{gauge_fixing}
    \langle \phi^0 \rangle = \frac{1}{r} = \frac{1}{g^2}\ , \qquad 
\langle\chi^{0\,a} \rangle = \langle Y^0_{ab}\rangle  =\langle A^0_\mu \rangle =0 \ ,
\eeq
where $g$ is the gauge coupling of the five-dimensional Yang-Mills theory. 
It is important to stress that imposing the condition \eqref{gauge_fixing}
corresponds to a restriction of the theory. Indeed not 
all values of $\chi^{0\,a},Y^0_{ab}$ and $A^0_\mu$ can be mapped by a superconformal transformation to zero. Nevertheless 
we will show below that a Poincar\'e supersymmetric theory arises after imposing \eqref{gauge_fixing}.
Moreover, the Weyl rescaling \eqref{WeylRescaling} of $r$, as dictated by the six-dimensional conformal symmetry, 
is precisely compatible with the Weyl weight of $\phi^0$ in the identification \eqref{gauge_fixing}.
In fact, we will show that the five-dimensional action still retains a scaling symmetry if the radius is 
rescaled as in \eqref{WeylRescaling}.

In the broken phase of conformal symmetry determined by \eqref{gauge_fixing}
the hypermultiplets fields $q^{I\, a \dbi}_n, \zeta^{I\, \da}_n$
are not convenient variables, since their mass dimensions
are not canonical. As a consequence, we define the rescaled fields
\begin{equation} \label{rescaled_hypers}
 h^{I\, a \dbi}_n = g \, q^{I\, a \dbi}_n \ , \qquad
 \psi^{I\, \da}_n = g \, \zeta^{I\, da}_n \ , \qquad
 n \ge 0 \ ,
\end{equation}
in such a way that all scalars have mass dimension and scaling weight $1$,
and all fermions have mass dimension and scaling weight $3/2$.

%%%%%%%%%%%%%%%%%%%%%%%%%%%%%%%%%%%%%%%%%%%%%%%%%%%%%%%%%%%%%%%%%

\section{Supersymmetric actions and conformal invariance} \label{sec:susyactions}

In this section we introduce a five-dimensional non-Abelian $\cN=2$ supersymmetric action 
for the Kaluza-Klein spectrum obtained in section \ref{N=2spectrum}. Our 
theory will include couplings which are only specified in terms 
of group theoretical constants and the Kaluza-Klein levels. An $\cN=2$ superconformal
action is presented in section \ref{N=2superconf_action}, while the 
Kaluza-Klein sums are made explicit in a restricted action in section \ref{sec:KKaction}. 
We propose to interpret this theory as an $\cN=2$ subsector
of a dimensionally reduced $(2,0)$ theory.

\subsection{An $\cN=2$ superconformal action for the Kaluza-Klein spectrum} \label{N=2superconf_action}

In the following we introduce an $\cN=2$ superconformal action for the spectrum discussed in 
section \ref{N=2spectrum}. Superconformal invariance is retained since we will 
include the additional vector multiplet $\widehat \cV^0$, defined in \eqref{def-cV0},
containing the radius and the Kaluza-Klein vector.
It will be necessary to introduce some additional notation in order to 
make contact with the general $\cN=2$ superconformal actions introduced in \cite{Bergshoeff:2002qk}.
The fields identified with the Kaluza-Klein zero modes are denoted as in Table \ref{N2_table}:
\beq \label{zero-modes_list}
   \text{vector multiplets:}\quad (A^0_\mu ,\phi^0, \chi^{0\,a}),\ \ (A^I_\mu ,\phi^I, \chi^{I\,a})\qquad \quad   
   \text{hypermultiplets:} \quad (q^{I a\dbi},\zeta^{I\dbi})\ .
\eeq
For the fields identified with excited Kaluza-Klein modes it will be 
convenient to use the notation \eqref{compl_splitting} for complex fields
introducing the $SO(2)_{\rm KK}$ index $\alpha$. This leads us 
to the following excited spectrum: 
\bea \label{excited-modes_list}
   \text{tensor multiplets:} && \big( F^{\{I \alpha n \} }_{\mu\nu}, \phi^{\{ I \alpha n \} }, \chi^{\{I \alpha n \} a} \big) \equiv 
  \big(F^{I \alpha}_{n\, \mu\nu}, \phi^{I \alpha}_n, \chi^{I \alpha\, a}_n \big)\nn \\[.1cm]
   \text{hypermultiplets:} && \big(q^{\{ I \alpha n \} a\dbi},\zeta^{\{I \alpha n\}\dbi} \big) \equiv \big(q^{I \alpha\, a\dbi}_n,\zeta^{J \alpha\, \dbi}_n \big) \ .
\eea
The main complication in the notation arises from the multi-index $\{I\alpha n \}$ which 
labels simultaneously the non-Abelian components $I,J =1,\ldots, |G|$, the 
$SO(2)_{\rm KK}$ labels $\alpha,\beta=1,2$, and the Kaluza-Klein levels $n,m \ge 1$. 
To avoid cluttering of indices in the following expressions we will 
denote this multi-index by 
\beq \label{multi-index}
   M = \{I\alpha n \}\ , \qquad N = \{J \beta m\}\ .
\eeq
A summation over $M,N$ then always amounts to summing over all indices including the 
infinite tower of Kaluza-Klein modes.
We will present the superconformal action as function of the four types of multiplets in 
\eqref{zero-modes_list} and \eqref{excited-modes_list}. 
To do that in an efficient way it is useful to introduce the following index combinations
\beq
    \widehat I \equiv (0,I) \quad \qquad \Lambda \equiv (0, I, M )\ ,\qquad \cI = (I, M)
\eeq
This means that $\widehat I,\widehat J,...$ label all vector multiplets and run over $|G|+1$ values,
$\Lambda, \Sigma,...$ run over all tensor and 
vector multiplets including the Kaluza-Klein tower. 
The indices $\cI,\cJ,...$ label all hypermultiplets, or vectors and tensor multiplets without $\mathcal{V}^0$. Finally, we also define 
\beq
   F^{\Lambda}_{\mu\nu} \equiv  (F^{\widehat I}_{\mu\nu}, F^{M}_{\mu\nu} ) \equiv 
 (F^{\widehat I}_{\mu\nu},  n \epsilon_{\beta \gamma} \delta^{\gamma \alpha} B^{I \beta}_{n\, \mu\nu} )\ .
\eeq
where we have recalled the definition of $F^M_{\mu\nu} = F^{I\alpha }_{n\, \mu\nu}$ as given in \eqref{def-Fn}. 
It is crucial to stress that the Kaluza-Klein interpretation dictates this non-trivial identification 
of $F^M_{\mu\nu}$ with $B_{n\, \mu\nu}^{I\alpha}$.
The important point is that while the $F^M_{\mu\nu}$ admit a rescaling with the Kaluza-Klein level compared to $B_{n\, \mu\nu}^{I\alpha}$, the 
scalars and fermions in the same multiplet are trivially matched with the $\cN=2$ formalism of \cite{Bergshoeff:2002qk}.\footnote{This implies that compared to 
\cite{Bergshoeff:2002qk} one has to adjust the notation, since there $F^M_{\mu\nu}$ and $B^M_{\mu\nu}$ are trivially identified.} The non-trivial 
rescaling of $B_{n\, \mu\nu}^{I\alpha}$ turns out to be consistent with the dimensional reduction
of the supersymmetry variations as can be checked for the Abelian six-dimensional theory recorded in Appendix \ref{ab_tensors_app}.

We are now in the position to discuss the Lagrangian in detail. The vector-tensor sector of 
an $\cN=2$ superconformal theory can be specified
by introducing a constant symmetric object $C_{\Lambda \Sigma \Theta}$, a constant anti-symmetric 
matrix $\Omega_{MN}$, and 
the gauge parameters $t_{\widehat K \Lambda }{}^{\Sigma}$ \cite{Bergshoeff:2002qk}. The
gauge parameters appear in the covariant derivatives 
\begin{align}  \label{N=2gaugings}
 \mathcal{D}_\mu \phi^{\Sigma} &=  \partial_\mu \phi^{\Sigma} 
+  t_{\widehat K \Lambda}{}^{\Sigma}  \ A_\mu^{\widehat K}  \ \phi^{\Lambda}\ , \nonumber\\[.1cm]
\mathcal{D}_\mu \chi^{a \Sigma} &= \partial_\mu \chi^{a \Sigma}  
+ t_{\widehat K \Lambda}{}^{\Sigma}  \ A_\mu^{\widehat K}\ \chi^{a \Lambda},\nonumber\\[.1cm]
\cD_\mu  F^N_{\nu\rho} &=  \partial_\mu F_{\nu \rho}^N +  t_{\widehat K M}{}^N \ A_\mu^{\widehat K} \ F_{\nu\rho}^M \ .
\end{align}
Note that strictly speaking only $C_{\widehat I\widehat J\widehat K}$ encodes extra information in addition to $\Omega_{MN}, t_{\widehat K \Lambda }{}^{\Sigma}$.
This is due to the fact that $C_{M \Lambda \Sigma}$ are given by
\beq \label{Ctensor}
   C_{M \Lambda \Sigma} =  t^{\phantom{M}}_{(\Lambda \Sigma)}{}^{N} \Omega^{\phantom{ M}}_{NM}\ , 
\eeq
where one symmetrizes in the indices $\Lambda , \Sigma$ including the usual factor $1/2$.
Here we have extended the range of indices on generators $t^{\phantom{M}}_{\Lambda \Sigma}{}^{\Theta}$ with 
the constraints
\beq
    t_{(\Lambda \Sigma)}{}^{\widehat I} = 0 \ ,\qquad t_{M \Sigma}{}^{\Theta} =0\ ,
\eeq
implying the absence of gaugings with a tensor index $M$.

Since we will later propose to use the $\cN=2$ superconformal theory to describe the dimensional reduced $(2,0)$ action 
we aim to use only couplings which are of group theoretic origin. We like to identify a subsector of 
the theory as $\cN=4$ super-Yang-Mills theory. This implies that components of 
$C_{\widehat I\widehat J\widehat L}$ have to encode the trace $d_{IJ}=C_{0IJ}$. The coupling $C_{000} = k_{\rm c}$ will determine the 
kinetic term of the auxiliary vector multiplet $\widehat \cV^0$, and will be left undetermined at the moment. 
We choose $C_{00I}=0$. More interesting are the couplings of the tensor multiplets. Here we are guided by \eqref{Ctensor}. 
To determine the gaugings we first note that the fields in $\widehat{\cV}^0$ cannot be gauged, such that 
$t_{\widehat I\, 0}{}^{\, \Lambda}= t_{\widehat I \Lambda}{}^{0}=0$.
Comparing the gaugings \eqref{SO2_covariant_derivative}, \eqref{covariant_derivative_gen} with \eqref{N=2gaugings}, we consider the following 
identification:
\beq \label{def-tI}
    t_{K\cI}{}^{\cJ} =   \left(\begin{array}{cc}
                                     t_{KI}{}^J & 0\\
                                    0 &  t_{K M}{}^{N}
                                  \end{array}\right)  =
                          \left(\begin{array}{cc}
                                     t_{KI}{}^J & 0\\
                                    0 &  t_{K \{I\alpha n\}}{}^{\{J \beta m\}}
                                  \end{array}\right) 
                              = \left(\begin{array}{cc}
                                     f_{KI}{}^J & 0\\
                                      0 & f_{KI}{}^J \delta^\beta_\alpha \delta_n^m
                                  \end{array}\right)\ ,
\eeq
and
\beq \label{def-t0}
   t_{0\cI}{}^{\cJ}=   \left(\begin{array}{cc}
                                 0 & 0\\
                                    0 &  t_{0M}{}^{N}
                                  \end{array}\right) =
                               \left(\begin{array}{cc}
                                 0 & 0\\
                                    0 &  t_{0\{I \alpha n\}}{}^{\{J \beta m\}}
                                  \end{array}\right)=   \left(\begin{array}{cc}
                                   0 & 0\\
                                      0 & \KKcharge \, \delta^{J}_{I}\epsilon_{\alpha \gamma} \delta^{\gamma \beta} \delta_n^m
                                  \end{array}\right)\ ,
\eeq
%with $M=\{ I\alpha n \}, N=\{ J\beta m\}$, and 
where $SO(2)_{\rm KK}$ indices have been raised and lowered using $\delta_{\alpha\beta}$. Here $t_{K I}{}^J, t_{I M}{}^N$ parametrize the non-Abelian 
gaugings with the vector zero modes and are thus given by the structure constants of 
$G$. The matrix $t_{0 \cI}{}^{\cJ}$
encodes the gauging of the massive tensor multiplets with $A^0$, which is interpreted 
as charge under the Kaluza-Klein vector.
In addition the anti-symmetric matrix $\Omega_{MN}$ can be read off from the Chern-Simons type kinetic terms 
of the tensors $F^M$ in \eqref{purely_bosonic_action_alphabeta}, and is given by 
\beq \label{def-OmegaMN}
   \Omega_{M N} = \Omega_{\{ I\alpha n \} \{J\beta m \}}= -\frac{2}{\KKcharge} \, d_{IJ} \epsilon_{\alpha \beta} \delta_{nm}\ ,
\eeq
where $n,m \ge 1$ as in the range of the multi-indices \eqref{multi-index}.
As we can see, $U(1)_{KK}\sim SO(2)_{\rm KK}$ plays a key role in the construction of this object. While the trace $d_{IJ}$ is symmetric, one can use the 
indices $\alpha,\beta$ and the anti-symmetric $\epsilon_{\alpha \beta}$, corresponding 
to the complex number $i$, to introduce $\Omega_{MN}$. Using 
\eqref{Ctensor} this will also allow us to introduce the symmetric tensor $C_{M \Lambda \Sigma }$ 
in terms of the anti-symmetric structure constants $f_{IJK} = d_{IL} f^{L}{}_{JK}$. 
To display the result, we introduce the matrix
\beq \label{def_CcIcJ}
    C_{\cI \cJ} = 
                            \left(\begin{array}{cc}
                                      C_{I J} & 0\\
                                      0 &  C_{MN }
                                  \end{array}\right)
                            = \left(\begin{array}{cc}
                                      C_{I J} & 0\\
                                      0 &  C_{\{I\alpha n\} \{ J\beta m \}}
                                  \end{array}\right) =
                                  \left(\begin{array}{cc}
                                      d_{I J} & 0\\
                                      0 &  d_{I J} \delta_{\alpha \beta} \delta_{nm}
                                  \end{array}\right) \ .
\eeq
In summary, taking into account the total symmetry in all three indices, 
all components of $C_{\Lambda \Sigma \Theta}$ are determined by 
\bea \label{form_of_cC}
   && C_{0 \cI \cJ} = C_{\cI \cJ}\ , \qquad C_{000}=k_{\rm c}\ ,\qquad 
 C_{MN K} = C_{\{ I\alpha n \} \{ J\beta m \} K}  =- \frac{1}{\KKcharge} f_{IJK}\, \epsilon_{\alpha \beta} \delta_{nm}\ , \nn \\[.1cm] 
   && C_{00\cI}=C_{IJ\cK} =C_{MNP} = 0\ . 
\eea
In evaluating these expressions we have used that $\epsilon_{\alpha \gamma} \delta^{\gamma \delta} \epsilon_{\delta \beta} = - \delta_{\alpha \beta}$.

Let us now include the hypermultiplets into the discussion. 
In a general $\cN=2$ superconformal theory the hypermultiplets span a hypercomplex manifold.
We choose the geometry of the hypercomplex manifold appearing in the reduction 
to be locally flat space. Since the dimension of this manifold is related to the 
dimension of the gauge group $G$, it posses sufficiently many isometries to
implement a gauging compatible with \eqref{N=2gaugings}.  
In coordinates $q^{a \da\cI}$ the metric is given by $C_{\cI \cJ} \epsilon_{ab} \epsilon_{\da \dbi}$,
with $C_{\cI \cJ}$ as defined in \eqref{def_CcIcJ}.\footnote{The three complex structures on this hypercomplex manifold
are encoded in the SU(2) triplet $J^{c \dc \cI}{}_{d \ddi \cJ}{}_{(ab)}$, where
 $J^{c \dc \cI}{}_{d \ddi \cJ}{}^{a}{}_{b} = \delta_\cJ^\cI \delta_{\ddi}^{\dc} (2  \delta_d^a  \delta_b^c   - \delta_d^c  \delta_b^a)$. }
The kinetic term of the fermionic partners $\zeta^{\da \cI}$ is simply given by $C_{\cI \cJ} \epsilon_{\da \dbi}$.
The gauging of the hyperscalars and fermions is 
\beq \label{q_cov}
  \cD_{\mu} q^{ a\dbi \cJ } =  \partial_\mu q^{a\dbi \cJ } +   t_{\widehat K \cI}{}^{\cJ}\,  A^{\widehat K}_\mu \, q^{a\dbi \cI } \ , \qquad \quad
 \cD_{\mu} \zeta^{\da \cJ }=  \partial_\mu \zeta^{\da \cJ } %+ \partial_\mu q^X \omega_{XB}{}^A
                       +   t_{\widehat K \cI}{}^{\cJ} \, A^{\widehat K}_\mu\, \zeta^{\da \cI }\ ,
\eeq
with constant $ t_{\widehat K \cI}{}^{\cJ}$ given in \eqref{def-tI} and \eqref{def-t0}.\footnote{The moment maps generating these gaugings are given by 
    $P_{\widehat K}{}_{ab} 
              = \tfrac12 C_{\cI \cJ} \, t_{\widehat K \cL}{}^\cJ  \epsilon_{c(a} \  q^{\cI }_{b) \dc}  \,  q^{c \dc\cL} $.}

Using these definitions we can now display the complete non-Abelian $\cN=2$ superconformal Lagrangian
\begin{align}  \label{ActionComplete}
\cL = & \;
 \phi ^{\Theta} C_{ \Theta \Lambda \Sigma} 
\left(- \tfrac 14 F_{\mu \nu}^{\Lambda} F^{ \Sigma \, \mu \nu }
 - \tfrac12 \bar{\chi}^{ \Lambda\, a} \slashed{\cal D} \chi^{ \Sigma}_a 
 - \tfrac12 \cD_\mu \phi^{ \Lambda} \cD^\mu \phi^{\Sigma} +
Y_{ab}^{\Lambda} Y^{{\Sigma}\, ab}  \right)
\nn \\[.1cm]
& + \tfrac{1}{16}  \epsilon^{\mu \nu \lambda \rho \sigma} \Omega_{MN} F_{\mu \nu}^M \cD _\lambda F_{\rho\sigma}^N 
   -\tfrac1{24} \epsilon^{\mu \nu \lambda \rho \sigma} C_{\widehat I \widehat J \widehat K} A_\mu^{\widehat I} F_{\nu \lambda}^{\widehat J} F_{\rho \sigma}^{\widehat K}  \nn \\[.1cm]
& -\tfrac i 8  C_{ \Lambda \Sigma \Theta} 
\Big(  
\bar{\chi}^{\Lambda\, a} \gamma^{\mu \nu} F^{\Sigma}_{\mu \nu} \chi_a^{\Theta} 
     +4  \bar{\chi}^{\Lambda\, a} \chi^{b \Sigma} Y_{ab}^{\Theta}
  \Big)
\nn \\[.1cm]
&  + \tfrac i 4 \,
  \phi^{\Theta} C_{ \Theta \Lambda \Sigma} 
\Big(
 t_{[ \Upsilon \Omega ]}{}^{\Lambda} \bar{\chi}^{\Upsilon\, a}\chi ^{\Omega}_a
\phi^{\Sigma}
-4  t_{( \Upsilon \Omega )}{}^{\Lambda} \bar{\chi}^{\Upsilon\,a} \chi ^{\Sigma}_a
\phi^{\Omega}
\Big) \nn \\[.1cm]
& - \tfrac12 \phi^{\widehat K}  C_{\widehat KMN} t_{\widehat I P}{}^M   t_{\widehat J Q}{}^N  \phi^{\widehat I} \phi ^{\widehat J} \phi ^{P} \phi^{Q}
\nn \\[.1cm]
& + C_{\cI \cJ} \Big (
-\tfrac12  \cD_\mu q^{\cI\, a \dbi} \cD^\mu q^{\cJ}_{\;a \dbi}
 -  \bar \zeta^{\cI\, \dbi } {\slashed \cD} \zeta^\cJ_{\;\dbi} 
\Big)
\nn \\[.1cm]
 &  +  C_{\cI \cJ} \Big( 
  2 i t_{\widehat K \cL}{}^{\cI} q^{\cL\, a \dbi } \bar\chi^{\widehat K }_a \zeta^{\cJ}_{\;\dbi} 
   + i \phi^{\widehat K} t_{\widehat K \cL}{}^\cI \bar \zeta^{\cJ\, \da} \zeta^\cL_{\da} 
\Big)
\nn \\[.1cm]
& + C_{\cI \cJ} \Big( 
  \, t_{\widehat K \cL}{}^\cJ  q^{\cL \,a \dc}  \,  q^{\cI \, b}{}_{\dc}   Y^{\widehat K}_{ab} 
  - \tfrac12 t_{\widehat I \cK}{}^{\cI} t_{\widehat J \cL}{}^{\cJ}   \phi^{\widehat I} \phi^{\widehat J}  q^{\cK a \dbi}  q^{\cL}_{\;a \dbi} 
\Big) \ .
\end{align}
This Lagrangian transforms with weight $5$ under Weyl rescalings of the fields 
with weights listed in Table \ref{N2_table}. Since the line element has Weyl weight $-2$ as in \eqref{WeylRescaling} 
this implies invariance of the 
five-dimensional action. Furthermore, the Lagrangian \eqref{ActionComplete} is invariant 
under the supersymmetry transformations parametrized by $\epsilon_a$ and the special supersymmetry transformations 
parametrized by $\eta_a$ given by \footnote{
The expression for $\delta F^{\Lambda}_{\mu\nu}$ with $\Lambda = \widehat I$ is 
not independent from the expression for $\delta A^{\widehat I}_\mu$. 
To check their compatibility, note that the second term 
in $\delta F^{\widehat I}_{\mu\nu}$ vanishes thanks to $t_{(\Lambda \Sigma)}{}^{\widehat I} = 0$.
In order to get
the third term in $\delta F^{\widehat I}_{\mu\nu}$,
one has to promote $\delta A^{\widehat I}_\mu$ to its full $x$-dependent form  before
taking the covariant derivative. As explained in \cite{Bergshoeff:2002qk}, this is done by means of the prescription 
$\epsilon^a \mapsto \epsilon^a + i x^\rho \gamma_\rho \eta^a$. The covariant derivative can thus act on an $x$-linear
term in $\delta A^{\widehat I}_\mu$ and produce the $\eta$-term in $\delta F^{\widehat I}_{\mu\nu}$.
}\begin{align}  \label{susy_trafo}
  \delta \phi^{\Lambda} &= \tfrac i2 \, \bar \epsilon^a \chi^{\Lambda}_a\ ,
\nn \\[.1cm]
   \delta  A^{\widehat I}_{\mu} &= \tfrac12 \, \bar \epsilon^a \gamma_{\mu} \chi^{\widehat I}_a \ ,
  \nn \\[.1cm]
   \delta  F^{\Lambda}_{\mu \nu} &= - \bar \epsilon^a \gamma_{[\mu} \cD_{\nu]} \chi^{\Lambda}_a
                   + { i}   t_{(\Sigma \Theta)}{}^{\Lambda} \phi^{\Sigma} \,  \bar \epsilon^a \gamma_{\mu \nu} \chi^{\Theta}_a
                   + { i} \bar \eta^a \gamma_{\mu \nu} \chi^{\Lambda}_a  \ ,
\nn \\[.1cm]
   \delta \chi^{\Lambda \, a} &= - \tfrac14 \gamma^{\mu \nu} F^{\Lambda}_{\mu \nu} \epsilon^a
                   - \tfrac i2  {\slashed \cD} \phi^{\Lambda} \epsilon^a - Y^{ \Lambda \, ab} \epsilon_b
                   + \tfrac12\, t_{(\Sigma \Theta)}{}^{\Lambda} \phi^{\Sigma} \phi^{\Theta} \epsilon^a + \phi^{\Lambda} \eta^a \ ,
 \nn \\[.1cm]
    \delta Y^{\Lambda \, ab} &= -\tfrac12 \bar \epsilon^{(a|} {\slashed \cD} \chi^{\Lambda|b)}
                 - \tfrac i2
                 \Big( t_{[\Sigma \Theta]}{}^{\Lambda} - 3   t_{(\Sigma \Theta)}{}^{\Lambda} \Big)
                 \phi^{\Sigma} \,  \bar \epsilon^{(a|} \chi^{\Theta |b) }
                 + \tfrac i2 \bar \eta^{(a} \chi^{\Lambda| b)}\ ,
\nn \\[.1cm]
  \delta  q^{\cI \, a \dbi} & = - { i} \bar \epsilon^a \zeta^{\cI \, \dbi} \ ,
\nn \\[.1cm]
   \delta  \zeta^{\cI \, \dbi} &= \tfrac i2  {\slashed \cD} q^{\cI\, a \dbi} \epsilon_a
                    - \tfrac12  \phi^{\widehat K} t_{\widehat K \cJ}{}^{\cI} q^{\cJ \, a \dbi} \epsilon_a
                    - \tfrac32 q^{\cI \, a \dbi} \eta_a  \ .
\end{align}
These transformation rules are consistent with Weyl rescalings of Table \ref{N2_table}, if one assigns 
Weyl weight $-1/2$ to the parameter $\epsilon^a$, and the weight $+1/2$ to $\eta^a$. Note that the gamma-matrices
with lower indices $\gamma_{\mu_1 \ldots \mu_k}$ scale with weight $-k$.

This completes the specification of the five-dimensional superconformal action in terms of the group theory invariants $d_{IJ}$, $f_{IJK}$, and 
the tensors $\delta_{\alpha \beta},\epsilon_{\alpha \beta}$ for complex fields parameterizing the full Kaluza-Klein tower. The crucial insight 
is that it is possible to combine the symmetric $d_{IJ}$ and the anti-symmetric $\epsilon_{\alpha \beta}$
to define the anti-symmetric $\Omega_{MN}$ as in \eqref{def-OmegaMN} for the massive 
Kaluza-Klein modes which naturally are complex fields.  This also permits us 
to combine the totally anti-symmetric $f_{IJK}$ and the anti-symmetric $\epsilon_{\alpha \beta}$ to define components of 
the totally symmetric $C_{\Lambda \Sigma \Theta}$. This implies that 
the non-Abelian version of the Kaluza-Klein theory fits naturally in the framework of $\cN=2$ supersymmetry. 
Furthermore, superconformal invariance can be implemented by introducing the 
vector multiplet $\widehat \cV^0$ defined in \eqref{def-cV0}. 

To close this section let us comment on the role of the additional multiplet $\widehat \cV^0$ in 
more detail. We have found that its kinetic term is determined by the constant $k_{\rm c}$. 
Identifying $\phi^0$ with the radius $r$ as in \eqref{gauge_fixing}, one 
can derive the kinetic term of $r$ after dimensional reduction of a six-dimensional gravity theory.
This is complicated by the fact that the proper supersymmetric fields in five dimensions
involve rescalings with $r$ as described in detail in \cite{Bonetti:2012fn}. However, the 
choice 
\beq 
  k_{\rm c}=0 
\eeq 
is natural from the point of view of $\cN=4$ supersymmetry, since 
a Chern-Simons term $k_{\rm c}\, A^0 \wedge F^0 \wedge F^0$ is absent in this case. Moreover,
$k_c=0$ is consistent with non-dynamical gravity in six dimensions.
In the following discussion we work in the phase 
with \eqref{gauge_fixing} implying that $k_{\rm c}$ drops from the action.

\subsection{Supersymmetric Kaluza-Klein Lagrangian in the broken phase} \label{sec:KKaction}

We are now in the position to present the $\cN=2$ action including all 
Kaluza-Klein levels. This amounts to restoring the Kaluza-Klein indices 
for the fields and summing up an infinite tower of multiplets $(B^{I\alpha}_{n\, \mu\nu}, \phi^{I\alpha}_n, \chi^{I\alpha\, a}_n )$ 
and $(q^{I\alpha\, a\dbi}_n,\zeta^{I\alpha\, \dbi}_n) $ in \eqref{ActionComplete}.
The resulting action is straightforwardly obtained but rather lengthy due to the fact 
that both $C_{\Lambda \Sigma \Theta}$ and $\Omega_{MN}$ 
appear in copies labeled by Kaluza-Klein indices. The result simplifies, however,
if we set $\widehat \cV^0$ to the values \eqref{gauge_fixing},
thus moving to the broken phase of conformal invariance.
Discussing the resulting action will be the task of this section.

As discussed already in section \ref{N=2spectrum}, the Abelian vector multiplet $\widehat \cV^0$
plays a special role in the $\cN=2$ spectrum. In a Kaluza-Klein theory $\widehat \cV^0$
has to be interpreted as part of the gravity multiplet with $A^0$ being the 
graviphoton under which all excited Kaluza-Klein modes are charged. 
We decouple gravity completely by imposing the 
condition \eqref{gauge_fixing}. As we will argue below, ordinary
$\cN=2$ supersymmetry is preserved despite the breaking of superconformal invariance. 
Furthermore, we make use of the rescaled hypermultiplet fields $h^{I\, a \dbi}_n, \psi^{I\, \da}_n$
defined in \eqref{rescaled_hypers}.

The resulting Lagrangian including all Kaluza-Klein modes listed in Table \ref{N2_table}
takes the form
\beq \label{cL_split}
\cL = \cL_0 + \sum_{n=1}^{\infty} \Re \cL_n \ ,
\eeq
where $\cL_0$ only involves massless multiplets, while $\cL_n$ collects
all terms constructed with the $n$th excited modes. We discuss $\cL_0$ and $\cL_n$ 
in turn.

To begin with, let us display the zero mode Lagrangian
\begin{align} \label{cL0-aux}
g^2 \cL_ 0  = & \;
 d_{IJ} \Big[ 
- \tfrac 14   F^{I\, \mu\nu} F^J_{\mu\nu}
- \tfrac 12   \cD^\mu \phi^I \cD_\mu \phi^J 
- \tfrac 12    \cD^\mu h^{I\, a \dbi} \cD_\mu h^J_{\, a \dbi} 
- \tfrac 12   \bar\chi^{I\,a} \slashed{\cD} \chi^J_{\, a} 
-   \bar \psi^{I\, \da} \slashed{\cD} \psi^J_{\, \da} 
+   Y^{I\,ab} Y^J_{\, ab} \Big] \nn \\[.1cm]
& +  f_{IJK} \Big[
 +   \tfrac{i}{2}   \phi^I    \bar\chi^{J\, a} \chi^K_{\,a} 
- i  \phi^I \bar\psi^{J\, \da} \psi^K_{\, \da} 
- 2 i  h^{I\, a \dbi} \bar\chi^J_a \psi^K_{\, \dbi} 
 +   h^{I\,a \dc} h^{J\, b}{}_{\dc} Y^{K}_{\,ab} \Big] \nn \\[.1cm]
& - \tfrac 12  f_{IJ}{}^{H} f_{H KL} \phi^I \phi^K h^{J\, a \dbi} h^L_{\,a \dbi}\ .  
\end{align}
We recognize that the terms contracted with the trace $d_{IJ}$ are the kinetic terms of the massless vectors, scalars and fermions, as well 
as the quadratic term for the auxiliary field. The terms involving the structure constants $f_{IJK}$ are Yukawa-type couplings and 
a scalar potential quartic in the fields $\phi^I,h^{I\, a \dbi}$. 
We stress that for the massless fields such quartic coupling are only possible if they also include scalars $h^{I\, a\dbi}$
due to the asymmetry of $f_{IJK}$. In section \ref{Integrating_out} we will discuss the properties of \eqref{cL0-aux} in more 
detail and relate it to $\cN=4$ supersymmetric Yang-Mills theory.

Let us now turn to the discussion of the Lagrangians $\cL_n$ in \eqref{cL_split} for the Kaluza-Klein tower. 
We insert \eqref{def-tI}-\eqref{form_of_cC} into the action \eqref{ActionComplete}, impose 
the condition \eqref{gauge_fixing}, 
and extract the terms for the Kaluza-Klein level $n$ to find 
\begin{align} \label{cLn_action}
g^2 \cL_n   = & \;  
 d_{IJ} \Big[
- \frac 12  \bar F^{I\, \mu\nu}_n F^J_{n\, \mu\nu}  
 + \frac {i}{4 m_n}  \epsilon^{\mu\nu\rho\lambda\sigma} \bar F^I_{n\,\mu\nu} \cD_\rho F^J_{n\, \lambda \sigma} 
 \nn \\[.1cm]
 & \qquad
-   \cD^\mu \bar \phi^I_n \cD_\mu \phi^J_n
  -  \cD^\mu \bar h_n^{I\, a \dbi} \cD_\mu h^J_{n\, a \dbi}
  -   \bar\chi^{I\, a}_n \slashed{\cD} \chi^J_{n\,a}
- 2 \bar\psi^{I \,\da}_n \slashed{\cD} \psi^J_{n\, \da} 
\nn \\[.1cm]
& \qquad
 - m_n^2   \bar\phi^I_n \phi^J_n 
 - m_n^2  \bar h^{I\, a \dbi}_n h^J_{n\, a \dbi}
- m_n \bar\chi^{I\,a}_n \chi^J_{n\,a} 
- 2 m_n \bar\psi^{I\, \da}_n \psi^J_{n\, \da}
+ 2  \bar Y^{I\, ab}_n Y^J_{n\,ab}
\Big ]
\nn \\[.1cm]
&  +  \frac{1}{m_n}\, f_{IJK} \Big[ 
  - \frac{i}{2}  \phi^K \bar F^{I\, \mu\nu}_n F^J_{n\, \mu\nu}  
   + i  \bar \phi^K_n F^{I\, \mu\nu} F^J_{n\, \mu\nu}  
   - i  \phi^K \cD^\mu \bar \phi^I_n \cD_\mu \phi^J_n
   +  2i  \bar \phi^K_n \cD^\mu \phi^I \cD_\mu \phi^J_n
\nn \\[.1cm]
&  \qquad
- i \phi^K \bar\chi^{I\, a}_n \slashed{\cD} \chi^J_{n\,a}
 +  2i  \bar \phi^K_n \bar\chi^{I\, a} \slashed{\cD} \chi^J_{n\,a}
+ 2i  \phi^K \bar Y^{I\, ab}_n Y^J_{n\,ab}
-  4i  \bar \phi^K_n Y^{I\, ab} Y^J_{n\,ab}
\nn \\[.1cm]
& \qquad
+ \frac {1}{4}  F^I_{\mu\nu} \bar\chi^{J\,a}_n \gamma^{\mu\nu} \chi^K_{n\,a} 
- \frac {1}{2}   \bar F^I_{n\, \mu\nu} \bar\chi^{J\,a} \gamma^{\mu\nu} \chi^K_{n\,a} 
+    Y^{I\,ab} \bar\chi^{J}_{n\,a} \chi^K_{n\,b}
- 2    \bar Y^{I\, ab}_n \bar\chi^J_a \chi^K_{n\,b} 
\nn \\[.1cm]
&\qquad
- 2 i m_n \phi^K \bar\chi^{I\,a}_n \chi^J_{n\,a}
+ 3 i m_n \bar\phi^K_n \bar\chi^{I\,a} \chi^J_{n\,a}
- 4 i  m_n \bar h^{I\, a \dbi}_n \bar\chi^J_a \psi^K_{n\, \dbi} 
 -2 i  m_n \phi^I \bar\psi^{J\, \da}_n \psi^K_{n\, \da} 
\nn \\[.1cm]
&\qquad
+ 2 m_n \bar h^{I\, a \dc}_{n} h^{J\, b}_n{}_{\dc} Y^{K}_{ ab}
- 3 i m_n^2 \phi^I \bar\phi^J_n \phi^K_n
- 2i m_n^2 \phi^I \bar h ^{J\, a \dbi}_{n} h^K_{n\, a \dbi}
\Big ]
\nn \\[.1cm]
& +\frac{1}{m_n}\, f_{IJ}{}^{H} f_{HKL} \Big[
- 3 m_n \phi^I \phi^K  \bar\phi^J_n \phi^L_n
- m_n \phi^I \phi^K    \bar h^{J\, a \dbi}_n h^L_{n\, a \dbi} 
-  \phi^I \phi^K \bar \chi^{J\, a}_n \chi^L_{n\,a} 
\nn \\[.1cm]
& \qquad
 + \bar\phi^I_n \phi^J \bar\chi^{K\,a} \chi^L_{n\,a}
 + 2 \bar\phi^I_n \phi^K \bar\chi^{J\,a} \chi^L_{n\,a}
 -   \bar\phi^I_n \phi^K_n \bar\chi^{J\, a} \chi^L_a 
  - \tfrac 12  \bar\phi^I_n \phi^J_n \bar\chi^{K\, a} \chi^L_a 
 \Big ]
\nn \\[.1cm]
& - \frac{i}{m_n}\, f_{IH}^{\phantom{AA}I_1} f_{JL}^{\phantom{AA}I_2} f_{KI_1 I_2} \phi^I \phi^J \phi^K \bar\phi^H_n \phi_n^L \ .
\end{align}
The terms contracted with the trace $d_{IJ}$ are kinetic terms and mass terms for all Kaluza-Klein excited modes. We note 
that the tensors $B_{n\, \mu\nu}^I= - \frac{i}{n}F_{n\, \mu\nu}^I$ have Chern-Simons kinetic terms and a mass term proportional to $n^2$.
Consistent with a Kaluza-Klein reduction all complex scalars $\phi^I_n,\, h^{I\, a\dbi}_n$ with $n>0$ have mass terms 
proportional to $n^2$, and all fermions $\chi^{I\, a}_n,\psi^{I \, \dbi}_n$ with $n>0$ have mass terms proportional to $n$.
More interestingly, this Lagrangian contains various terms at the non-Abelian level 
containing $f_{IJK}$. These include new kinetic terms for all singlets under the second $SU(2)$
in \eqref{R_breaking}, Pauli terms coupling 
the tensors and gauge fields to the fermions, Yukawa type couplings, and a complicated scalar potential. 
The full scalar potential and four Fermi terms can only be determined after eliminating the auxiliary fields 
$Y^{I\, a b}_n$.
We will discuss this elimination process in section \ref{Integrating_out}.

It is important to stress that the action \eqref{cLn_action} preserves $\cN=2$ supersymmetry but breaks 
the special supersymmetries parametrized by $\eta^a$ in \eqref{susy_trafo}. This can be seen straightforwardly
by inspecting the superconformal variations of the fermion in $\widehat \cV^0$: 
\beq \label{cV0-trafo}
   \delta \chi^{0\,a} = - \tfrac14 \gamma^{\mu \nu} F^{0}_{\mu \nu} \epsilon^a 
                         - \tfrac i2  {\slashed \cD} \phi^{0} \epsilon^a 
                         - Y^{0\, ab} \epsilon_b  
                         + \phi^{0} \eta^a\, .  \qquad
\eeq
Using the condition \eqref{gauge_fixing} we realize that the supersymmetry parameter $\epsilon^a$ drops
from \eqref{cV0-trafo} which implies that the restricted action is 
still $\cN=2$ supersymmetric. In contrast $\eta^a$ appears after imposing \eqref{gauge_fixing} 
in the transformation $\delta \chi^{0\,a}=g^{-2} \eta^a$, which implies that $\chi^{0\,a}$
is needed to ensure invariance of the action under special supersymmetry transformations. In other 
words, the condition \eqref{gauge_fixing} will break the special supersymmetry transformations
parametrized by $\eta^a$. 
The ordinary supersymmetry transformations in the restricted phase are given by
\begin{align}  \label{susy_trafo_gaugefixed}
   \delta  A^{I}_{\mu} &= \tfrac12 \, \bar \epsilon^a \gamma_{\mu} \chi^{ I}_a \ , \nn \\[.1cm]
  \delta \phi^{ I}_n & = \tfrac i2 \, \bar \epsilon^a \chi^{ I}_{n\,a}\ , \nn \\[.1cm]
   \delta  F^{I}_{n\,\mu\nu} &= - \bar \epsilon^a \gamma_{[\mu} \cD_{\nu]} \chi^I_{n\,a}
                      - \tfrac i2 f_{JK}{}^I \phi^J_n \, \bar\epsilon^a \gamma_{\mu\nu} \chi^K_a
                      + \tfrac i2 f_{JK}{}^I \phi^J \, \bar\epsilon^a \gamma_{\mu\nu} \chi^K_{n\,a}
                     - \tfrac 12 m_n \, \bar\epsilon^a \gamma_{\mu\nu} \chi^I_{n\,a}    \ ,
\nn \\[.1cm]
\delta \chi^{ I\, a}_n &= - \tfrac14 \gamma^{\mu \nu} F^{ I}_{n\, \mu \nu} \epsilon^a
                   - \tfrac i2  {\slashed \cD} \phi^{ I}_n \epsilon^a - Y^{  I \, ab}_n \epsilon_b
                   + \tfrac 12 f_{JK}{}^I \phi^J \phi^K_n \epsilon^a
                   + \tfrac i2 m_n \phi^I_n \epsilon^a \ ,
 \nn \\[.1cm]
  \delta Y^{ I \, ab}_n &= -\tfrac12 \, \bar \epsilon^{(a|} {\slashed \cD} \chi^{ I|b)}_n
                - i f_{JK}{}^I \phi^J_n \, \bar \epsilon^{(a|} \chi^{K|b)}
               + \tfrac i2  f_{JK}{}^I \phi^J \, \bar \epsilon^{(a|} \chi^{K|b)}_n
               - \tfrac 12 m_n \, \bar  \epsilon^{(a|} \chi^{K|b)}_n \ ,
\nn \\[.1cm]
   \delta  h^{I \, a \dbi}_n &= - { i} \, \bar \epsilon^a \psi^{I \, \dbi}_n\ ,
\nn \\[.1cm]
\delta  \psi^{I \, \dbi}_n &= \tfrac i2  {\slashed \cD} h^{I \, a \dbi}_n \epsilon_a
                  - \tfrac 12 f_{JK}{}^I \phi^J h^{K\, a \dbi}_n \epsilon_a
                  - \tfrac i2 m_n h^{I\, a \dbi}_n \epsilon_a    \ ,
\end{align}
where $n \geq 0$ labels both zero and excited modes. We close this subsection 
by pointing out that the Lagrangian \eqref{cL_split} posses a scaling symmetry
when using the Weyl weights of Table \ref{N2_table} and additionally 
assigning scaling weight $-1/2$ to the gauge coupling constant $g$,
in such a way that $m_n$ has weight $+1$ for any $n>0$. This can be interpreted 
as a remnant of the full six-dimensional $(2,0)$ conformal symmetry as 
discussed in section \ref{sec:conformal_symmetry}.

This concludes our discussion of the general $\cN=2$ action for the Kaluza-Klein 
tower. Our approach can be summarized as follows. While an action for  
full six-dimensional non-Abelian $(2,0)$ theories is unknown the Abelian free
six-dimensional $(2,0)$ theory admits a six-dimensional pseudoaction. 
It can be compactified on a circle with arbitrary radius yielding a 
five-dimensional action with $\cN=4$ supersymmetry. We proposed a gauged version 
of this theory preserving only half, namely $\cN=2$, supersymmetry, by
interpreting the zero mode vectors $A^I$ as gauge potentials for the whole Kaluza-Klein tower. 
In order to argue for a six-dimensional origin of this theory all higher-dimensional
symmetries need to be realized or appear in a gauge-fixed phase. Our five-dimensional 
actions \eqref{ActionComplete} and \eqref{cL_split}, however, clearly only realize part of the 
six-dimensional superconformal $(2,0)$ symmetries manifestly. In particular, we
have singled out the zero modes for gauging which seems naively incompatible with 
six-dimensional Poincar\'e invariance. 

It is precisely the non-Abelian gauging that 
prevents us to write down an $\cN=4$ action. Nevertheless, we regard
our Lagrangians as the starting point to give a lower-dimensional 
Lagrangian formulation for $(2,0)$ theories. The next step in the construction 
must be the restoration of the full set of six-dimensional symmetries. In section 
\ref{sec:N=4completion} have a closer look at supersymmetry  
and suggest a strategy to implement its enhancement. In the next subsection \ref{Integrating_out}
we concentrate on two special cases in which partial symmetry restoration is achieved.

\subsection{Special cases and integrating out auxiliary fields $Y$} \label{Integrating_out}

We have just proposed a Lagrangian for all Kaluza-Klein modes in an 
$\cN=2$ supersymmetric framework. In particular we made use of complete  
$\cN=2$ vector and tensor multiplets including auxiliary fields $Y^{I\, ab}_n$. 
These fields appear only algebraically in the Lagrangian and can be eliminated 
consistently by using their equations of motion. While the action \eqref{cL_split}
is a sum of terms $\cL_n$ only involving fields at the Kaluza-Klein level $n$ and 
zero modes, the elimination of auxiliary fields will induce a non-trivial 
mixing among exited modes. Despite the fact that it is interesting to investigate this
structure in more detail, we will focus here on only two special cases where the 
computation is straightforward and the lift to $\cN=4$ can be performed explicitly.

As a first special case we study the zero mode Lagrangian $\cL_0$ given in \eqref{cL0-aux}, and 
drop all massive modes. This is motivated physically with the dimensional reduction 
argument for small radius $r$ where massive Kaluza-Klein modes are dropped, or rather integrated out, that are 
above a certain energy scale. 
The equation of motion for the auxiliary fields then 
simply reads
\beq \label{massless_Y}
Y^{I\, ab}  =  - \tfrac 12  f^I_{\phantom{A}JK}  h^{J\,a \da} h^{K\, b}{}_{\da} \ .
\eeq
Inserting \eqref{massless_Y}
into \eqref{cL0-aux} a quartic potential in $h$ is generated, and the zero mode Lagrangian $\cL_0$ takes the form
\begin{align} \label{cL0-elim}
g^2 \cL_ {\rm YM}  = & \;
 d_{IJ} \Big[ 
- \tfrac 14   F^{I\, \mu\nu} F^J_{\mu\nu}
- \tfrac 12   \cD^\mu \phi^I \cD_\mu \phi^J 
- \tfrac 12    \cD^\mu h^{I\, a \dbi} \cD_\mu h^J_{\, a \dbi} 
- \tfrac 12   \bar\chi^{I\,a} \slashed{\cD} \chi^J_{\, a} 
-   \psi^{I\, \da} \slashed{\cD} \psi^J_{\, \da} \Big] \nn \\[.1cm]
& +  f_{IJK} \Big[
 +   \tfrac{i}{2}   \phi^I    \bar\chi^{J\, a} \chi^K_{\,a} 
- i  \phi^I \bar\psi^{J\, \da} \psi^K_{\, \da} 
- 2 i  h^{I\, a \dbi} \bar\chi^J_a \psi^K_{\, \dbi}  \Big] \nn \\[.1cm]
& + f_{IJ}{}^{H} f_{H KL}  \Big[ 
%\tfrac{1}{16 \ell}  q^I_{a \da} q^{J\, b \da} q^K_{b \dbi} q^{L\, a \dbi} 
- \tfrac{1}{4}  h^{I\, a \da} h^{J\, b}{}_{\da} h^{K\,\phantom{a}\dbi}_{\phantom{K}a} h^L{}_{b \dbi}
 - \tfrac 12   \phi^I \phi^K h^{J\, a \dbi} h^L_{\,a \dbi} \Big ] \ .   
\end{align}
This Lagrangian is a simple rewriting of $\cN = 4$ super Yang-Mills theory in 
terms of $\cN=2$ multiplets, as can be checked by inserting (\ref{schematic_sigma}), (\ref{gauge_fixing}) and (\ref{def-Omega}) 
into the $\cN=4$ super Yang-Mills Lagrangian (see eqn.~(\ref{SYMaction}) later on). Therefore it 
possess enhanced supersymmetry which is not present in the massive Kaluza-Klein tower.

As a second special case we consider the Abelian truncation of the full Lagrangian \eqref{cL_split}. 
This is achieved by dropping all terms constructed with the structure constants $f_{IJK}$. The equations 
of motions for the auxiliary fields read simply $Y^{I\, ab}_n=0$, such that they can be trivially dropped from the 
Lagrangian. The resulting theory is free and given by
\begin{align} \label{cL-free}
g^2 \cL_{\rm free}  = & \;
 d_{IJ} \Big[ 
- \tfrac 14   F^{I\, \mu\nu} F^J_{\mu\nu}
- \tfrac 12   \partial^\mu \phi^I \partial_\mu \phi^J 
- \tfrac 12    \partial^\mu h^{I\, a \dbi} \partial_\mu h^J_{\, a \dbi} 
- \tfrac 12   \bar\chi^{I\,a} \slashed{\partial} \chi^J_{\, a} 
-   \psi^{I\, \da} \slashed{\partial} \psi^J_{\, \da} \Big] \nn \\[.1cm]
& + \sum_{n=1}^\infty
 d_{IJ} \Big[
- \tfrac 12  \bar F^{I\, \mu\nu}_n F^J_{n\, \mu\nu}  
 + \tfrac {i}{4 m_n}  \epsilon^{\mu\nu\rho\lambda\sigma} \bar F^I_{n\,\mu\nu} \partial_\rho F^J_{n\, \lambda \sigma} 
 \nn \\[-.1cm]
 & \phantom{+ \sum_{n=1}^{\infty}   d_{IJ} \Big[}
-   \partial^\mu \bar \phi^I_n \partial_\mu \phi^J_n
  -  \partial^\mu h_n^{I\, a \dbi} \partial_\mu h^J_{n\, a \dbi}
  -   \bar\chi^{I\, a}_n \slashed{\partial} \chi^J_{n\,a}
- 2 \bar\psi^{I \,\da}_n \slashed{\partial} \psi^J_{n\, \da} 
\nn \\[-.1cm]
 & \phantom{+ \sum_{n=1}^{\infty}   d_{IJ} \Big[}
 - m_n^2   \left( 
 \bar\phi^I_n \phi^J_n 
+  \bar h^{I\, a \dbi}_n h^J_{n\, a \dbi}
\right)
- m_n \left( 
 \bar\chi^{I\,a}_n \chi^J_{n\,a} 
+  2 \bar\psi^{I\, \da}_n \psi^J_{n\, \da}
\right) \Big ] \ .
\end{align}
This Lagrangian is the $\cN=2$ supersymmetric extension of the purely bosonic Lagrangian \eqref{purely_bosonic_action} 
in the gauge \eqref{gauge_fixing}. In fact, this theory is actually  $\cN=4$
supersymmetric. Furthermore, it can be obtained by a compactification of the full $(2,0)$ Abelian 
pseudoaction \eqref{6Dab_action} on a circle and therefore admits non-manifest six-dimensional Poincar\'e invariance.
Five-dimensional Kaluza-Klein actions arising from such a compactification have been 
considered before in \cite{Lee:2000kc}.
We stress that it is hard to interpret the action \eqref{cL-free} with 
the full Kaluza-Klein tower as an effective action for the Coulomb branch of the five-dimensional 
theory. This is due to the fact that it contains modes of arbitrary high mass $m_n$ that rather should be 
integrated out above the cutoff scale.

%%%%%%%%%%%%%%%%%%%%%%%%%%%%%%%%%%%%%%%%%%%%%%%%%%%%%%%%%%%%%%%%%%%%%%%%%%%%%%%%%%%%%%
\section{Harmonization and $USp(4)$ covariant action}  \label{sec:N=4completion}

As we have discussed in the previous section the Lagrangian \eqref{cL_split} 
does not possess all the symmetries expected from a 
circle compactification of a theory with $(2,0)$ superconformal invariance. 
For example the compactification on a circle is not expected to break supersymmetry, in particular 
the extended $USp(4)$ R-symmetry. In this section we concentrate on the breaking 
of the R-symmetry group $USp(4)_R \rightarrow SU(2)_R \times SU(2)$ and we propose a way to restore it. The key 
idea is to parametrize the embeddings of $SU(2)_R \times SU(2)$ into 
$USp(4)_R$ by additional bosonic variables and integrate
over such embeddings.  This construction depends crucially
on two special features of the Lagrangian \eqref{cL_split}.  Firstly, its spectrum
can be rearranged into full $\cN=4$ supermultiplets, as shown in section \ref{N=2spectrum}. Secondly,
the global symmetry of the theory includes not only the $SU(2)_R$ R-symmetry
of the $\cN=2$ algebra, but also the additional $SU(2)$ symmetry. 
In the following, we use techniques inspired by the construction 
of harmonic superspaces for theories with extended supersymmetry. 
Note in particular that conformal
symmetry plays no role in our discussion.\footnote{In appendix \ref{Sect:Superconformal5},
however, we consider possible extensions of the formalism,
to combine higher supersymmetry with scale invariance.}

Let us outline our strategy in more detail. As a first step
towards an $\cN=4$ Lagrangian, the R-symmetry group
of the $\cN=2$ Lagrangian \eqref{cL_split} is enhanced to $USp(4)_R$.
This is achieved by techniques familiar in harmonic superspace \cite{Galperin:1984av, Ivanov:1984ut, Galperin:1984bu,Hartwell:1994rp,Howe:1995md,HSS}. 
They are introduced in subsection \ref{Sect:CosetConstruction} and are applied to the Lagrangian \eqref{cL_split} 
in subsection \ref{Sec:Haction}. For the precise practical realization we will follow closely \cite{Antoniadis:2007cw} (see also \cite{Antoniadis:2009nv}). 
In subsection \ref{Sec:coset} we present a realization of the $\cN= 4$ supersymmetry
algebra on a suitable harmonized Grassmann analytic coset.
The key point of the construction is the possibility
of modding out analytically half of the 16 supercharges of the
$\cN = 4$ algebra. We claim that the invariance
of the original Lagrangian \eqref{cL_split} under $\cN=2$ supersymmetry,
together with $USp(4)_R$ symmetry and closure of the full algebra
is a strong hint of invariance of the harmonized Lagrangian 
under $\cN=4$ supersymmetry.
Furthermore, harmonization techniques
can be applied to study how the vector/tensor and hypermultiplet
moduli space of the $\cN=2$ Lagrangian \eqref{cL_split}
combine into the moduli space of the harmonized $USp(4)_R$ invariant
Lagrangian. This is also addressed in subsection~\ref{Sec:coset}.

In this work we refrain from a full harmonic superspace construction, and in particular
we do not discuss superfields. As a consequence,
supersymmetry is realized non-linearly and the representation of
the algebra on fields closes only up
to equations of motion. Many of our claims concerning restoration of
$\cN=4$ supersymmetry are most directly addressed in a full superspace 
analysis and we hope to return to these issues in the future.

\subsection{$\cN=2$ formulation and coset construction}\label{Sect:CosetConstruction}

In section~\ref{N=2spectrum} we have discussed the decomposition of the spectrum of 
the $\cN=4$ supersymmetric theory into $\cN=2$ language on a group-theoretic level as representations 
of the respective R-symmetry groups. In order to write the action (\ref{ActionComplete}) in 
an $USp(4)_R$ covariant manner we will now make this decomposition more precise and give an explicit parametrization. 
To this end we introduce harmonic variables \cite{Galperin:1984av, Ivanov:1984ut, Galperin:1984bu,Hartwell:1994rp,Howe:1995md,HSS} on the coset manifold
\beq\label{CosetDef}
\frac{USp\,(4)_R}{SU(2)_R\times SU(2)} \cong \frac{SO(5)}{SO(4)} \cong S^4\sim (u_i^a,u_i^{\da})\ .
\eeq
where we have introduced the representatives $(u_i^a,u_i^{\da})$ as a particular fixed $USp\,(4)$ matrix. 
Recall that the latter are $4 \times 4$ unitary matrices
that preserve the symplectic form $\Omega$, corresponding to $USp(4)=U(4)\cap Sp(4,\bbC)$. 
As a result, $u_i^a , u_i^{\da}$
satisfy the pseudo-reality constraints
\begin{equation} \label{HarmonicConjugation}
 \bar u^i_a \equiv (u_i^a)^* = \Omega^{ij} \epsilon_{ab} u_j^b \ , \qquad
  \bar u^i_{\da} \equiv (u_i^{\da})^* = \Omega^{ij} \epsilon_{\da \dbi} u_j^{\dbi} \ ,
\end{equation}
along with 
\begin{eqnarray} \label{unit}
 && \Omega^{ij} u_i^a u_j^b = \epsilon^{ab} \ , \hspace{2cm}
  \Omega^{ij} u_i^a u_j^{\dbi} = 0 \ , \hspace{2cm}
  \Omega^{ij} u_i^{\da} u_j^{\dbi} = \epsilon^{\da \dbi} \ , \\
&& \epsilon_{ab} u_i^a u_j^b + \epsilon_{\da \dbi} u_i^{\da} u_j^{\dbi} = \Omega_{ij} \ . \nn
\end{eqnarray}
As a by product, we have also the unimodularity condition
\begin{equation}
 \epsilon^{ijkl} u_i^a u_j^b u_k^{\da} u_l^{\dbi} = \epsilon^{ab} \epsilon^{\da \dbi} \ .
\end{equation}
It follows from \eqref{unit}, taking into account that the antisymmetric invariant $\epsilon^{ijkl}$
is not primitive, but is derived from $\Omega^{ij}$ according to
\begin{equation}
 \epsilon^{ijkl} = 3 \Omega^{[ij} \Omega^{kl]} \ ,
\end{equation}
in our conventions. A more intuitive way to think about $(u_i^a,u_i^{\da})$ is to regard them as spherical 
harmonics on $S^4$. In particular, we can write arbitrary functions on this manifold as a series 
expansion in $(u_i^a,u_i^{\da})$.

With these objects we can parametrize the decompositions \eqref{rep_decomp} of the scalars $\sigma^{I\,ij}_n$ which are in the 
$\mathbf{5}$ of $USp(4)$ into representations of $SU(2)_R\times SU(2)$ as 
\beq \label{Harmonization}
   \left(\begin{array}{cc}
      u_i^a u^b_j & u_i^{a} u^{\dbi}_j  \\
      u_i^{\da} u^b_j  & u_i^{\da} u^{\dbi}_j 
   \end{array} \right) \sigma^{I\, ij}_n = \left(\begin{array}{cc}
      \tfrac{1}{\sqrt{2}} \epsilon^{ab} \phi^{I}_n& h^{I\, a\dbi}_n  \\
       - h^{I\,  b \da}_n  & - \tfrac{1}{\sqrt{2}} \epsilon^{\da \dbi}   \phi^{I}_n 
   \end{array} \right)\ ,
\eeq
where we have used equation \eqref{unit} and the fact that $\sigma^{I\, ij}_n$ are anti-symmetric traceless.
Similarly, we can write for the decomposition of the fermions $\lambda^{Ii}_n$, which form a $\mathbf{4}$ of $USp\,(4)$
\beq \label{ProjectionFerm}
\left(\begin{array}{c} u_{i}^a \\ u_{i}^{\da} \end{array} \right) \lambda^{I\, i}_n = 
 \left(\begin{array}{c}  \chi^{I a}_n \\  \sqrt{2} \,\psi^{I \da}_n \end{array} \right) \ .
\eeq
We can think of these identifications as refined versions of \eqref{rep_decomp} and \eqref{schematic_sigma}, where 
in addition we have also imposed the condition \eqref{gauge_fixing} (and thus set $\phi^0=g^{-2}$) 
and used the rescaled hypermultiplet fields introduced in \eqref{rescaled_hypers}. This is due to the fact that 
in this section we will not deal with questions of conformal invariance. We, however, refer the reader 
to appendix~\ref{Sect:Superconformal5} for a brief discussion on this topic. 

As we can see, from an $\cN=4$ point of view, the coset (\ref{CosetDef}) allows us to organize the 
spectrum of physical fields with respect to a particular $\cN=2$ subalgebra of the full $\cN=4$ supersymmetry. 
Different choices of the coset representatives $(u_i^a,u_i^{\da})$ correspond to different choices of this subalgebra. 
From a purely $\cN=2$ point of view reparametrizations of the coset (\ref{CosetDef}) act like 
external automorphisms, which mix the $SU(2)_R$-symmetry group non-trivially with the second $SU(2)$ to enhance 
it to $USp(4)_R$. With this observations in mind, we propose to promote the $\cN=2$ supersymmetric action 
described in the previous section to $\cN=4$ by `harmonizing' this choice of subalgebra. In more technical 
terms, we consider an infinite set of $\cN=2$ actions of the type (\ref{ActionComplete}) parametrized by 
the coset representatives $(u_i^a,u_i^{\da})$ and integrate the latter in an $USp(4)$ covariant manner. 
We will make this idea more precise in the following.

Let us close this subsection by pointing out that the identifications \eqref{Harmonization} and \eqref{ProjectionFerm}
suggest that it is natural to enlarge the auxiliary fields in the vector 
multiplets $Y^{I\,ab}_n$
to a full $\mathbf{10}$ of $USp\,(4)_R$. This amounts to defining $Y^{I\, ij}_n$ that is symmetric in $i,j$ and specified by
\begin{align}
\left(\begin{array}{cc}
      u_i^a u^b_j & u_i^{a} u^{\dbi}_j  \\
      u_i^{\da} u^b_j  & u_i^{\da} u^{\dbi}_j 
   \end{array} \right) Y^{I\, ij}_n =\left(\begin{array}{cc}Y_n^{I\,ab} & Y_n^{I\,a\dbi} \\ Y_n^{I\,b \da} & Y_n^{I\,\da\dbi}\end{array}\right)\,. 
\end{align}
We stress that this requires the introduction of further auxiliary fields $Y_n^{I\,a\dbi}$ and $Y_n^{I\,\da \dbi}$ that 
transform in the $\mathbf{(2,2)}$ and $\mathbf{(1,3)}$ of $SU(2)_R\times SU(2)$ respectively. It would be interesting
to relate these new fields to off-shell formulations of $\cN=2$ and $\cN=4$ supersymmetric theories.

%%%%%%%%%%%%%%%%%%%%%%%%%%%%%%%%%%%%%%%%%%%%%%%%%%%%%%%%%%%%
\subsection{Harmonic action and coset integration} \label{Sec:Haction}

In this section we will explain the idea of harmonizing the $USp(4)_R$ R-symmetry group in the rigid case in more detail. 
While our methods might be more generally applicable, 
our main concern will be the harmonization of the Lagrangian \eqref{cL_split}. 
More precisely, our starting point will be the Lagrangian \eqref{cL_split} after 
integrating out the auxiliary fields $Y^{I\, ab}_n$ resulting in an expression of the 
form 
\footnote{By a slight abuse of notation, we do not indicate in this expression that the Lagrangian 
depends explicitly on the zero mode vectors $A^I=A^I_0$ and not only on their field strength $F^I_{0}$.}
\begin{align} \label{SartingN=2} 
\mathcal{L}_{\cN=2} \Big[ (h^{I\, a\da}_n,\psi^{I\, \da}_n) ;(\phi^I_n,\chi^{I\, a}_n,F_{n}^I) \Big]\,, \qquad n \geq 0\ ,
\end{align}
which is a functional of the components of the hyper-, vector- and tensor multiplets respectively, as discussed in section~\ref{sec:KKaction}. 
We have also used the rescaled fields $h^{I\, a\da}_n,\psi^{I\, \da}_n$ as introduced in \eqref{rescaled_hypers}.
Let us stress that in the following two ingredients are crucial: an $\cN=2$ supersymmetric action manifestly invariant under $SU(2)_R\times SU(2)$
together with a spectrum which furnishes full $\cN=4$ representations as described in section~\ref{Sect:CosetConstruction}. 

We can now use the coset parameters introduced in (\ref{CosetDef}) to rewrite the dependence on 
the fields in $\cN=2$ in terms of $\cN=4$ degrees of freedom. Moreover, all fields are promoted to
depend on the coordinates $(u^a_i,u^{\da}_i)$ on the $S^4$ given by the coset \eqref{CosetDef}.
The decomposition \eqref{Harmonization} and \eqref{ProjectionFerm} can be inverted to infer
\begin{align}
  \phi^{I}_n(u) &=\tfrac{1}{\sqrt{2}} \epsilon_{ab} u^a_i u^b_j\ \sigma^{I\, ij}_n(u)\ , & \chi^{I\, a}_n(u) & = u^a_i \lambda^{I\, i}_n(u)\ , \\
   h^{I a\dbi}_n(u) &= u^a_i u^{\dbi}_j\ \sigma^{I\, ij}_n(u) \ ,    & \psi^{I\, \da}_n(u) & = \tfrac{1}{\sqrt{2}} u^{\da}_i \lambda^{I\, i}_n(u)\ . \nn
\end{align}
Using these identifications the $\cN=2$ Lagrangian becomes a functional on $S^4$ and reads
\beq \label{L(u)}
   \mathcal{L}_{\cN=2}\Big[u,(\sigma_n^{I\, ij}(u),\lambda^{I\, i}_n(u ),F_{n}^I(u)) \Big]\ 
   \equiv\ \mathcal{L}_{\cN=2} \Big[ (h^{I\, a\da}_n(u),\psi^{I\, \da}_n(u)) ;(\phi^I_n(u),\chi^{I\, a}_n(u),F_{n}^I(u)) \Big]\,. 
\eeq
This action is still only invariant under $SU(2)_R\times SU(2)$ but 
not under $USp(4)_R$. The reason is that for the moment the quantities $(u_i^a,u_i^{\da})$ are fixed 
and do not transform under generic $USp(4)_R$ transformations. In order to obtain $USp(4)_R$ invariance
we integrate  $\mathcal{L}_{\cN=2}$ given in \eqref{L(u)} over the sphere $S^4$ as
\beq \label{USp4action}
\mathcal{L}_{USp(4)}\big[\sigma^{I\, ij}_n,\lambda^{I\, i}_n,F_{n}^I\big]=
\int_{S^4} d u\,   \mathcal{L}_{\cN=2}\Big[u,(\sigma_n^{I\, ij}(u),\lambda^{I\, i}_n(u ) ,F_{n}^I(u) ) \Big]\ ,
\eeq
where $du$ is the integration measure on $S^4$ defined through 
the following rules (see e.g.~\cite{Buchbinder:2008ub} and also \cite{HSS})
\beq
 \int_{S^4}du\,1=1\,, \qquad \quad \int_{S^4}du\,(\text{irred. tensor of }USp(4)_R)=0\,.\label{HarmonicIntegration}
\eeq
In other words, after expanding the various terms in irreducible representations of $USp(4)_R$
the integration over the harmonic parameters $(u_i^a,u_i^{\da})$ extracts the $USp(4)_R$-singlet component of the integrand.
A helpful way to think about this integration is to imagine an expansion of the integrand in 
spherical harmonics of $S^4$ and to eliminate all higher excitation modes.

We claim that (\ref{USp4action}) is $USp(4)$ invariant. Indeed, this action is invariant under all 
transformations $g\in SU(2)_R\times SU(2)\subset USp(4)$, since this is a property of our starting 
action (\ref{SartingN=2}), which is valid for any (fixed) choice of $(u_i^a,u_i^{\da})$. Therefore 
$\mathcal{L}_{USp(4)}$ is invariant under $g$ pointwise on $S^4$. Transformations 
$g^\perp\in USp(4)/(SU(2)_R\times SU(2))$ in the $USp(4)$ complement do not have this property, 
but on the contrary correspond to a non-trivial transformation on the sphere. Invariance under the 
latter, however, in (\ref{USp4action}) is achieved through the integration over $S^4$, since the 
action of $g^{\perp}$ can be absorbed by a simple coordinate reparametrization of the integral. 
Thus we conclude that the action (\ref{USp4action}) is invariant under $USp(4)$. We stress once more 
that this $USp(4)$ is not just some global symmetry of the action but arises through a non-trivial 
extension of the $\cN=4$ R-symmetry group via the enlargement of the five-dimensional space-time with an additional $S^4$.

We note that we have also allowed for an implicit harmonic-dependence of the fields themselves 
in (\ref{USp4action}). A priori, this dependence on the harmonic coordinates is equivalent to adding infinitely 
many additional degrees of freedom, which correspond to higher modes in a decomposition in spherical harmonics 
on $S^4$. As we will explain in section~\ref{Sec:coset}, not all of these modes are necessarily physical 
degrees of freedom. Indeed, we interpret the action as living on a particular Grassmann-analytic coset realization of the 
$\cN=4$ Poincar\'e superalgebra. Within this framework we expect that suitable constraints 
(so-called H-analyticity conditions) can be imposed on all fields, which truncate their harmonic 
expansion. In the simplest case, these constraints amount to dropping all internal $u$-dependence 
of the component fields $(\sigma^{I\,,ij}_n,\lambda_n^{I\,i},F_n^I)$ appearing in the harmonized 
Lagrangian~(\ref{USp4action}). However, a general study of these conditions is beyond the scope 
of this work. Indeed, such questions are more efficiently attacked within a full superfield 
approach which will be attempted elsewhere.

\subsection{Two special cases} \label{Sec:two_special}

This subsection is devoted to the study of the two special cases
discussed in subsection \ref{Integrating_out}.
Firstly, we demonstrate explicitly that
the harmonization formalism 
of the previous subsections can be applied to the zero mode part
and give rise to maximally supersymmetric Yang-Mills theory. 
This will give us a good opportunity 
to explicitly demonstrate the integration over the harmonic $u$-variables. 
Secondly, the Abelian action \eqref{cL-free} for the full
Kaluza-Klein tower is shown to lift to 
the Abelian six-dimensional $(2,0)$ pseudoaction \eqref{6Dab_action} 
compactified on a circle.

The starting point for the discussion of the zero mode Lagrangian
in the action (\ref{cL0-aux}) where auxiliary fields $Y^{I\,ab}$
have been integrated out. When the harmonization prescription given in
\eqref{USp4action} is applied to \eqref{cL0-aux}, the 
on-shell zero mode Lagrangian takes the form
\begin{align} \label{HarmonizedZeroModes}
g^2 \cL_{0} & =  \int_{S^4} du \Big \{d_{IJ} \Big[
- \tfrac 14 F^{I\, \mu\nu} F_{\mu\nu}^J
- \tfrac 14 \cD^\mu \sigma^{I\, ij} \cD_\mu \sigma^{J\, kl} \left( - u_{ij} \hat u_{kl} + 2 u_{ik} \hat u_{jl} \right)
-\tfrac 12  \bar \lambda^{I\,i} \slashed{\cD} \lambda^{J\,j} \left( - u_{ij} - \hat u_{ij} \right)
 \Big] \nn
\\[.1cm]
&- \tfrac{i}{\sqrt 2} f_{IJK} \sigma^{I\, ij} \bar \lambda^{J\,k} \lambda^{K\,l} 
\left( -\tfrac 12 \hat u_{ij} u_{kl} - \tfrac 12 u_{ij} \hat u_{kl} + 2  u_{ik} \hat u_{jl} \right) \nn \\[.1cm]
& - \tfrac{1}{16} f_{IJ}{}^H f_{HKL} \sigma^{I\,ij} \sigma^{K\, kl} \sigma^{J\,mn} \sigma^{L\,pq} 
\left( 4 u_{ik} \hat u_{jn} u_{mp} \hat u_{lq} + 4 u_{ij} \hat u_{kl} u_{mp} \hat u_{nq} \right) \Big \} \ ,
\end{align}
where we have introduced the following shorthand notation
\begin{align}
&u_{ij}=u_i^{a}u_j^b\epsilon_{ab}\,,&&\hat{u}_{ij}=u_i^{\da}u_j^{\dbi}\epsilon_{\da\dbi}\,.
\end{align}
For performing the harmonic integral, we will drop the explicit 
harmonic dependence of all the fields in the following, i.e.~we 
will assume that the H-analyticity relations (that we have mentioned earlier) 
are imposed on all fields. By definition the integral over $u$ will 
simply pick the $USp\,(4)_R$ singlet component of the integrand, according 
to (\ref{HarmonicIntegration}). Let us make a few comments about some of 
the terms involved. Since the $\sigma^{ij}$ transform in the 
$\mathbf{5}$ of $USp\,(4)_R$ we have the following channels appearing 
in the scalar bilinear term
\beq
\mathbf{5}\otimes \mathbf{5}=\mathbf{1}\oplus \mathbf{10}\oplus\mathbf{14}\,.
\eeq
Of these, only the $\mathbf{1}$, which corresponds to the `trace' 
in the $USp\,(4)_R$ sense, can be combined with the harmonics to yield 
a singlet expression, i.e.
\beq \label{eq:sigmasigma}
\sigma^{I\,ij}\sigma^{J\,kl} =\left(-\tfrac {1}{12} \sigma^{I\,mn}\sigma^{J}_{mn} \right) \epsilon^{ijkl}+\text{higher irreps.}\,.
\eeq
Indeed, the $\epsilon^{ijkl}$ combines with the harmonics in 
(\ref{HarmonizedZeroModes}) to form a singlet combination. In 
the same manner, we can deal with the fermion bilinear term. 
Here the decomposition for the $\mathbf{4}$ reads
\beq
\mathbf{4}\otimes\mathbf{4}=\mathbf{1}\oplus\mathbf{5}\oplus\mathbf{10}\,.
\eeq
As in the scalar case, only the singlet term can give rise to 
a non-vanishing contribution in the harmonic integral, namely
\beq \label{eq:lambdalambda}
\bar{\lambda}^{I\,i}\lambda^{J\,j} =  \left( - \tfrac 14 \bar{\lambda}^{I\,k}\lambda^{J}_k \right)  \Omega^{ij}+\text{higher irreps.}\,.
\eeq
This essentially accounts for the first line in (\ref{HarmonizedZeroModes}). 
In the second line, the only contribution comes from the $\mathbf{5}$ 
channel in the decomposition of the two fermions, which combines with 
the $\mathbf{5}$ of the scalar to form a singlet, i.e.
\beq
\sigma^{I\,ij}\bar{\lambda}^{J\,k}\lambda^{K\,l} =
\left( - \tfrac {1}{12} \sigma^{I\,mn}\bar{\lambda}^{J}_{m}\lambda^{K}_n \right) \epsilon^{ijkl}+\text{higher irreps.}\,.
\eeq
Finally, the most difficult contribution is the last line 
in (\ref{HarmonizedZeroModes}). However, the simplest way 
to disentangle the latter is to recall that this term originally 
arose from integrating out the auxiliary component of the $\cN=2$ 
vector multiplet, which transforms in the 
$(\mathbf{3},\mathbf{1})$ in $SU(2)_R\times SU(2)$. The latter 
comes from the reduction of a field that transforms in the 
$\mathbf{10}$ of $USp\,(4)_R$ which hints to the fact, that the 
only relevant channel  in the decomposition of the four scalar 
fields in the last line of (\ref{HarmonizedZeroModes}) is
\begin{align}
\mathbf{5}\otimes \mathbf{5}\otimes \mathbf{5}\otimes \mathbf{5} = (\mathbf{1}\oplus \mathbf{10}\oplus\mathbf{14})\otimes(\mathbf{1}\oplus \mathbf{10}\oplus\mathbf{14})\ \rightarrow\ \mathbf{10}\otimes \mathbf{10}\,.
\end{align}
This leads us to consider
\begin{align}
\sigma^{I\,ij} \sigma^{K\,kl} \sigma^{J\, mn} \sigma^{L\,pq}  = &  
\left( - \tfrac{1}{16} \sigma^{I\,h_1 h_2} \sigma^{K}_{\,h_1 h_2} \sigma^{J\, h_3 h_4} \sigma^{L}_{\,h_3 h_4} \right)
\left[ \epsilon^{ijkl} \epsilon^{mnpq} - \epsilon^{ijpq} \epsilon^{klmn} + \dots \right] \nn \\
& + \text{other irreps.}\,,\label{Decompose10}
\end{align}
where the dots denote additional terms constructed with $\Omega^{ij}$ that
vanish upon contraction with $u_{ij},\hat u_{ij}$ but are needed
to make the square bracket a $USp(4)_R$ irreducible tensor.  
With these results, the harmonized Lagrangian (\ref{HarmonizedZeroModes})
takes the form
\begin{align}
g^2 \cL_{\cN=4}^{\rm SYM} =& d_{IJ} \Big[
- \tfrac 14 F^{I\, \mu\nu} F_{\mu\nu}^J
- \tfrac 14 \cD^\mu \sigma^{I\, ij} \cD_\mu \sigma^J{}_{ij}
-\tfrac 12  \bar \lambda^{I\,i} \slashed{\cD} \lambda^J_{\,i}
 \Big] \nn
\\[.1cm]
&- \tfrac{i}{\sqrt 2} f_{IJK} \sigma^{I\, ij} \bar \lambda^J_{\,i} \lambda^K_{\,j}
- \tfrac{1}{16} f_{IJKL} \sigma^{I\,ij} \sigma^K{}_{ij} \sigma^{J\,kl} \sigma^L{}_{kl} \ ,\label{SYMaction}
\end{align}
which is the maximally supersymmetric on-shell Yang-Mills action in five dimensions.
It is invariant under the following $\cN=4$ supersymmetry transformations:
\begin{align}
 \delta A^I_\mu &= \tfrac 12 \bar\epsilon^i \gamma_\mu \lambda^I_i \ , \nn \\[.1cm]
 \delta \sigma^{I\,ij} &= - i \sqrt 2  \Big( \bar\epsilon^{[i|} \lambda^{J|j]} 
 + \tfrac 14 \Omega^{ij} \bar \epsilon^k \lambda^I_k \Big) \nn \\[.1cm]
 \delta \lambda^{I\,i} & = -\tfrac 14 F^I_{\mu\nu} \gamma^{\mu\nu} \epsilon^i
 - \tfrac{i}{\sqrt 2}  \slashed{\cD} \sigma^{I\,ij} \epsilon_j
 + \tfrac 12  f^I_{\phantom{I}JK} \sigma^{J\, ij} \lambda^{K}_{\,jk} \epsilon^k \ .
\end{align}

In a completely analogous way we can treat the second special case of section \ref{Integrating_out},
the Abelian version of the full Kaluza-Klein tower given by the Lagrangian \eqref{cL-free}.
We will assume that the same H-analyticity relations as in the zero mode case
are imposed on all fields.  
The integration over harmonic variables is carried out
taking into account the expressions \eqref{eq:sigmasigma} and \eqref{eq:lambdalambda}
for the singlet component in the $\mathbf 5 \otimes \mathbf 5$ and $\mathbf 4 \otimes \mathbf 4$
channels, respectively. The result
 is the $\cN=4$ free theory of one
massless vector multiplet along with a tower of 
massive tensor multiplets,
\begin{align} 
g^2 \cL_{\cN=4}^{\rm free}  = & \;
 d_{IJ} \Big[ 
- \tfrac 14   F^{I\, \mu\nu} F^J_{\mu\nu}
- \tfrac 14   \partial^\mu \sigma^{I\, ij} \partial_\mu \sigma^J_{\,ij} 
- \tfrac 12   \bar\lambda^{I\,i} \slashed{\partial} \lambda^J_{\, i} 
\Big]
 \nn \\[.1cm]
& + \sum_{n=1}^\infty
 d_{IJ} \Big[
- \tfrac 12  \bar F^{I\, \mu\nu}_n F^J_{n\, \mu\nu}  
 + \tfrac {i}{4 m_n}  \epsilon^{\mu\nu\rho\lambda\sigma} \bar F^I_{n\,\mu\nu} \partial_\rho F^J_{n\, \lambda \sigma} 
 \nn \\[-.1cm]
 & \phantom{+ \sum_{n=1}^{\infty}   d_{IJ} \Big[}
-  \tfrac 12  \partial^\mu \bar \sigma^{I\,ij}_n \partial_\mu \sigma^J_{n\, ij}
-   \bar\lambda^{I\, i}_n \slashed{\partial} \lambda^J_{n\,i}
 - \tfrac 12 m_n^2 \bar\sigma^{I\, ij}_n \sigma^J_{n\,ij}
- m_n  \bar\lambda^{I\,i}_n \lambda^J_{n\,i}  \Big ] \ .
\end{align}
It is straightforward to check that this Lagrangian 
corresponds to the circle compactification of the six-dimensional 
$(2,0)$ pseudoaction for non-interacting 
tensor multiplets, given in \eqref{6Dab_action}.

%%%%%%%%%%%%%%%%%%%%%%%%%%%%%%%%%%%%%%%%%%%%%%%%%%%%%%%%%%%%%
\subsection{Grassmann analytic coset construction} \label{Sec:coset}

It is an interesting question in which sense the action discussed in the previous section 
is indeed $\cN=4$ supersymmetric. As we have pointed out repeatedly, what 
we have achieved is, starting from an $\cN=2$ supersymmetric action with 
manifest $SU(2)_R\times SU(2)$ symmetry, to enhance $SU(2)_R$ to $USp(4)_R$ with 
the help of additional compact coordinates and present a manifestly invariant action 
formulated as a group integral over the latter. In this section we wish to make the supersymmetric 
properties more transparent by interpreting (\ref{USp4action}) as a Grassmann analytic 
formulation of a fully $\cN=4$ supersymmetric action. There are two issues which we 
need to discuss: First of all, we need to construct explicitly a G-analytic space 
and recast (\ref{USp4action}) in this form, which will make the supersymmetry properties 
manifest. Secondly, concerning the coset of physical scalars, we need to discuss how the vector 
multiplet K\"ahler cone and the hyper K\"ahler manifold of the hypermultiplet scalar 
fields arise naturally as coset constructions in this framework, with the harmonic 
coordinates as a common ingredient to both of them. 

%%%%%%%%%%%%%%%%%%%%%%%%%%%%%%%%%%%%%%%%%%%%%%%%%%%%%%%%%%%%%
We first want to interpret the (projected) multiplets in (\ref{Harmonization}) 
and (\ref{ProjectionFerm}) as representations of a particular $\cN=4$ superconformal 
algebra, which we will now construct as a so-called Grassmann-analytic (or G-analytic) 
coset. Our discussion will be strongly inspired by similar constructions in four 
dimensions with $\cN=2$ and $\cN=4$ supersymmetry (see particularly \cite{HSS,Antoniadis:2007cw,Antoniadis:2009nv}). 
When we denote the 5-dimensional Poincar\'e superalgebra with 16 supercharges by 
$\mathfrak{P}^{(5|16)}$ then the $\cN=4$ action is formulated on the coset 
\begin{align}
\mathbb{R}^{(1,4|16)}\sim\frac{\mathfrak{P}^{(5|16)}}{(M_{\mu\nu},T_{ij})}\,.\label{FlatSuper}
\end{align}
Here we have introduced explicit generators for various symmetries, according to the following table
\begin{center}
\begin{tabular}{ccc}\hline
&&\\[-12pt]
\textbf{symmetry} & \textbf{generators} & \textbf{indices}\\\hline
&&\\[-10pt]
Lorentz transformations & $M_{\mu\nu}$ & $\mu,\nu=0,1,\ldots,4$ \\[2pt] 
translations & $P_\mu$ & \\[2pt]
$USp\,(4)_R$ R-symmetry & $T_{ij}$ & $i,j=1,\ldots,4$\\[2pt]
Poincar\'e supersymmetry & $Q_i$ & $i=1,\ldots,4$\\[2pt]\hline
\end{tabular}
\end{center}
We note, however, that in general (due to the absence of appropriate systems of 
full $\cN=4$ auxiliary fields) Poincar\'e  supersymmetry transformations will be 
realized in a non-linear manner in the action.
For the sake of simplicity, no central charges are considered in the following
construction. Nonetheless, we are confident that our formalism 
can be suitably generalized to accommodate them. Such extension 
could be helpful in the analysis of a possible six-dimensional
origin of the five-dimensional symmetry algebra.

To formulate the action (\ref{USp4action}), however, we want to have an alternative 
coset, in which half the supergenerators, for example $Q_{i=3,4}$, 
have been moved to the coset denominator. However, this would lead to inconsistencies, 
since the $USp(4)_R$ generators $T_{ij}$, which are already present in the denominator 
in (\ref{FlatSuper}) will require adding $Q_{i=1,2}$ as well, to 
form a closed algebra. To remedy this problem, we will now discuss a G-analytic coset 
obtain by harmonization of (\ref{FlatSuper}). Indeed, we can consider a particular semi-direct 
product of (\ref{FlatSuper}) with the coset space (\ref{CosetDef})
\begin{align}
\frac{\mathfrak{P}^{(5|16)}}{(M_{\mu\nu},T_{ij})}\ltimes \frac{\widehat{USp\,(4)}}{\parbox{2.8cm}{\vspace{0.1cm}$\widehat{SU(2)}_R\times \widehat{SU(2)}$}}=
\frac{\mathfrak{P}^{(5|16)}\ltimes \widehat{USp\,(4)}}{\parbox{6.4cm}{\vspace{0.1cm}$(M_{\mu\nu},Q_{\da},T_{ab},T_{\da\dbi},T_{a\da}-Z_{a\da},Z_{ab},Z_{\da\dbi})$}}\,,\label{SemiDirectCoset}
\end{align}
where in addition we have introduced the generators
\begin{center}
\begin{tabular}{ccc}\hline
&&\\[-12pt]
\textbf{symmetry} & \textbf{generators} & \textbf{indices}\\\hline
&\\[-8pt]

$\widehat{SU(2)}_R$ & $Z_{ab}$ & $a,b=1,2$ \\[2pt]
$\widehat{SU(2)}$ & $Z_{\da\dbi}$ & $\da,\dbi=1,2$ \\[2pt]
$\widehat{USp(4)}$ & $Z_{ab}\,,Z_{a\da}\,,Z_{\da\dbi}$ & \\[2pt]\hline
\end{tabular}
\end{center}
where the supergenerators appear in a harmonically projected form
\begin{align}
&\begin{array}{l}Q^a=u_i^a Q^i\\[6pt] Q^{\da}=u_i^{\da} Q^i\end{array}\,,&&\text{with}&& \frac{\widehat{USp\,(4)}}{\parbox{2.8cm}{\vspace{0.1cm}$\widehat{SU(2)}_R\times \widehat{SU(2)}$}}\sim(u_i^a,u_i^{\da})\,.\label{SuperProjection}
\end{align}
In these expressions we have used a hat on the groups appearing in the coset to distinguish them from their counterparts in section \ref{Sect:CosetConstruction}. 
Notice that in the denominator of (\ref{SemiDirectCoset}) only the combination $T_{a\da}-Z_{a\da}$ is present. While each 
of the $USp(4)_R$ generator $T_{a\da}$ and the $\widehat{USp(4)}$ generator $Z_{a\da}$ separately, would transform 
$Q_{\da}$ into $Q_{a}$, the difference of both acts trivially on 
all supergenerators. In this way, it is consistent to put only half of the (harmonically projected) 
supergenerators in the coset denominator. Notice that this coset realization makes precise the 
discussion towards the end of section~\ref{Sect:CosetConstruction}: for any fixed choice of harmonic 
variables $(u_i^a,u_i^{\da})\in\widehat{USp(4)}/(\widehat{SU(2)}_R\times \widehat{SU(2)})$ eqn.~(\ref{SemiDirectCoset}) 
describes the $\cN=2$ coset $\mathfrak{P}^{(5|8)}/(M,Z_{ab})$. However, the interplay with the harmonic 
variables discussed in (\ref{CosetDef}) allows us to interpret eqn.~(\ref{SemiDirectCoset}) as a 
Grassmann-analytic version of an $\cN=4$ algebra.

Before explicitly constructing the coset space (\ref{SemiDirectCoset}) we comment that from here 
it becomes transparent why the fields in the proposed action (\ref{USp4action}) have an 
implicit harmonic dependence: Indeed, since $\cL_{\cN=2}$ in  (\ref{USp4action}) is understood to vary over 
(\ref{SemiDirectCoset}) the fields entering it are required to be representations of 
the latter. Thus, once we choose an explicit parametrization, all multiplets will be 
non-trivial functions of these parameters.\footnote{We remind the reader again that 
the supersymmetry transformations are realized in a non-linear manner in (\ref{USp4action}) 
and thus the corresponding Grassmann coordinates have already been integrated out.} This 
particularly means that all fields generically have a non-trivial expansion in the spherical harmonics on $S^4$.

What remains to be done from a purely algebraic point of view is to give some hints 
on the actual construction of the coset, in particular the harmonic coordinates. 
Our considerations will mirror a similar discussion in four dimensions~\cite{Antoniadis:2007cw} 
(see also \cite{Antoniadis:2009nv}). Indeed, in addition to the coset 
parameters $(u_i^a,u_i^{\da})$ of eqn.~(\ref{SuperProjection}) we can introduce 
another set of harmonic coordinates ${\kappa_i}^{\hat{\imath}}$ of $USp(4)$. 
They satisfy the constraints
\begin{align}
&{\kappa_{\hat{\imath}}}^i{\kappa_i}^{\hat{\jmath}}=\delta^{\hat{\imath}}_{\hat{\jmath}}\,,&& 
  {\kappa_i}^{\hat{\imath}}{\kappa_{\hat{\imath}}}^j=\delta^{j}_{i}\,,&&\epsilon_{\hat{\imath}\hat{\jmath}\hat{k}\hat{l}}\,
  {\kappa_{i}}^{\hat{\imath}}{\kappa_{j}}^{\hat{\jmath}}{\kappa_{k}}^{\hat{k}}{\kappa_{l}}^{\hat{l}}=\epsilon_{ijkl}\,,
\end{align}
where for convenience of the reader we have denoted $\widehat{USp\,(4)}$ indices in this section 
with a hat to avoid confusion. Indeed, these harmonics transform in a two-fold manner
\begin{align}
\delta{\kappa_i}^{\hat{\imath}}={\eta_i}^j\,{\kappa_j}^{\hat{\imath}}+{\kappa_i}^{\hat{\jmath}}\,{\tau_{\hat{\jmath}}}^{\hat{\imath}}
\end{align}
under rigid $\widehat{USp\,(4)}$ transformations (with parameter ${\tau_{\hat{\jmath}}}^{\hat{\imath}}$) 
and local $USp(4)$ (with parameter ${\eta_i}^j$). In order to construct the coset (\ref{SemiDirectCoset}) 
we will have to make a change of variables from $(u_{\hat{\imath}}^a,u_{\hat{\imath}}^{\da},\kappa_i^{\hat{\imath}})$ 
to new harmonics $(w_i^{a},w_i^{\da},z_i^{\hat{\imath}})$ which are inert under rigid $\widehat{USp\,(4)}$ 
transformations and transform in a simple manner under local $USp\,(4)$, such that we can covariantly 
impose the constraint
\begin{align}
(T_{a\da}-Z_{a\da})F=0\,,\label{CosetConstraint}
\end{align}
on covariant objects $F$. The first step is to introduce harmonic projections of $\kappa$
\begin{align}
&{\kappa_a}^b=u_a^i{\kappa_i}^{\hat{\jmath}}u_{\hat{ \jmath}}^{b}\,,&& {\kappa_{\da}}^{\dbi}= 
   u_{\da}^i{\kappa_i}^{\hat{\jmath}}u_{\hat{\jmath}}^{\dbi}\,,&& {\kappa_{a}}^{\dbi}=
   u_{a}^i{\kappa_i}^{\hat{\jmath}}u_{\hat{\jmath}}^{\dbi}\,,&& {\kappa_{\da}}^{b}=u_{\da}^i{\kappa_i}^{\hat{\jmath}}u_{\hat{\jmath}}^{b}\,.
\end{align}
We then perform the following non-linear change of variables
\begin{align}
&{z_a}^b={\kappa_a}^b\,,&&{z_{\da}}^{\dbi}={\kappa_{\da}}^{\dbi}\,,&&{z_{a}}^{\da}={\kappa_{a}}^{\dbi}{(\kappa^{-1})_{\dbi}}^{\da}\,,&&{z_{\da}}^{a}={\kappa_{\da}}^{b}{(\kappa^{-1})_{b}}^{a}\,,\\
&w_i^{a}=u_i^a+u_i^{\dbi}{z_{\dbi}}^a\,,&&w_i^{\da}=u_i^{\da}\,,&&w^i_{a}=u^i_a-{z_a}^{\dbi}u_{\dbi}^i\,,&&w_{\da}^i=u_{\da}^i\,.
\end{align}
We notice that the new harmonics $(w_i^a,w_i^{\da})$ are no longer unitary, 
since they do not satisfy the analogue of (\ref{HarmonicConjugation}). 
However, upon introducing the parameter
\begin{align}
&{\hat{\eta}_a}\phantom{}^b=w_a^i\,{\eta_i}^j\,w_j^b\,,&&{\hat{\eta}_a}\phantom{}^{\dbi} = 
  w_a^i\,{\eta_i}^j\,w_j^{\dbi}\,,&&{\hat{\eta}_{\da}}\phantom{}^b = 
  w_{\da}^i\,{\eta_i}^j\,w_j^b\,,&&{\hat{\eta}_{\da}}\phantom{}^{\dbi}=w_{\da}^i\,{\eta_i}^j\,w_j^{\dbi}\,,
\end{align}
the new harmonic variables transform in the following manner under local $USp\,(4)$ transformations
\begin{align}
&\delta w_i^a=w_i^{\dbi}\,{\hat{\eta}_{\dbi}}\phantom{}^a\,,&&\delta w_i^{\da}=0\,,&&\delta w^i_{a}=-{\hat{\eta}_{a}}\phantom{}^{\da}\,w^i_{\da}\,,&&\delta w^i_{\da}=0\,.
\end{align}
As we can see, these new variables no longer mix under local $USp\,(4)$, 
therefore allowing us to covariantly introduce the constraint (\ref{CosetConstraint}).

After discussing the G-analytic coset, we now also have 
to discuss the geometrical aspects of this construction. We want to understand 
how the $\cN=2$ vector and hypermultiplet moduli spaces discussed in section~\ref{N=2spectrum}, 
fit together with the harmonic variables. Following \cite{Cremmer:1980gs}, the structure of 
the space of physical scalar fields in $\cN=4$ supergravity in five dimensions takes the form 
of a coset $\mathfrak{G}^{\text{SUGRA}}/USp\,(4)$. In the rigid limit this structure is 
preserved with $USp\,(4)$ playing the role of the five-dimensional R-symmetry 
group.\footnote{In the Abelian case the manifold is simply $SO(5,n)/(SO(5)\times SO(n))\times SO(1,1)$ 
with the last factor representing the graviscalar, which simply factorizes out in the rigid limit.} 
We will therefore consider the $\cN=4$ moduli space in the following to be of the form 
\begin{align}
\frac{\mathfrak{G}}{\parbox{1.3cm}{\vspace{0.1cm}$\widehat{USp\,(4)}$}}\sim (\sigma^{I \, ij},\sigma^{M \, ij})\,,\label{N4ModuliSpace}
\end{align} 
where we particularly note the absence of the scalar fields of the additional multiplet $\widehat{\mathcal{V}}^0$ 
on the right hand side. Moreover, as in section~\ref{sec:susyactions}, 
we will drop the index of the Kaluza-Klein mode in this section to avoid cluttering 
of the formulas. In equation (\ref{N4ModuliSpace}) $\mathfrak{G}$ is a (possibly non-compact) 
group manifold which explicitly depends on the choice of gauge-group $G$. Following the logic 
of section~\ref{Sect:CosetConstruction} we will have to reformulate this manifold in terms of 
$\cN=2$ language in an $USp\,(4)$ covariant manner in order to make contact with the geometrical 
discussion in section~\ref{Gaugings_and_bosonic_action}. We indeed expect that the $\cN=2$ vector- 
and hypermultiplet moduli spaces can also be understood as particular harmonizations of the 
coset (\ref{N4ModuliSpace}) for fixed values of the harmonic variables.

We will begin by considering the $\cN=2$ vector multiplet sector discussed in 
section~\ref{Gaugings_and_bosonic_action} by introducing the following harmonized coset
\begin{align}
\mathcal{M}_{\text{VT}}
   =\frac{\mathfrak{G}}{USp\,(4)}\ltimes\frac{\widehat{USp\,(4)}}{\parbox{2.8cm}{\vspace{0.1cm}$\widehat{SU(2)}_R\ltimes \widehat{SU(2)}$}}
   =\frac{\mathfrak{G}\ltimes \widehat{USp\,(4)}}{\parbox{5.45cm}{\vspace{0.1cm}$(L^{\mathcal{I}}_{a\da},T_{ab},T_{\da\dbi},T_{a\da}-Z_{a\da},Z_{ab},Z_{\da\dbi})$}}
   \sim (\epsilon_{ab}w_i^{a}w_{j}^b\sigma^{\mathcal{I} \, ij},w_i^a,w_i^{\da})\label{HarmonizedCosetVector}
\end{align}
where we recall that
\begin{align}
\epsilon_{ab}w_i^{a}w_{j}^b\sigma^{\mathcal{I} \, ij}=-\epsilon_{\da\dbi}w_i^{\da}w_{j}^{\dbi}\sigma^{\mathcal{I} \, ij}\,,
\end{align}
since $\sigma^{\mathcal{I} \, ij}$ transforms in the $\mathbf{5}$ of $USp\,(4)$. 
Moreover, in (\ref{HarmonizedCosetVector}) we have explicitly introduced coset 
representatives for the manifold in equation (\ref{N4ModuliSpace})
\begin{center}
\begin{tabular}{ccc}\hline
&&\\[-12pt]
\textbf{object} & \textbf{generators} & \textbf{indices}\\\hline
&&\\[-10pt]
$\mathfrak{G}/USp\,(4)$ & $L^\mathcal{I}_{a\da}\,,L^{\mathcal{I}}_{ab}\,,L^\mathcal{I}_{\da\dbi}$ & $\mathcal{I}=(I,M)$\\[2pt]\hline
\end{tabular}
\end{center}
In the same manner, we can deal with the hypermultiplet moduli space, which 
we can realize as the coset
\begin{align}
\mathcal{M}_{\text{H}}
   =\frac{\mathfrak{G}}{USp\,(4)}\ltimes\frac{\widehat{USp\,(4)}}{\parbox{2.8cm}{\vspace{0.1cm}$\widehat{SU(2)}_R\ltimes \widehat{SU(2)}$}}& 
   =\frac{\mathfrak{G}\ltimes \widehat{USp\,(4)}}{\parbox{6.2cm}{\vspace{0.1cm}$(L^{\mathcal{I}}_{ab},L^{\mathcal{I}}_{\da\dbi},T_{ab},T_{\da\dbi},T_{a\da}-Z_{a\da},Z_{ab},Z_{\da\dbi})$}}\nonumber\\
   &\sim (w_i^{a}w_{j}^{\da}\sigma^{\mathcal{I} \, ij},w_i^a,w_i^{\da})\label{HarmonizedCosetHyper}
\end{align}
As we can see, the moduli spaces of $\cN=2$ vector/tensor- and hypermultiplets 
can be understood as particular G-analytic coset representations in the limit 
of fixed harmonic variables. Taking them rather to be non-trivial coordinates 
in the $S^4$ manifold (\ref{CosetDef}) combines them into the $\cN=4$ manifold (\ref{N4ModuliSpace}).

We note that in this discussion, we have left the $\cN=4$ moduli space $\mathfrak{G}$ undetermined and we have constructed 
the $\cN=2$ vector- and hyper multiplet moduli spaces from it. It is an interesting question, whether the process can 
be reversed and knowledge of the latter allows to construct $\mathfrak{G}$.

\section{Six-dimensional theory and one-loop Wess-Zumino terms}\label{Sect:LoopCheck}

In the previous sections we have presented a five-dimensional action that encodes the dynamics 
of an infinite tower of massive non-Abelian tensors coupled to Yang-Mills theory. We 
have proposed that this action describes a six-dimensional $(2,0)$ theory compactified 
on a circle. While it remains to be understood how precisely the former arises through a 
Kaluza-Klein reduction from six dimensions, there is a slightly different question one can 
pose: To what extent is it possible to extract information about the $(2,0)$ theory from 
the five-dimensional action (\ref{cL_split})? In particular, since all our discussion so far 
has been purely on a classical level, one might worry that quantum effects could be spoiled 
through our particular lower-dimensional treatment. In this section we want to present a brief 
one-loop check, which hints to the fact that this is not the case and indeed information about 
the six-dimensional theory can in principle be reliably computed in the compactified theory. 
Indeed, in the recent work \cite{Bonetti:2012fn} we have demonstrated in a non-supersymmetric 
context that anomalies are accessible through lower dimensional theories obtained through circle 
compactifications.\footnote{See also \cite{Czech:2011dk} for a recent analysis of scattering amplitudes in theories for multiple M5 branes.} In the same spirit we therefore expect that it should be possible to make 
non-trivial statements about anomalies of the six-dimensional $(2,0)$ theory by studying quantum 
effects (i.e.~one-loop corrections) of (\ref{cL_split}). The quantity we want to discuss in this 
context is the conformal anomaly of multiple M5 branes, which has 
attracted a lot of interest (see \cite{Henningson:1998gx,Harvey:1998bx,Intriligator:2000eq,Bastianelli:2000hi,Yi:2001bz,Maxfield:2012aw}). 
We want to understand whether or not this anomaly is in 
principle accessible from our proposed five-dimensional 
action (\ref{cL_split}).

To this end, we will closely follow a discussion in \cite{Intriligator:2000eq}, where a stack of 
$N$ M5 branes is considered and $SO(5)_R$ R-symmetry is gauged with the help of a background 
connection. A comparison is made between the full non-Abelian phase and the broken phase where 
one M5-brane has been separated from the stack. Indeed, a deficit in the quantum anomalies is 
found which is argued to be counterbalanced by adding an anomalous Wess-Zumino term $S_{\rm WZ}$ 
to the classical action with a characteristic scaling behavior in the number of M5 branes $N$. 
In this section, we will not discuss how $S_{\rm WZ}$ is generated in our five-dimensional approach, 
but we will test whether the scaling behavior is washed out through the five-dimensional effective 
treatment of the theory. To be precise, we will assume the presence of $S_{\rm WZ}$ in the action 
and show that changing the radius $r$ (and thus integrating out massive Kaluza-Klein modes in the 
effective action) will not modify its scaling behavior for large $N$.

%%%%%%%%%%%%%%%%%%%%%%%%%%%%%%%
\subsection{Coupling to background fields}
Our discussion follows closely the proposal made in \cite{Intriligator:2000eq} and 
we will therefore use $SO(5)_R$ notation rather than $USp\,(4)_R$. 
To this end we introduce the indices $A=1,\ldots,5$ in the representation $\mathbf{5}$ of 
$SO(5)_R$. All our expressions, however, can be converted into $USp(4)_R$ in a straightforward manner. 
In \cite{Intriligator:2000eq} the anomalies of $(2,0)$ theories are studied by 
gauging the R-symmetry $SO(5)_R$ with a background connection $\mathbf{\Upsilon}^{AB}_{\boldsymbol{{\mu}}}=-\mathbf{\Upsilon}^{BA}_{\boldsymbol{\mu}}$ 
with field strength $\boldsymbol{\mathfrak{F}}^{AB}_{\boldsymbol{{\mu}}\boldsymbol{{\nu}}}$, where 
$\boldsymbol{{\mu}},\boldsymbol{{\nu}}=0,...,5$ are six-dimensional space-time indices.

We consider an A-D-E gauge group $G$ and pick a Cartan generator $\mathfrak{T}$. 
Let us denote $\boldsymbol{\Phi}^A$ the linear combination of the
scalars $\boldsymbol{\sigma}^{A\,I}$ along the direction of $\mathfrak T$.
Giving $\boldsymbol{\Phi}^A$ a non-vanishing vacuum expectation value $\langle \boldsymbol{\Phi}^A \rangle $
breaks the gauge group to the little group of $\mathfrak T$,
\beq \label{Gbreaking}
H \times U(1) \subset G \ ,
\eeq
where we have singled out the $U(1)$ factor associated to $\mathfrak T$ itself.
In the following, it proves convenient to define the constrained field
\beq \label{def-varphi}
\boldsymbol{\varphi}^A =  \boldsymbol{\Phi}^A  / \sqrt{  \boldsymbol{\Phi}^B   \boldsymbol{\Phi}_B  } \ .
\eeq
It satisfies $\boldsymbol{\varphi}^A \boldsymbol{\varphi}_A = 1$ identically. Introduction of the  vacuum 
expectation value $\langle \boldsymbol{\Phi}^A \rangle $ also breaks the $SO(5)_R$ down to $SO(4) \cong SU(2)_R\times SU(2)$. While the scalar fields 
decompose as described in (\ref{rep_decomp}), the vector fields $\mathbf{\Upsilon}^{AB}_{\boldsymbol{\mu}}$ 
decompose in the following manner
\begin{align}
\mathbf{10}\,&\rightarrow\, (\mathbf{3},\mathbf{1})+(\mathbf{1},\mathbf{3})+(\mathbf{2},\mathbf{2})\,,\\
\mathbf{\Upsilon}^{AB}_{\boldsymbol{{\mu}}}\,&\rightarrow \hspace{0.1cm}\left(\mathbf{\Upsilon}^{(ab)}_{\boldsymbol{{\mu}}},
\hspace{0.05cm}\mathbf{\Upsilon}^{(\da\dbi)}_{\boldsymbol{{\mu}}},\mathbf{\Upsilon}^{a\da}_{\boldsymbol{{\mu}}}\right)\,,
\end{align}
into irreducible $SU(2)_R\times SU(2)$ representations.

The key point of the discussion of \cite{Intriligator:2000eq} is that even in this broken phase the anomaly
originally induced by the full massless spectrum of the theory with group $G$ has a non-trivial effect. In 
the broken phase one expects some states to acquire a mass. For sufficiently large masses these states 
have to be integrated out to formulate the new effective action for the unbroken theory with group $H \times U(1)$. 
Since the massless spectrum has be reduced in this breaking, a deficit in quantum anomalies is generated 
with respect to the unbroken phase. This deficit has to be counterbalanced by adding an anomalous  
Wess-Zumino term $S_{\rm WZ}$ to the classical action. Applying this logic to the background 
gauge field $\mathbf{\Upsilon}^{AB}_{\boldsymbol{{\mu}}}$ amounts to studying the analog of the 
't Hooft anomaly matching \cite{tHooft}.

To give the explicit form of $S_{\rm WZ}$ we introduce an auxiliary seven-dimensional manifold $\Sigma_7$
whose boundary is the six-dimensional space-time. All fields, including the 
background gauge fields $\mathbf{\Upsilon}^{AB}_{\boldsymbol{{\mu}}}$, are extended to $\Sigma_7$. 
The Wess-Zumino term takes the form \cite{Intriligator:2000eq} 
\beq \label{WZTerm}
S_{\text{WZ}}=\frac{c(G)-c(H)}{6}\int_{\Sigma_7} \mathbf{\Omega}_3({\boldsymbol{\varphi}},\mathbf{\Upsilon})\wedge d\mathbf{\Omega}_3({\boldsymbol{\varphi}},\mathbf{\Upsilon})\,,
\eeq
in order to balance the 't Hooft anomaly. 
On the right hand side of \eqref{WZTerm} we have introduced the three-form
$\mathbf{\Omega}_3({\boldsymbol{\varphi}},\mathbf{\Upsilon})$ that is determined locally by the condition
\begin{align} \label{eta4}
d\mathbf{\Omega}_3({\boldsymbol{\varphi}},\mathbf{\Upsilon}) \equiv  
\eta_4({\boldsymbol{\varphi}},\mathbf{\Upsilon}) = &
\frac{1}{64\pi ^2}  \epsilon_{A_1A_2A_3A_4A_5} \,
\Big[
D \boldsymbol{\varphi}^{A_1} \wedge D \boldsymbol{\varphi}^{A_2}
\wedge D \boldsymbol{\varphi}^{A_3} \wedge D \boldsymbol{\varphi}^{A_4} \\[.1cm]
& \qquad - 4 \; \boldsymbol{\mathfrak{F}}^{A_1 A_2} \wedge D \boldsymbol{\varphi}^{A_3} \wedge D \boldsymbol{\varphi}^{A_4}
+ 4 \; \boldsymbol{\mathfrak{F}}^{A_1 A_2} \wedge \boldsymbol{\mathfrak{F}}^{A_3 A_4} 
\Big] \boldsymbol{\varphi}^{A_5}\ . \nn
\end{align}
Of crucial interest are the group theoretical constants $c(G)$ and $c(H)$ appearing in 
\eqref{WZTerm}. A conjecture made in \cite{Intriligator:2000eq} is that $c(G) = c_2(G) |G|$, where $c_2(G)$ is the 
dual Coxeter number and $|G|$ is the dimension of $G$. For $A_{N}$ Lie algebras in the breaking $SU(N+1)\rightarrow SU(N) \times U(1)$ 
one has
\beq \label{cG-H}
 \frac{c(G)-c(H)}{6}=\frac{N(N+1)}{2}\ ,
\eeq
where we have inserted $c(SU(N))=N(N^2-1)$. The $A_N$ groups appear, for example, on the 
world-volume theory of a stack of M5-branes. For such setups the anomaly has been 
previously computed in \cite{Harvey:1998bx}. Since for theories with 16 supercharges the 
R-current and the stress-energy tensor are components of the same multiplet, the 't Hooft anomaly 
will be proportional to the conformal anomaly. Thus, the appearance of a term of the form (\ref{WZTerm}) 
allows us to draw conclusions about the conformal anomaly of the theory. 
We will investigate whether the scaling behavior of the prefactor of \eqref{WZTerm} is independent of 
an effective treatment of the five-dimensional compactified theory. To be precise, we will check whether 
a different scaling behavior is produced upon slightly changing the cut-off scale and thus integrating 
out five-dimensional Kaluza-Klein modes. If this was indeed the case, it would indicate that it is much 
harder to extract the conformal anomaly of six-dimensional $(2,0)$ theories from an effective 
five-dimensional treatment since it would be washed out by quantum effects. Fortunately, we will 
be able to argue in the following that this is not the case.

%%%%%%%%%%%%%%%%%%%%%%%%%%%%%%%%%%%%%%%%%%%%%%%%%%%%%%%
\subsection{One-loop correction of the five-dimensional action}

To infer the couplings to $\mathbf{\Upsilon}^{AB}_{\boldsymbol{{\mu}}}$ in the 
five-dimensional theory we first replace all covariant derivatives \eqref{covariant_derivative_gen} for fields charged under 
the R-symmetry group by their $SO(5)_R$ covariantizations.   
Since the tensors $B_n^I$ are neutral under $SO(5)_R$ their covariant derivatives will not 
admit the standard electric gauging. However, one expects that the correct 
$SO(5)_R$ covariantizations involves the five-dimensional Kaluza-Klein reductions $\Omega_{3\, n}$ of the 
three-form introduced in \eqref{eta4}. Such covariantizations have been discussed in six dimensions in \cite{Ganor:1998ve,Intriligator:2000eq}. For the tensors
corresponding to the U(1) in \eqref{Gbreaking} one has 
\beq \label{Y_cov}
  \widehat \cD B_n = dB_n + \alpha\, \Omega_{3 \, n }\ ,
\eeq
where $\alpha$ is a constant which we will discuss in more detail below. 
Due to the common six-dimensional origin of $\alpha$ in (\ref{Y_cov}), 
we will assume in the following that this charge is the same for all Kaluza-Klein modes $B_n$.
To determine the form of $\Omega_{3\, n}$ we make the following Kaluza-Klein 
reduction ansatz 
\beq
   \mathbf{\Upsilon}^{A}= \Upsilon^{A}+\mathfrak{y}^{A} (dy-A^0)\ . 
\eeq
Note that in contrast to the reduction \eqref{basic_KK} of dynamical fields of the six-dimensional theory
we have only kept the zero modes in Kaluza-Klein expansion. This is analogous to the reduction 
of the six-dimensional metric presented in \eqref{KK_metric}.
We will denote the Kaluza-Klein modes of the scalar fields $\boldsymbol{\varphi}^{A}$ by $\varphi_n^{A}$.

The modification \eqref{Y_cov} of the covariant derivatives will give rise to new interaction terms 
in the Lagrangian (\ref{cL_split}). In particular one finds the modified 
Chern-Simons term for the massive tensors of the form
\beq
    \epsilon^{\mu \nu \rho \lambda \sigma} \bar F_{n \mu \nu}\widehat \cD_{\rho}  F_{n \lambda \sigma}\ .
\eeq
One expects that similar to the arguments in \cite{Ganor:1998ve} the correction term involving $\Omega_3$ 
is obtained as a one-loop correction when moving to the broken phase \eqref{Gbreaking}.
A complete analysis of these terms will be performed elsewhere, but for the moment 
we want to focus on two particular additional structures
\begin{align} \label{SkyInteractions}
\mathcal{L}_{\text{bgd}}=\alpha\, \sum_{n=0}^\infty\epsilon_{A_1A_2A_3A_4A_5}\left[in B_n\wedge \Upsilon^{A_1A_2}\wedge \mathfrak{F}^{A_3A_4}\,\bar{\varphi}^{A_5}_n+B_n\wedge \mathfrak{F}^{A_1A_2}\wedge 
d\mathfrak{y}^{A_3A_4}\bar{\varphi}_n^{A_5}\right]+\ldots\,.
\end{align}
Using (\ref{SkyInteractions}) to compute one-loop corrections to the classical 
action (\ref{cLn_action}) we encounter new interaction vertices of the form
 \begin{align}
 &\parbox{2.4cm}{\epsfig{file=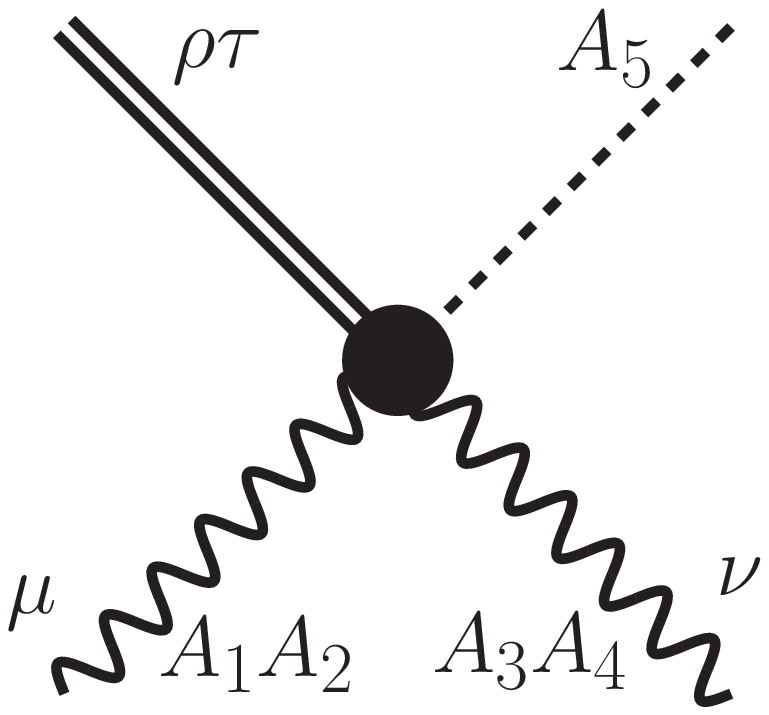,width=2.4cm}}\hspace{0.5cm}\sim\hspace{0.5cm} \alpha\, n \epsilon^{\mu\nu\rho\tau\sigma}p_{\sigma}\,\epsilon_{A_1A_2A_3A_4A_5}\,,\label{Vertex1}\\[10pt]
 &\parbox{2.4cm}{\epsfig{file=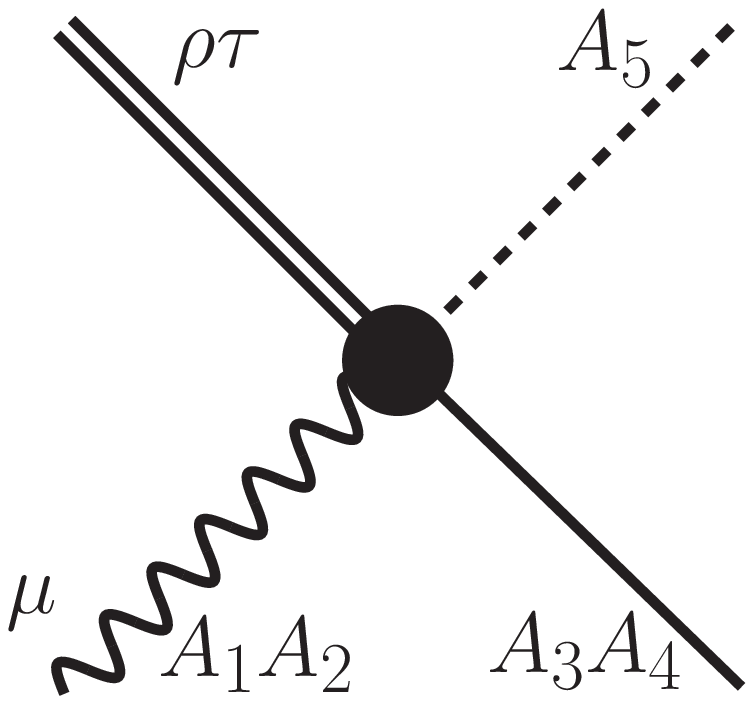,width=2.4cm}}\hspace{0.5cm}\sim\hspace{0.5cm} \alpha\, \epsilon^{\mu\rho\tau\nu\sigma}p^{(1)}_{\nu}p^{(2)}_\sigma\,\epsilon_{A_1A_2A_3A_4A_5}\,.\label{Vertex2}
 \end{align}
Here we have denoted the field $\Upsilon$  with wiggly lines, the scalar field $\mathfrak{y}$ 
with solid lines, the scalar field $\varphi$ with dashed lines, and the tensor field $B_n$ with 
double lines. From the Lagrangian \eqref{cL_split} one infers that the propagators for the dynamical fields are given schematically by
 \begin{align}
 \parbox{1.9cm}{\epsfig{file=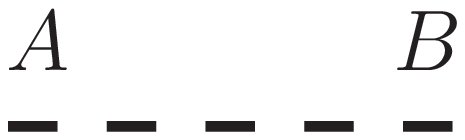,width=1.9cm}}\hspace{1cm}&\sim\hspace{1cm}\frac{\delta_{AB}}{p^2+m_n^2}\,,\\
 \parbox{2cm}{\epsfig{file=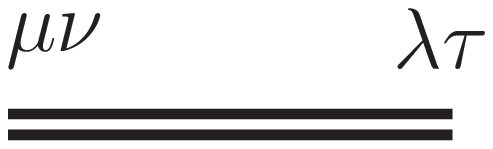,width=2cm}}\hspace{1cm}&\sim\hspace{1cm}  \frac{ 2 \delta_{[\mu}^{[\lambda} \delta_{\nu]}^{\tau]} 
  - 4 m_n^{-2} p_{[\mu} \delta_{\nu]}^{[\lambda} p^{\tau]} + m_n^{-1} \epsilon_{\mu \nu \rho}^{\phantom{\mu \nu \rho}\lambda \tau} p^\rho }{{p^2+m_n^2}}\,.\label{Prop2}
 \end{align}
We note that these quantities are still formulated in an $SO(5)_R$ covariant form. However, 
as already remarked previously, the introduction of the scalar vacuum expectation values 
breaks this to $SU(2)_R\times SU(2)$. In this way all tensor structures in  (\ref{Vertex1})---(\ref{Prop2}) should be decomposed accordingly. 
Since this will not be important for our argument, we will not perform this task here, but we 
will rather focus on one particular $SU(2)_R\times SU(2)$ channel.\footnote{We can think of this 
as considering one particular projection with harmonic variables according to the logic 
of section~\ref{sec:N=4completion}. }

\begin{figure}[h!]
 \begin{center}
 \rotatebox{270}{\parbox{2cm}{\epsfig{file=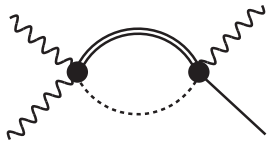,width=2cm}}}
 \caption{One-loop four-point diagram with external background fields.}
 \label{Fig:Loop1}
 \end{center}
 \end{figure}

With these ingredients, we can construct the 4-point diagram at one-loop level depicted 
in figure~\ref{Fig:Loop1}.
This diagram corresponds to a term in the effective action of the form
\begin{align}
\alpha^2\,\mathfrak{y}^{a\da}\Upsilon_{a\da}\wedge \mathfrak{F}^{b\dbi}\wedge \mathfrak{F}_{b\dbi}+\ldots\,,
\end{align}
where the dots denote further $SU(2)_R\times SU(2)$ channels. 
However, we would like to point out that this term is proportional to $\alpha^2$ due to the 
fact that figure~\ref{Fig:Loop1} contains two vertices of the type (\ref{Vertex1}) and (\ref{Vertex2}). 
In \cite{Intriligator:2000eq} it has been proposed that $\alpha$ is of order $N$ for $N\gg 1$, which leads 
to a scaling behavior of the form $N^2$. Moreover, comparing to (\ref{WZTerm}) we see that 
these are precisely the same terms that arise upon dimensional reduction to five dimensions once some of the scalar 
fields $\boldsymbol{\varphi}$ in the definition of $\eta_4$ in eqn.~(\ref{eta4}) are replaced by 
their vacuum expectation values and the normalization \eqref{def-varphi} is taken into account. 
We thus conclude that the scaling behavior of the Wess-Zumino term (\ref{WZTerm}), predicted for reasons of anomaly 
cancellation in \cite{Intriligator:2000eq}, is not modified through integrating out Kaluza-Klein modes in the 
effective five-dimensional action. Any such contributions will only modify the numerical prefactor 
consistently with a change of the cut-off scale. 

We repeat again that this analysis only indicates that information of the six-dimensional theory at the 
quantum level is not spoiled through an effective treatment of the compactified theory. In particular, 
the calculations of this section do not allow us to draw direct information about the conformal 
anomaly of $(2,0)$ theories. The computation required to analyze the conformal anomaly involves
integrating out all massive modes in the breaking \eqref{Gbreaking} starting from our non-Abelian action 
and will be the subject of a future publication.

\newpage

%%%%%%%%%%%%%%%%%%%%%%%%%%%%%%%%%%%%%%%%%%%%%%%%%%%%%%%%%%%%%%%%%%%%%%%%%%%%%%%%%%%%%%

%%%%%%%%%%%%%%%%%%%%%%%%%%%%%%%%%%%%%%%%%%%%%%%%%%%%%%%%%%%%%%%%%%%%%%%%%%%%%%%%%%%%%%
\section{Discussions and conclusions}
%%%%%%%%%%%%%%%%%%%%%%%%%%%%%%%%%%%%%%%%%%%%%%%%%%%%%%%%%%%%%%%%%%%%%%%%%%%%%%%%%%%%%%

With the motivation to describe the dynamics of $(2,0)$ self-dual non-Abelian tensors 
in six dimensions, we have studied five-dimensional supersymmetric 
Kaluza-Klein inspired actions. 
Our starting point is the five-dimensional action \eqref{purely_bosonic_action}
for the Kaluza-Klein tower of modes of six-dimensional Abelian self-dual tensors 
coupled to the Kaluza-Klein vector. 
Our efforts have been directed towards a generalization of this action
that can accommodate many relevant features   
expected from non-Abelian $(2,0)$ theories. 
We propose the superconformal action \eqref{ActionComplete}, 
together with \eqref{cL_split} in which conformal symmetry is fixed,
as a promising candidate of a five-dimensional theory 
encoding non-trivial information about the physics of $(2,0)$
non-Abelian tensors. 

Indeed, the action \eqref{ActionComplete} possesses some key features that make it 
particularly attractive in this respect.
Firstly, its spectrum
contains the Kaluza-Klein towers of all bosonic and fermionic degrees of freedom 
that are expected upon circle compactification of six-dimensional $(2,0)$ tensor multiplets.
Secondly, these Kaluza-Klein towers are gauged under a non-Abelian group $G$. 
The zero modes of six-dimensional tensors play a distinct role, as they are
used as gauge connections. All other bosonic and fermionic fields, however,
are treated on the same footing and transform in the adjoint representation
of $G$.
Furthermore, all couplings of the action \eqref{ActionComplete}
are determined exclusively in terms of group theoretical invariants and are written as sums over all Kaluza-Klein levels. 
Finally, the action \eqref{ActionComplete} is invariant under five-dimensional superconformal symmetry with eight 
supercharges. Note also that in addition to the $\cN=2$ $R$-symmetry group $SU(2)_R$
there is a further global $SU(2)$ symmetry. Superconformal symmetry can be restricted to Poincar\'e
supersymmetry yielding the action \eqref{cL_split}. 

By means of a five-dimensional formulation 
we have been able to avoid several complications present in six dimensions.
The six-dimensional self-duality was imposed on the level of the action just as in the Abelian 
case \cite{Bonetti:2012fn}. In this process zero modes and 
excited modes have to be treated differently. For the massless modes one can 
remove the tensor duals and package the degrees of freedom into vector multiplets. 
This is not possible for the excited modes where it is necessary to 
keep the tensor degrees of freedom. In fact massive tensors are generated by absorption
of a massless vector by a massless tensor, according to a 
 St\"uckelberg-like mechanism. Any attempt to dualize the 
massive tensor would require dualizing both the massless vector and the massless tensor,
with no net effect on the spectrum of the theory.

Building on this undemocratic treatment of massless and massive degrees of 
freedom we have implemented the non-Abelian gauging. 
The characteristic data defining the $\cN=2$ superconformal theories are 
given in terms of the group invariants $d_{IJ}$ and $f_{IJK}$, i.e.~the 
trace and the structure constants of $G$, and the Kaluza-Klein levels. For this identification
to work it was crucial that the excited modes at each Kaluza-Klein level 
are naturally complex and thus correspond to a real $SO(2)_{\rm KK}$-doublet. 
As a consequence, the $SO(2)_{\rm KK}$-epsilon invariant can be combined 
with the totally antisymmetric structure constants $f_{IJK}$ to yield
symmetric composite invariants that can enter the
$\cN=2$ data specifying the dynamics of the vector and tensor sector
of the model.
It is also interesting to point out that the superconformal formalism
in five dimensions naturally accommodates the compactification radius and the 
Kaluza-Klein vector into an additional vector multiplet coupled to all excited multiplets. 
This is essential in order to get mass terms for these Kaluza-Klein modes
once conformal symmetry is fixed.

In two special cases the connection between the five-dimensional 
action \eqref{cL_split} and six-dimensional physics becomes transparent
and the number of supercharges is automatically enhanced from eight to sixteen. 
On the one hand, in the regime in which all excited Kaluza-Klein modes can be neglected
the massless sector of \eqref{cL_split} is shown 
to be a rewriting of $\cN=4$ super Yang-Mills theory. This is precisely
the expected outcome of dimensional reduction of a non-Abelian $(2,0)$ theory on a circle.
On the other hand, if the non-Abelian gauging is turned off the full action
\eqref{cL_split} for both zero modes and excited modes coincides 
with the action for a set of Abelian $(2,0)$ tensor multiplets 
on a circle of arbitrary radius. Note that in this second special
case we also recover six-dimensional Poincar\'e invariance in the 
limit of infinite radius, even though it is realized in a non-manifest way
in the five-dimensional action.

We are confident that our actions \eqref{ActionComplete} and \eqref{cL_split}
can be used to extract information about six-dimensional physics also
away from the two special situations mentioned above. In order to substantiate
our claim, it is essential to study in more detail the status 
of global and local symmetries of \eqref{ActionComplete}, \eqref{cL_split}
from a six-dimensional perspective.
The global symmetry algebra of $(2,0)$ theories is given by the superconformal algebra  $OSp(8^*|4)$
\cite{Kac:1977em,Nahm:1977tg, Claus:1997cq, Bergshoeff:1999db}.\footnote{For a constructive classification of lowest weight unitary irreducible representations of the six-dimensional conformal algebras $OSp(8^*|2N)$ see \cite{Dobrev:2002dt}.}
On the one hand, invariance under five-dimensional Poincar\'e symmetry, $R$-symmetry and $Q$-supersymmetry
is expected to be unaltered by compactification on a circle. 
On the other hand, symmetry under the remaining generators of $OSp(8^*|4)$, including 
Poincar\'e transformations involving the circle direction, dilatations, conformal boosts, $S$-supersymmetry,
is expected to be restored only in the decompactification limit.

In addition to the global symmetries listed above, $(2,0)$ theories
are expected to possess a local symmetry that generalizes the 
usual Abelian gauge symmetry of tensor fields, $\delta B = d \Lambda$.
On general
grounds one can argue that  it yields an infinite tower of local symmetries 
once the theory is compactified on a circle. This corresponds to
an expansion of the six-dimensional gauge parameter in Kaluza-Klein modes.
It is conceivable that our action can be regarded as the outcome 
of a suitable gauge-fixing of such an infinite tower of local invariances.
From this perspective, the non-Abelian gauge transformations \eqref{gauge_trans}
are interpreted as the residual local symmetry of the theory.
The undemocratic treatment of zero modes might be thus
a by-product of this gauge-fixing procedure.
A better understanding of these issues is desirable
and left for future research.

In this work we have started to address global symmetry restoration
by analyzing in detail the possibility of $R$-symmetry and $Q$-supersymmetry
enhancement. 
By allowing a functional dependence of all fields on an additional $S^4$ and integrating over the latter we 
enhance the $SU(2)_R \times SU(2)$ symmetry of our $\cN=2$ action \eqref{ActionComplete} to $USp\,(4)_R$,
which is the R-symmetry group compatible with $\cN=4$ supersymmetry in five dimensions. 
The $\cN=2$ multiplets  can thus be understood as particular decompositions of full $\cN=4$ multiplets with the help 
of additional bosonic coordinates that parametrize $S^4$. Upon rewriting the action using this language we 
restore the full $\cN=4$ R-symmetry group by integrating over the latter in a well-defined manner. For example, 
we have demonstrated explicitly, that upon imposing particular H-analyticity conditions on the various fields, 
thereby constraining their functional dependence on $S^4$ and performing explicitly the integration over the 
$u$-coordinates, we recover maximally symmetric Yang-Mills theory for the massless sector of the theory. 
In its manifestly $USp\,(4)$ invariant form, the full action can be used to quantize the theory keeping the dependence of the field on $S^4$.

The $(2,0)$ non-Abelian theories we are aiming at are supposed
to describe, in particular, the world-volume 
theory of a stack of M5-branes. To provide further evidence for such an identification
we have briefly analyzed some quantum aspects of our action. 
Similar to the discussion in \cite{Bonetti:2012fn}, one can ask if the theory captures information about the
six-dimensional anomalies. While in \cite{Bonetti:2012fn} we have focused on 
aspects of six-dimensional gravitational anomalies, we here were considering terms required to 
cancel the six-dimensional conformal anomaly in the breaking $SU(N+1) \rightarrow SU(N) \times U(1)$.
We have argued that due to the tower of massive modes new one-loop Feynman diagrams are allowed
which induce a correction to a Wess-Zumino type coupling of an external $USp\,(4)_R$ gauge field. 
Our findings support the claim that the large-$N$ scaling behavior of the Wess-Zumino term can 
indeed be extracted by computing five-dimensional one-loop diagrams.
%The large-$N$ scaling was argued to be consistent with the expected characteristic behavior of 
%the world-volume theory of multiple M5-branes.

Some remarks are in order about the interpretation
of classical actions like \eqref{cL_split} with an infinite tower of massive Kaluza-Klein modes.
Indeed, there seems to be a tension between 
a spectrum with arbitrarily high masses and the standard effective field
theory paradigm, according to which massive degrees of freedom have to be 
integrated out. More precisely, if the compactification radius is $r$
and the fixed energy scale of reference is $E$,
excited modes with level $n \gg rE$ should not be included in the low energy spectrum.
We suggest that our actions for Kaluza-Klein
towers can be considered as  valid effective actions in the decompactification regime $r \gg E^{-1}$,
since in this limit an arbitrarily high number of excited modes can  be
integrated in.
This set-up can be contrasted with the set-up of refs.~\cite{Douglas:2010iu,Lambert}.
In that case, the opposite situation $r \ll E^{-1}$ is considered,
in such a way that all excited modes should be integrated out. As conjectured in \cite{Douglas:2010iu,Lambert},
in this limit all higher modes are still accessible as non-perturbative 
excitations of super Yang-Mills fields. Our Kaluza-Klein description 
can perhaps be considered as a complementary treatment of the same degrees of freedom
in a different regime.

In order to link our proposed action \eqref{ActionComplete} with the $(2,0)$ theories
we have demonstrated that it has many highly appealing 
features, both in terms of symmetries of the classical action as well as 
properties at the quantum level. Many questions are still left open for investigation. 
First of all, restoration of six-dimensional 
global symmetries and especially Poincar\'e symmetry in the decompactification limit is desirable.
To this end, it might be possible to use a procedure similar to the one we applied to restore 
R-symmetry, by formulating the theory more abstractly in terms of a fixed embedding 
tensor and summing over all possible embeddings. This would ensure a more democratic 
treatment of the $A_n^I$ and $B_n^I$ and might result in an interesting effective dynamics. 
Secondly, in terms of geometry, our construction gave an explicit proposal for the scalar 
moduli space in $\cN=2$ language, i.e.~for the scalars in the tensor, vector and hypermultiplets separately. 
It is desirable to study in more detail how the harmonization procedure allows us to 
combine them into a single moduli space for the $\cN=4$ theory.
Besides this, it would be interesting to analyze further remnants of the 
six-dimensional superconformal symmetry. In order to do that it appears to be crucial  
to introduce additional compensator multiplets and investigate their relations with the 
supergravity multiplet. It would also be 
interesting to study the scalar potential of our theory in more detail, which could lead, for example, to further 
insights into the M5-brane dynamics. Finally, we have presented a preliminary discussion of 
some quantum properties of five-dimensional action.
It is a very interesting question to continue this 
investigation and to extract the conformal anomaly in the case of breaking 
the gauge group $SU(N)\rightarrow U(1)^N$. 
We hope to return to a more thorough investigation of these aspects in the future.

%%%%%%%%%%%%%%%%%%%%%%%%%%%%%%%%%%%%%%%%

\vspace*{1cm}
\noindent
{\bf Acknowledgments}: 
We would like to thank Gianguido Dall'Agata, Michael Douglas, Davide Gaiotto, Olaf Hohm, Denis Klevers, Noppadol Mekareeya,
Tom Pugh, Raffaele Savelli, Maximilian Schmidt-Sommerfeld, Emery Sokatchev, Washington Taylor, and Marco Zagermann for interesting discussions. 
We are grateful to the Simons Center for Geometry and Physics (Stony Brook University) for hospitality.
SH would like to thank the Hausdorff Research Institute for Mathematics (University of Bonn) for kind hospitality during completion of this work.
This work was supported by a research grant of the Max Planck Society.

\vspace*{2cm}

%%%%%%%%%%%%%%%%%%

\appendix

\noindent {\bf \LARGE Appendices}

\section{Notations, conventions and useful identities} \label{appendix_conventions}

Five-dimensional flat spacetime indices $\mu,\nu,\dots$ run from 0 to 4 and are raised and
lowered with the mostly plus Minkowski metric $g_{\mu\nu} = {\rm diag}(-,+,+,+,+)$ and its
inverse $g^{\mu\nu}$.
The Levi-Civita tensor
$\epsilon_{\mu\nu\rho\sigma\tau}$ has components
\begin{equation}
\epsilon_{01234} = +1 = - \epsilon^{01234} \ .
\end{equation}
Five-dimensional gamma matrices $\gamma^\mu$ are constant, complex-valued $4 \times 4$ matrices
satisfying the anticommutation relation
\begin{equation}
\{ \gamma^\mu , \gamma^\nu \} = 2 g^{\mu\nu} \, \mathbb I\ .
\end{equation} 
We use the shorthand notation $\gamma^{\mu_1 \dots \mu_p} = \gamma^{[\mu_1} \dots \gamma^{\mu_p]}$,
and we choose a representation of gamma matrices such that
\begin{equation}
\gamma^{\mu\nu\rho\sigma\tau} = i\, \epsilon^{\mu\nu\rho\sigma\tau} \, \mathbb I\ .
\end{equation}
We further assume the hermiticity property
\begin{equation}
\gamma^0 \gamma^\mu (\gamma^0)^{-1} = -(\gamma^\mu)^\dagger \ .
\end{equation}
The charge conjugation matrix $C$ in five dimensions
 acts on gamma matrices according to
\begin{equation}
C \gamma^{\mu} C^{-1} = + (\gamma^\mu)^{\sf T} \ .
\end{equation}
We use a representation such that $C$ is real and satisfies
\begin{equation}
C^{\sf T} = -C = C^{-1} \ .
\end{equation}

In our work we encounter three different kinds of symplectic indices.
First of all, we have indices $i,j=1,\dots,4$ of the $\mathbf 4$ representation of $USp(4)_R$. 
Secondly we find two different copies of the $\mathbf 2$ representation of $SU(2)_R$,
labeled by indices $a,b=1,2$ and $\da , \dbi = 1 , 2$. Each symplectic 
group is endowed with a primitive antisymmetric invariant: $\Omega_{ij}$ for $USp(4)_R$ and 
$\epsilon_{ab}, \epsilon_{\da \dbi}$ for the two copies of $SU(2)$.

For all symplectic groups we adopt the same conventions
regarding the inverse of the antisymmetric invariant, the raising and lowering of indices, 
and the reality properties.
For definiteness, we write down the conventions for $USp(4)_R$. The inverse $\Omega^{ij}$
of $\Omega_{ij}$ is defined by the relation
\begin{equation}
 \Omega_{ik} \Omega^{jk} = \delta^j_i \ .
\end{equation}
Given any object $T^i$ with (at least) one symplectic index, raising and lowering 
of $i$ are performed   according to the NW-SE convention:
\begin{equation}
T^i = \Omega^{ij}T_j \ , \qquad
T_i = T^j \Omega_{ji} \ .
\end{equation}
Complex conjugation interchanges upper and lower symplectic indices.
The antisymmetric invariant satisfies the reality property
\begin{equation}
(\Omega_{ij})^* = \Omega^{ij} \ .
\end{equation}
An explicit realization of the invariants $\Omega_{ij}, \epsilon_{ab}$ with all required properties  is
furnished by
\begin{equation} \label{def-Omega}
 \Omega_{ij} = 
\begin{pmatrix}
 0 & 1 & & \\
 -1 & 0 & & \\
 & & 0 & 1 \\
 & & -1 & 0
\end{pmatrix} = \Omega^{ij} \ , \qquad
\epsilon_{ab} = 
\begin{pmatrix}
 0 & 1 \\
 -1 & 0 
\end{pmatrix}
= \epsilon^{ab} \ .
\end{equation}
The second expression can also be applied to $\epsilon_{\da \dbi}, \epsilon^{\da \dbi}$.

Let us now discuss in more detail symplectic spinors, i.e.~spinors
carrying one of the three kinds of symplectic indices listed above. For definiteness, we 
write down equations with $i,j$ indices, but the same 
conventions apply to $a,b$ and $\da , \dbi$ indices. 
The Dirac bar of a symplectic spinor $\lambda^{ i}$ is defined according to
\begin{equation}
\bar\lambda^i = (\lambda_{ i})^\dagger \gamma^0 \ .
\end{equation}
If symplectic indices are omitted in a spinor bilinear, a NW-SE contraction is understood,
\begin{equation}
\bar\lambda \chi = \bar\lambda^i \chi_{ i} \ .
\end{equation}
The Fierz rearrangement formula for anticommuting spinors in five dimensions reads
\begin{equation}
(\bar\psi_1 \psi_2)(\bar \psi_3 \psi_4) = - \tfrac 14 (\bar \psi_1 \psi_4) (\bar \psi_3 \psi_2)
- \tfrac 14 (\bar \psi_1 \gamma^\mu \psi_4) (\bar \psi_3 \gamma_\mu \psi_2)
+ \tfrac 18 (\bar \psi_1 \gamma^{\mu\nu} \psi_4) (\bar \psi_3 \gamma_{\mu\nu} \psi_2) \ ,
\end{equation}
where spinors $\psi_1, \psi_2, \psi_3,\psi_4$ can carry arbitrary indices 
and Kaluza-Klein levels.
In our conventions, complex conjugation acting on the product 
of anticommuting variables does not change their order. Therefore,
the reality of bilinears is determined by the basic relation
\footnote{Care has to be taken 
in raising/lowering  indices with $\Omega$ in equations involving complex conjugation. For example,
moving the index $j$ in \eqref{spinor_bilinear_reality} gives 
$(\bar\lambda^i \chi^j)^* =- \bar\chi_j \lambda_{ i}$.
}
\begin{equation} \label{spinor_bilinear_reality}
 (\bar\lambda^i \chi_{ j})^* = \bar\chi^j \lambda_{ i} \ .
\end{equation}

The Majorana condition for a symplectic spinor $\lambda^i$ reads
\begin{equation}
 \bar\lambda^i = \Omega^{ij} (\lambda_j)^{\sf T} C \ .
\end{equation}
As a result, if $\lambda^i, \chi^j$ 
are Majorana, we have the flip property
\begin{equation}
\bar\lambda^i \gamma^{\mu_1 \dots \mu_p} \chi^j = \bar\chi^j \gamma^{\mu_p \dots \mu_1} \lambda^i \ .
\end{equation}
Note that an extra minus sign is needed if the $USp(4)_R$ indices $i,j$ are
contracted on both sides according to the NW-SE convention. 
This implies that $\bar\lambda^i \chi_i$ is purely imaginary for real $\lambda^i, \chi^i$.
Any symplectic spinor $\lambda^i$ can be decomposed in a $SO(2)$ 
doublet of Majorana symplectic spinors $\lambda^{i\, \alpha}$, $\alpha = 1,2$:
\begin{equation}
 \lambda^i = \tfrac{1}{\sqrt{2}} \big( \lambda^{i\, \alpha = 1} 
+ i \lambda^{i\, \alpha = 2}  \big)\ , \qquad
 \bar\lambda^{i\, \alpha} = \Omega^{ij} (\lambda^\alpha_j)^{\sf T} C \ .
\end{equation}
Multiplication of $\lambda^i$ by a $U(1)$ phase
is equivalent to an $SO(2)$ rotation of the doublet $\lambda^{i\, \alpha}$. 
With this understanding, equations \eqref{SO2_action}, \eqref{SO2_covariant_derivative} 
hold also if $X$ is a symplectic spinor.

With the definitions \eqref{compl_splitting} and \eqref{def-epsilon_alphabeta} one infers the the following identities
to match the $SO(2)$ and the complex notations. They are written
with $SU(2)_R$ indices for definiteness, but they
hold for arbitrary symplectic indices. One has   
\begin{align} \label{rewrite_alphabeta}
\delta_{\alpha \beta} x^\alpha y^\beta &= 2 \Re (\bar x y) \ , \nn &
\epsilon_{\alpha \beta} x^\alpha y^\beta &= 2 \Im (\bar x y) \ ,\\[.1cm]
\delta_{\alpha \beta} \bar \chi^{a\,\alpha} \lambda^\beta_a &= 2 i \Im( \bar\chi^a \lambda_a) \ ,& 
\epsilon_{\alpha \beta} \bar \chi^{a\,\alpha} \lambda^\beta_a &= - 2 i \Re( \bar\chi^a \lambda_a) \ , \nn \\[.1cm]
\delta_{\alpha \beta} \bar\psi^a x^\alpha \lambda^\beta_a &= 2 i \Im (\bar\psi^a \bar x \lambda_a)\ , &
\epsilon_{\alpha \beta} \bar\psi^a x^\alpha \lambda^\beta_a &= - 2 i \Re (\bar\psi^a \bar x \lambda_a) \ ,  
\end{align}
where $x,y$ are complex bosonic fields, $\chi, \lambda$ are complex spinors, $\psi$ is a Majorana spinor. 
The same identities hold when $SU(2)_R$ indices are contracted with a tensor that satisfies a pseudo-reality condition (e.g.~$Y^{I\,ab}$).

%%%%%%%%%%%%%%%%%%%%%%%%%%%%%%%%%%%%%%%%

\section{Summary of indices} \label{index-appendix}

In this appendix we summarize the index conventions used throughout this work. We stress that we 
sometimes also distinguish indices with letters appearing at different positions in the alphabet. 
For example, we consider indices $I,J,...$ and $M,N,...$ to label different objects. 
The complete list of indices reads:\\[-.1cm]

\begin{tabular}{ll}
 Spacetime indices  & \\[.1cm]
 $\mu,\nu=0,\dots, 4$ & five-dimensional indices \\
 $\boldsymbol{\mu} , \boldsymbol{\nu} = 0, \dots , 5$ & six-dimensional indices \\[.1cm]
Symplectic indices & \\[.1cm]
 $i,j=1,\dots, 4$ & indices of the $\mathbf 4$ representation of $USp(4)_R$ \\
 $A,B=1, \dots, 5$ & indices of the $\mathbf 5$ representation of $SO(5)_R$ \\
 $a,b=1,2$ & indices of the $\mathbf 2$ representation of $SU(2)_R$ \\
 $\da,\dbi= 1, 2$ & indices of the $\mathbf 2$ representation of $SU(2)$ \\[.1cm]
Other indices & \\[.1cm]
 $n,m \ge 1$ & Kaluza-Klein level for excited modes \\
 $\alpha,\beta = 1,2$ & indices of the $\mathbf 2$ representation of $SO(2)_{\rm KK}$ \\
 $I,J=1,\dots, |G|$ & indices in the adjoint representation of the gauge group $G$ \\
 $M = \{I\alpha n\}, N = \{J\beta m\}$ & multi-index labeling massive $\cN=2$ tensor multiplets \\
 $\widehat I = (0,I)$ & collective index for all $\cN=2$ vector multiplets, including $\widehat \cV^0$ \\
 $\cI = (I,M)$ & collective index useful in the discussion of $\cN=2$ hypermultiplets \\
 $\Lambda = (0,I,M)$ & collective index useful in the superconformal phase
\end{tabular}

%%%%%%%%%%%%%%%%%%%%%%%%%%%%%%%%%%%%%%%%

\section{Six-dimensional $(2,0)$ pseudoaction for Abelian tensors} \label{ab_tensors_app}

In this appendix we review the supersymmetry transformations and the
associated supersymmetric action for a 
collection of non-interacting 
six-dimensional tensor multiplets $\sixcT^I$, labelled by the degeneracy index $I$.
The content of tensor multiplets $\sixcT^I$ is summarized in Table \ref{6d_table}.
In order to improve the readability,
we refrain from using boldface symbols for six-dimensional quantities
in the following expressions.

The linearized Poincar\'e $(2,0)$ supersymmetry transformations
read \cite{Bergshoeff:1999db}\footnote{
Compared to reference \cite{Bergshoeff:1999db}, the fields and the supersymmetry parameter
have been rescaled by suitable factors to achieve canonical normalization in the pseudoaction below.}
\begin{align}
 \delta(\epsilon) B^I_{\mu\nu} &= - \bar \epsilon^i \gamma_{\mu\nu} \lambda_i \ , \nn \\[.1cm]
 \delta(\epsilon) \lambda^{I\, i} & = \tfrac 16 \cH^I_{\mu\nu\rho} \gamma^{\mu\nu\rho} \epsilon^i 
+ 2 \slashed{\partial} \sigma^{I\, ij} \epsilon_j \ ,\nn \\[.1cm]
 \delta(\epsilon) \sigma^{I\, ij} & = - 4 \left( \bar\epsilon^{[i} \lambda^{I\, j]} 
+ \tfrac 14 \Omega^{ij} \bar\epsilon^k \lambda^I_k \right) \ .
\end{align}
Recall that the tensor field strength is defined as 
$\cH^I_{\mu\nu\rho} = 3 \partial_{[\mu} B^I_{\nu\rho]}$.
Note that contraction with $\gamma^{\mu\nu\rho} \epsilon^i$
automatically selects the anti-self-dual part of the field strength,
because $\epsilon^i$ is a left-handed Weyl spinor in our conventions. 
The supersymmetry algebra closes only up to the free-field equations of
motion for $B^I_{\mu\nu}, \lambda^{I\,i}, \sigma^{I\,ij}$. They can
be derived from the following supersymmetric pseudoaction:
\begin{align} \label{6Dab_action}
 S^{(6)} = \int d^6 x \; d_{IJ} \Big\{  - \tfrac {1}{12}  \, \cH^{I\, \mu\nu\rho} \cH^J_{\mu\nu\rho} 
- \tfrac 12  \, \partial^\mu \sigma^{I\, ij} \partial_\mu \sigma^J_{ij} 
- \tfrac 14  \, \bar\lambda^{I\,i} \slashed{\partial} \lambda^J_i \Big\}  \ .
\end{align}
We stress that this is not a proper action, since the self-duality
constraint on the field strengths of tensors
cannot be derived from it, and has to be imposed at the level of the equations of motion.
In order to write down kinetic terms, the symmetric, positive-definite, constant matrix
$d_{IJ}$ has been introduced.

\section{Weyl rescaling and $USp(4)$ R-symmetry in five dimensions}\label{Sect:Superconformal5}

In section~\ref{sec:N=4completion} we have discussed a proposal to 
enhance the R-symmetry group of the five-dimensional action (\ref{cL_split}) to $USp(4)_R$ 
compatible with $\cN=4$ supersymmetry. Throughout this discussion we have neglected the 
question of conformal invariance, and have worked in the restricted frame using (\ref{gauge_fixing}). 
Notice that indeed no conformal extension of the five-dimensional super Poincar\'e algebra with 
16 supercharges exists \cite{Seiberg:1997ax}. However, in this section we will briefly outline 
how at least five-dimensional scaling symmetry can be restored in the $USp(4)_R$ covariant form.

%%%%%%%%%%%%%%%%%%%%%%%%%%%%%%%%%%%%%%%%%%%%%%%%%%%%%%%%%%%%%%%%
\subsection{Additional multiplets}
The main problem in restoring scaling invariance in the $USp(4)_R$ covariant frame is the additional 
superfield $\widehat{\mathcal{V}}^0$ defined in (\ref{def-cV0}). Indeed, while from the very beginning we have 
chosen the spectrum of physical fields in a way which is compatible with $USp(4)_R$ R-supersymmetry 
(see section~\ref{N=2spectrum}), there is only a single $\cN=2$ vector multiplet (\ref{def-cV0}), 
whose scalar field $\phi^0$ is particularly important for scaling invariance. We now propose to 
upgrade the single $\cN=2$ multiplet (\ref{def-cV0}) to at least the field content of a full $\cN=4$ 
vector multiplet. 

It turns out, however, that this alone is not sufficient either but additional full $\cN=4$ multiplets 
need to be introduced. This is not surprising from the point of view of supergravity: In fact, as discussed 
in \cite{Howe:1981ev}, for five-dimensional (linearized) off-shell supergravity a total of five compensating 
vector multiplets need to be added. In the discussion of the rigid limit (which is relevant in section~\ref{sec:N=4completion}), 
gravity has been decoupled and any constraints on the compensators (apart from fixing Weyl invariance) 
stemming from the supergravity multiplet had already been trivially fulfilled. Moreover, by presenting 
the action with reference to a fixed $\cN=2$ subalgebra as in (\ref{cL_split}), local $USp(4)$ invariance 
was implicitly gauge-fixed (see previous section) and the corresponding constraints implicitly solved. 
This indeed leaves only the compensator (\ref{def-cV0}) that was added for scaling invariance in the 
rigid $\cN=2$ action (see section \ref{N=2spectrum}). However, when we attempt to restore the full $USp(4)$ 
R-symmetry group it is not surprising that part of the local gauge freedom is restored thereby requiring 
the introduction of additional multiplets.

In the following subsection we aim to rewrite
the action \eqref{ActionComplete} in a harmonized form incorporating
Weyl scaling symmetry. To this end, suitable harmonic variables
are introduced.  
Note that the construction of such variables can be performed making use of a single 
$\cN=4$ compensator multiplet. It will be denoted $\cV^0$
and is the minimal extension of the $\cN=2$ compensator multiplet $\widehat{\cV}^0$.
 Let us stress again, 
however, that one compensator multiplet is not sufficient
in complete treatment of the compensator sector of the theory, which
is beyond the scope of this appendix.

%%%%%%%%%%%%%%%%%%%%%%%%%%%%%%%%%%%%%%%%%%%%%%%%%%%%%%%%%%%%%%%%
\subsection{Scaling invariant action}\label{Sect:SuperConformalN4}

After introducing additional degrees of freedom, we will now briefly outline how to 
construct a five-dimensional scaling invariant action. As in the case of $\cN=2$ 
supersymmetry, the instrumental ingredient are the scalar fields. Our presentation 
will follow very closely a similar discussion in four-dimensions \cite{Antoniadis:2007cw}.

Under Weyl rescalings, every field transforms with a Weyl weight, according to the 
powers described in section~\ref{N=2spectrum}. A scaling invariant action is achieved 
by compensating this weight factor with the help of additional fields. Thus, our 
starting point is the following harmonic projection of the scalar component of the 
additional $\cN=4$ vector multiplet\footnote{We recall that we are only discussing 
Weyl invariance, which is why not all compensating multiplets are needed.} $\mathcal{V}^0$ 
which we denote $\sigma^{0 \, ij}$
\beq
 \phi^0=\tfrac{1}{\sqrt 2} \sigma^{0 \, ij}u_i^au_j^{b}\epsilon_{ab}\,,\qquad \qquad
 q^{0 \, a\da}=(\phi^0)^{1/2} \sigma^{0 \, ij}u_i^au_j^{\da}\,.
\eeq
Here the component $\phi^0$ transforms with a weight factor under Weyl rescalings with parameter $\rho$, 
while the $q^{0 \, a\da}$ components in addition undergo non-trivial local $USp\,(4)$ 
transformations with parameters $\lambda_i^j$ as
\beq
 \delta\phi^0=\rho \phi^0\,,\qquad \qquad \delta q^{0 \, a\da}=\epsilon^{ab}u_{b}^i{\lambda_i}^ju_j^{\da} 
+ \tfrac 32 \rho q^{0\, a \da} \,.
\eeq
As we can see, the combination
\beq
\widetilde{q}^{0 \, a\da}=\frac{q^{0 \, a\da}}{(\phi^0)^{3/2}}\,,\qquad \qquad \delta \widetilde{q}^{0 \, a\da}=\epsilon^{ab}u_{b}^i{\lambda_i}^ju_j^{\da}\,,
\eeq
is weightless under Weyl rescalings and can be used to define quantities which 
are inert under local $USp(4)$. In the discussion in section~\ref{sec:susyactions} 
the latter had been gauged by setting $\widetilde{q}^{0 \, a\da}=0$. As 
before (see e.g.~(\ref{ActionComplete})) we can use $\phi^0$ to compensate for Weyl rescalings. 
Thus, with the help of these fields, we can define new (field-dependent) harmonic variables
\begin{align}
&v_i^a=u_i^a+u_i^{\da}\,\epsilon_{\da\dbi}\,\widetilde{q}^{0 \, a\dbi}\,,&&v_i^{\da}=u_i^{\da}\,,\\
&\bar{v}^i_a=\bar{u}^i_a\,,&&\bar{v}^i_{\da}=\bar{u}^i_{\da}+\epsilon^{ab}\,\widetilde{q}^0_{b\da}\,\bar{u}_a^i\,,
\end{align}
which are inert under local  transformations ($\delta v=\delta \bar{v}=0$). With 
the help of these quantities we can now write the harmonic projection of the physical 
scalar fields 
\beq \label{Harmonization-w}
   \left(\begin{array}{cc}
      v_i^a v^b_j & v_i^{a} v^{\dbi}_j  \\
      v_i^{\da} v^b_j  & v_i^{\da} v^{\dbi}_j 
   \end{array} \right) \sigma^{I\, ij}_n = \left(\begin{array}{cc}
      \tfrac{1}{\sqrt{2}} \epsilon^{ab} \phi^{I}_n& (\phi^0)^{-1/2} q^{I\, a\dbi}_n  \\
       - (\phi^0)^{-1/2}  q^{I\,  b \da}_n  & - \tfrac{1}{\sqrt{2}} \epsilon^{\da \dbi}   \phi^{I}_n 
   \end{array} \right)\ ,
\eeq
which can be used to write the action in an $USp\,(4)_R$ covariant 
fashion\footnote{To avoid cluttering of this formula, we refrain from displaying 
the explicit $v$-dependence of all fields.}
\begin{align}
&\mathcal{L}_{USp(4)}[\sigma^{ij},\lambda^i,A,B]=\nonumber\\
&=\int_{S^4} du\,\mathcal{L}_{\cN=2}\left[\left(\tfrac{v^a_i v^{\da}_j\sigma^{ij}}{(\sigma^{0 \, ij}u_i^au_j^{b}\epsilon_{ab})^{1/2}},
\tfrac{v^{\da}_i\lambda^i}{(\sigma^{0 \, ij}u_i^au_j^{b}\epsilon_{ab})^{1/2}}\right);\left(v_i^a v_j^b\epsilon_{ab}\sigma^{ij},v^a_i\lambda^i,A\right);\left(v_i^a v_j^b\epsilon_{ab}\sigma^{ij},v_i^a\lambda^i,B\right)\right]\nonumber\\
&\hspace{1.5cm}+\text{higher terms}\,,\label{USp4actionComplete}
\end{align}
where the higher terms denote additional contributions including the superpartners 
of the compensating scalars, which are needed to achieve full local $USp\,(4)$ symmetry. 
%%%%%%%%%%%%%%%%%%%%%%%%%%%%%%%%%%%%%%%%%%%%%%%%%%%%%%%%%%%%%%%%%%%%%%%%%%%%%%%%%%%%%%%%%
%%%%%%%%%%%%%%%%%%%%%%%%%%%%%%%%%%%%%%%%%%%%%%%%%%%%%%%%%%%%%%%%%%%%%%%%%%%%%%%%%%%%%%%%%

%%%%%%%%%%%%%%%%%%%%%%%%%%%%%%%%%%%%%%%%%%%%%%%%%%%%%%%%%%%%%%%%%%%%%%


\begin{thebibliography}{99}
%%%%%%%%%%%%%%%%%%%%%%%%%%%%

\bibitem{Witten:1995zh}
  E.~Witten,
  ``Some comments on string dynamics,''
  In *Los Angeles 1995, Future perspectives in string theory* 501-523
  [hep-th/9507121].
  %%CITATION = HEP-TH/9507121;%%

%%%%%%%%
% Some approaches to non-Ab (2,0) 
%%%%%%%%

\bibitem{Indirect20}
An incomplete list includes:
E.~Witten,
  ``Five-branes and M theory on an orbifold,''
  Nucl.\ Phys.\ B {\bf 463} (1996) 383
  [hep-th/9512219];
  %%CITATION = HEP-TH/9512219;%%

N.~Seiberg,
  ``New theories in six-dimensions and matrix description of M theory on T**5 and T**5 / Z(2),''
  Phys.\ Lett.\ B {\bf 408} (1997) 98
  [hep-th/9705221];
  %%CITATION = HEP-TH/9705221;%%
  
  O.~Aharony, M.~Berkooz, S.~Kachru, N.~Seiberg and E.~Silverstein,
  ``Matrix description of interacting theories in six-dimensions,''
  Adv.\ Theor.\ Math.\ Phys.\  {\bf 1} (1998) 148
  [hep-th/9707079];
  %%CITATION = HEP-TH/9707079;%%
  
O.~Aharony, M.~Berkooz and N.~Seiberg,
  ``Light cone description of (2,0) superconformal theories in six-dimensions,''
  Adv.\ Theor.\ Math.\ Phys.\  {\bf 2} (1998) 119
  [hep-th/9712117];
  %%CITATION = HEP-TH/9712117;%%

  R.~G.~Leigh and M.~Rozali,
  ``The Large N limit of the (2,0) superconformal field theory,''
  Phys.\ Lett.\ B {\bf 431} (1998) 311
  [hep-th/9803068];\\[.1cm]
  %%CITATION = HEP-TH/9803068;%%  
and references therein. 


\bibitem{Ganor:1998ve}
  O.~Ganor and L.~Motl,
  ``Equations of the (2,0) theory and knitted five-branes,''
  JHEP {\bf 9805} (1998) 009
  [hep-th/9803108].
  %%CITATION = HEP-TH/9803108;%%

\bibitem{Harvey:1998bx}
  J.~A.~Harvey, R.~Minasian and G.~W.~Moore,
  ``NonAbelian tensor multiplet anomalies,''
  JHEP {\bf 9809} (1998) 004
  [hep-th/9808060].
  %%CITATION = HEP-TH/9808060;%%


\bibitem{Intriligator:2000eq}
  K.~A.~Intriligator,
  ``Anomaly matching and a Hopf-Wess-Zumino term in 6d, N=(2,0) field theories,''
  Nucl.\ Phys.\ B {\bf 581} (2000) 257
  [hep-th/0001205].
  %%CITATION = HEP-TH/0001205;%%

%%%%%%%%
% Reviews on (2,0) 
%%%%%%%%

\bibitem{Seiberg:1997ax}
  N.~Seiberg,
  ``Notes on theories with 16 supercharges,''
  Nucl.\ Phys.\ Proc.\ Suppl.\  {\bf 67} (1998) 158
  [hep-th/9705117].
  %%CITATION = HEP-TH/9705117;%%

\bibitem{Witten:2009at}
  E.~Witten,
  ``Geometric Langlands From Six Dimensions,''
  arXiv:0905.2720 [hep-th].
  %%CITATION = ARXIV:0905.2720;%%
  
\bibitem{Gaiotto:2010be}
  D.~Gaiotto, G.~W.~Moore and A.~Neitzke,
  ``Framed BPS States,''
  arXiv:1006.0146 [hep-th].
  %%CITATION = ARXIV:1006.0146;%%

%%%%%%%%%%%%%%%%%
%%%%%%%%%%%%%%%%%


\bibitem{Marcus:1982yu}
  N.~Marcus and J.~H.~Schwarz,
  ``Field Theories That Have No Manifestly Lorentz Invariant Formulation,''
  Phys.\ Lett.\ B {\bf 115} (1982) 111.
  %%CITATION = PHLTA,B115,111;%%

\bibitem{Siegel:1983es} An incomplete list includes:
  W.~Siegel,
  ``Manifest Lorentz Invariance Sometimes Requires Nonlinearity,''
  Nucl.\ Phys.\ B {\bf 238} (1984) 307;
  %%CITATION = NUPHA,B238,307;%%

  M.~Henneaux and C.~Teitelboim,
  ``Dynamics Of Chiral (selfdual) P Forms,''
  Phys.\ Lett.\ B {\bf 206} (1988) 650;
  %%CITATION = PHLTA,B206,650;%%

  B.~McClain, F.~Yu and Y.~S.~Wu,
  ``Covariant quantization of chiral bosons and OSp(1,1|2) symmetry,''
  Nucl.\ Phys.\ B {\bf 343} (1990) 689;
  %%CITATION = NUPHA,B343,689;%%

  P.~Pasti, D.~P.~Sorokin and M.~Tonin,
  ``On Lorentz invariant actions for chiral p forms,''
  Phys.\ Rev.\ D {\bf 55} (1997) 6292
  [hep-th/9611100];
  %%CITATION = HEP-TH/9611100;%%

  D.~Belov and G.~W.~Moore,
  ``Holographic Action for the Self-Dual Field,''
  hep-th/0605038.
  %%CITATION = HEP-TH/0605038;%%
  
\bibitem{Townsend:1983xs}
  P.~K.~Townsend, K.~Pilch and P.~van Nieuwenhuizen,
  ``Selfduality in Odd Dimensions,''
  Phys.\ Lett.\ B {\bf 136} (1984) 38
   [Addendum-ibid.\ B {\bf 137} (1984) 443].
  %%CITATION = PHLTA,B136,38;%%


\bibitem{Bonetti:2012fn}
F.~Bonetti, T.~W.~Grimm and S.~Hohenegger,
  ``A Kaluza-Klein inspired action for chiral p-forms and their anomalies,''
  arXiv:1206.1600 [hep-th].
  %%CITATION = ARXIV:1206.1600;%%
  
%%%%%%%%
% Recent progress on (2,0) action
%%%%%%%%

\bibitem{Lambert:2010wm}
 N.~Lambert and C.~Papageorgakis,
  ``Nonabelian (2,0) Tensor Multiplets and 3-algebras,''
  JHEP {\bf 1008} (2010) 083
  [arXiv:1007.2982 [hep-th]].
  %%CITATION = ARXIV:1007.2982;%%

\bibitem{Ho:2011ni}
  P.~-M.~Ho, K.~-W.~Huang and Y.~Matsuo,
  ``A Non-Abelian Self-Dual Gauge Theory in 5+1 Dimensions,''
  JHEP {\bf 1107} (2011) 021
  [arXiv:1104.4040 [hep-th]];
  %%CITATION = ARXIV:1104.4040;%% 
  
  K.~-W.~Huang,
  ``Non-Abelian Chiral 2-Form and M5-Branes,''
  arXiv:1206.3983 [hep-th].
  %%CITATION = ARXIV:1206.3983;%%

\bibitem{Lambert:2011gb}
  N.~Lambert and P.~Richmond,
  ``(2,0) Supersymmetry and the Light-Cone Description of M5-branes,''
  JHEP {\bf 1202} (2012) 013
  [arXiv:1109.6454 [hep-th]].
  %%CITATION = ARXIV:1109.6454;%%  

\bibitem{Linander:2011jy}
  H.~Linander and F.~Ohlsson,
  ``(2,0) theory on circle fibrations,''
  JHEP {\bf 1201} (2012) 159
  [arXiv:1111.6045 [hep-th]].
  %%CITATION = ARXIV:1111.6045;%%

 
\bibitem{Chu:2012um}
  C.~-S.~Chu and S.~-L.~Ko,
  ``Non-abelian Action for Multiple Five-Branes with Self-Dual Tensors,''
  JHEP {\bf 1205} (2012) 028
  [arXiv:1203.4224 [hep-th]].
  %%CITATION = ARXIV:1203.4224;%%
  
  C.~-S.~Chu, S.~-L.~Ko and P.~Vanichchapongjaroen, 
  ``Non-Abelian Self-Dual String Solutions,''  
  JHEP {\bf 1209} (2012) 018 [arXiv:1207.1095 [hep-th]].
  %%CITATION = ARXIV:1207.1095;%%
  
 
 \bibitem{Palmer:2012ya}
  S.~Palmer and C.~Saemann,
  ``M-brane Models from Non-Abelian Gerbes,''
  JHEP {\bf 1207} (2012) 010
  [arXiv:1203.5757 [hep-th]].
  %%CITATION = ARXIV:1203.5757;%%
 
  
\bibitem{Ho:2012nt}
  P.~-M.~Ho and Y.~Matsuo,
  ``Note on non-Abelian two-form gauge fields,''
  arXiv:1206.5643 [hep-th].
  %%CITATION = ARXIV:1206.5643;%%
  
%%%%%
%   5d, N=2
%%%%%
\bibitem{Bergshoeff:2001hc}
  E.~Bergshoeff, S.~Cucu, M.~Derix, T.~de Wit, R.~Halbersma and A.~Van Proeyen,
  ``Weyl multiplets of N=2 conformal supergravity in five-dimensions,''
  JHEP {\bf 0106} (2001) 051
  [hep-th/0104113].
  %%CITATION = HEP-TH/0104113;%%
  
\bibitem{Bergshoeff:2002qk}
  E.~Bergshoeff, S.~Cucu, T.~De Wit, J.~Gheerardyn, R.~Halbersma, S.~Vandoren and A.~Van Proeyen,
  ``Superconformal N=2, D = 5 matter with and without actions,''
  JHEP {\bf 0210} (2002) 045
  [hep-th/0205230].
  %%CITATION = HEP-TH/0205230;%%  

%%%

\bibitem{Cremmer:1980gs}
  E.~Cremmer,
  ``Supergravities In 5 Dimensions,''
  In *Salam, A. (ed.), Sezgin, E. (ed.): Supergravities in diverse dimensions, 
   vol. 1* 422-437. (In *Cambridge 1980, Proceedings, Superspace and supergravity* 267-282) and Paris Ec. Norm. Sup. - LPTENS 80-17 (80,rec.Sep.) 17 p. (see Book Index)

\bibitem{Gunaydin:1983bi}
  M.~Gunaydin, G.~Sierra and P.~K.~Townsend,
  %``The Geometry of N=2 Maxwell-Einstein Supergravity and Jordan Algebras,''
  Nucl.\ Phys.\ B {\bf 242} (1984) 244;
  %%CITATION = NUPHA,B242,244;%%

  M.~Gunaydin, G.~Sierra and P.~K.~Townsend,
  ``Gauging the d = 5 Maxwell-Einstein Supergravity Theories: More on Jordan Algebras,''
  Nucl.\ Phys.\ B {\bf 253} (1985) 573.
  %%CITATION = NUPHA,B253,573;%%

\bibitem{Gunaydin:1999zx}
  M.~Gunaydin and M.~Zagermann,
  ``The Gauging of five-dimensional, N=2 Maxwell-Einstein supergravity theories coupled to tensor multiplets,''
  Nucl.\ Phys.\ B {\bf 572} (2000) 131
  [hep-th/9912027].
  %%CITATION = HEP-TH/9912027;%%

\bibitem{Ceresole:2000jd}
  A.~Ceresole and G.~Dall'Agata,
  ``General matter coupled N=2, D = 5 gauged supergravity,''
  Nucl.\ Phys.\ B {\bf 585} (2000) 143
  [hep-th/0004111].
  %%CITATION = HEP-TH/0004111;%%

\bibitem{Bergshoeff:2004kh}
  E.~Bergshoeff, S.~Cucu, T.~de Wit, J.~Gheerardyn, S.~Vandoren and A.~Van Proeyen,
  ``N = 2 supergravity in five-dimensions revisited,''
  Class.\ Quant.\ Grav.\  {\bf 21} (2004) 3015
   [Class.\ Quant.\ Grav.\  {\bf 23} (2006) 7149]
  [hep-th/0403045].
  %%CITATION = HEP-TH/0403045;%%

%%%%%%%%%
%%%%%%%%%

%%%%%%%%%
%  6d, (1,0)
%%%%%%%%%

\bibitem{Samtleben:2011fj}
  H.~Samtleben, E.~Sezgin and R.~Wimmer,
  ``(1,0) superconformal models in six dimensions,''
  JHEP {\bf 1112} (2011) 062
  [arXiv:1108.4060 [hep-th]];
  %%CITATION = ARXIV:1108.4060;%%
  
  H.~Samtleben, E.~Sezgin, R.~Wimmer and L.~Wulff,
  ``New superconformal models in six dimensions: Gauge group and representation structure,''
  arXiv:1204.0542 [hep-th].
  %%CITATION = ARXIV:1204.0542;%%

\bibitem{Akyol:2012cq}
  M.~Akyol and G.~Papadopoulos,
  ``(1,0) superconformal theories in six dimensions and Killing spinor equations,''
  JHEP {\bf 1207} (2012) 070
  [arXiv:1204.2167 [hep-th]].
  %%CITATION = ARXIV:1204.2167;%%


%%%%%%%%%%%%
%%   (2,0) via 5d super Yang-Mills
%%%%%%%%%%%%


\bibitem{Douglas:2010iu}
  M.~R.~Douglas,
  ``On D=5 super Yang-Mills theory and (2,0) theory,''
  JHEP {\bf 1102} (2011) 011
  [arXiv:1012.2880 [hep-th]];
  %%CITATION = ARXIV:1012.2880;%%

\bibitem{Lambert}
  N.~Lambert, C.~Papageorgakis and M.~Schmidt-Sommerfeld,
  ``M5-Branes, D4-Branes and Quantum 5D super-Yang-Mills,''
  JHEP {\bf 1101} (2011) 083
  [arXiv:1012.2882 [hep-th]].
  %%CITATION = ARXIV:1012.2882;%%

%%%%%%%%%%%
%%  5d, N=4
%%%%%%%%%%%


\bibitem{Awada:1985ep}
  M.~Awada and P.~K.~Townsend,
  ``N=4 Maxwell-einstein Supergravity In Five-dimensions And Its Su(2) Gauging,''
  Nucl.\ Phys.\ B {\bf 255} (1985) 617.
  %%CITATION = NUPHA,B255,617;%%

\bibitem{Gunaydin:1985cu}
  M.~Gunaydin, L.~J.~Romans and N.~P.~Warner,
  ``Compact and Noncompact Gauged Supergravity Theories in Five-Dimensions,''
  Nucl.\ Phys.\ B {\bf 272} (1986) 598.
  %%CITATION = NUPHA,B272,598;%%

\bibitem{Dall'Agata:2001vb}
  G.~Dall'Agata, C.~Herrmann and M.~Zagermann,
  ``General matter coupled N=4 gauged supergravity in five-dimensions,''
  Nucl.\ Phys.\ B {\bf 612} (2001) 123
  [hep-th/0103106].
  %%CITATION = HEP-TH/0103106;%%
  
\bibitem{Schon:2006kz}
  J.~Sch\"on and M.~Weidner,
  ``Gauged N=4 supergravities,''
  JHEP {\bf 0605} (2006) 034
  [hep-th/0602024].
  %%CITATION = HEP-TH/0602024;%%



%%%%%%%%%%%%%%%%%%%%%%%%%%%%
%  Harmonic superspace in general
%%%%%%%%%%%%%%%%%%%%%%%%%%%

\bibitem{Galperin:1984av}
  A.~Galperin, E.~Ivanov, S.~Kalitsyn, V.~Ogievetsky and E.~Sokatchev,
  ``Unconstrained N=2 Matter, Yang-Mills and Supergravity Theories in Harmonic Superspace,''
  Class.\ Quant.\ Grav.\  {\bf 1} (1984) 469.
  %%CITATION = CQGRD,1,469;%%

\bibitem{Ivanov:1984ut}
  E.~Ivanov, S.~Kalitsyn, A.~V.~Nguyen and V.~Ogievetsky,
  ``Harmonic Superspaces Of Extended Supersymmetry. The Calculus Of Harmonic Variables,''
  J.\ Phys.\ A A {\bf 18} (1985) 3433.
  %%CITATION = JPAGB,A18,3433;%%

\bibitem{Galperin:1984bu}
  A.~Galperin, E.~Ivanov, S.~Kalitsyn, V.~Ogievetsky and E.~Sokatchev,
  ``Unconstrained Off-Shell N=3 Supersymmetric Yang-Mills Theory,''
  Class.\ Quant.\ Grav.\  {\bf 2} (1985) 155.
  %%CITATION = CQGRD,2,155;%%


\bibitem{Hartwell:1994rp}
  G.~G.~Hartwell and P.~S.~Howe,
  ``(N, p, q) harmonic superspace,''
  Int.\ J.\ Mod.\ Phys.\ A {\bf 10} (1995) 3901
  [hep-th/9412147].
  %%CITATION = HEP-TH/9412147;%%
  

\bibitem{Howe:1995md}
  P.~S.~Howe and G.~G.~Hartwell,
  ``A Superspace survey,''
  Class.\ Quant.\ Grav.\  {\bf 12} (1995) 1823.
  %%CITATION = CQGRD,12,1823;%%
  
\bibitem{HSS} A.~Galperin, E.~Ivanov, V.~Ogievetsky, E.~Sokatchev,  
   ``Harmonic Superspace,'' 
   UK: Cambridge Univ. Press, 2001,306p.
   
%%%%%%%%%%%%%%%%%%%%%%%
%%%%%%%%%%%%%%%%%%%%%%%

\bibitem{BonettiGrimmHohenegger2}
F.~Bonetti, T.~W.~Grimm and S.~Hohenegger, {\it to appear.}

\bibitem{tHooft}
G. 't Hooft, Recent Developments in Gauge Theories, eds. G. 't Hooft et. al., Plenum Press,
NY, 1980.


%%%%%%%%%
%%  N^3 (2,0) anomaly
%%%%%%%%%
\bibitem{Henningson:1998gx} M.~Henningson and K.~Skenderis, ''The Holographic Weyl anomaly,'' JHEP {\bf 9807} (1998) 023 [hep-th/9806087].
%%CITATION = HEP-TH/9806087;%%

\bibitem{Bastianelli:2000hi} F.~Bastianelli, S.~Frolov and A.~A.~Tseytlin, ``Conformal anomaly of (2,0) tensor multiplet in six-dimensions and AdS / CFT correspondence,''
JHEP {\bf 0002} (2000) 013 [hep-th/0001041].
%%CITATION = HEP-TH/0001041;%%

\bibitem{Yi:2001bz} P.~Yi, ``Anomaly of (2,0) theories,'' Phys.\ Rev.\ D {\bf 64} (2001) 106006 [hep-th/0106165].
%%CITATION = HEP-TH/0106165;%%

\bibitem{Maxfield:2012aw} 
     T.~Maxfield and S.~Sethi, 
     ``The Conformal Anomaly of M5-Branes," 
     JHEP {\bf 1206} (2012) 075
     [arXiv:1204.2002 [hep-th]].
%%CITATION = ARXIV:1204.2002;%%

\bibitem{Klebanov:1996un} I.~R.~Klebanov and A.~A.~Tseytlin, ``Entropy of near extremal black p-branes,'' Nucl.\ Phys.\ B {\bf 475} (1996) 164 [hep-th/9604089].
%%CITATION = HEP-TH/9604089;%%

\bibitem{Bolognesi:2011rq} S.~Bolognesi and K.~Lee, ``1/4 BPS String Junctions and $N^3$ Problem in 6-dim (2,0) Superconformal Theories,'' Phys.\ Rev.\ D {\bf 84} (2011) 126018
[arXiv:1105.5073 [hep-th]].
%%CITATION = ARXIV:1105.5073;%%

\bibitem{Bolognesi:2011nh} S.~Bolognesi and K.~Lee, ``Instanton Partons in 5-dim SU(N) Gauge Theory,'' Phys.\ Rev.\ D {\bf 84} (2011) 106001 [arXiv:1106.3664 [hep-th]].
%%CITATION = ARXIV:1106.3664;%%

\bibitem{Kim:2011mv} 
       H.~-C.~Kim, S.~Kim, E.~Koh, K.~Lee and S.~Lee, 
       ``On instantons as Kaluza-Klein modes of M5-branes,"
       JHEP {\bf 1112} (2011) 031 [arXiv:1110.2175 [hep-th]].
      %%CITATION = ARXIV:1110.2175;%%

\bibitem{Kim:2012av} H.~-C.~Kim and S.~Kim, ``M5-branes from gauge theories on the 5-sphere,'' arXiv:1206.6339 [hep-th].
%%CITATION = ARXIV:1206.6339;%%

\bibitem{Kallen:2012zn} J.~Kallen, J.~A.~Minahan, A.~Nedelin and M.~Zabzine, ``$N^3$-behavior from 5D Yang-Mills theory,'' arXiv:1207.3763 [hep-th].
%%CITATION = ARXIV:1207.3763;%%

%%%%%%%%%%%%%%%%%
% Refs for the (2,0) superconformal algebra and its application to field theory
%%%%%%%%%%%%%%%%%

\bibitem{Kac:1977em}
  V.~G.~Kac,
  ``Lie Superalgebras,''
  Adv.\ Math.\  {\bf 26} (1977) 8;
  %%CITATION = ADMTA,26,8;%%

%\bibitem{Kac:1977qb}
  V.~G.~Kac,
  ``A Sketch of Lie Superalgebra Theory,''
  Commun.\ Math.\ Phys.\  {\bf 53} (1977) 31.
  %%CITATION = CMPHA,53,31;%%

\bibitem{Nahm:1977tg}
  W.~Nahm,
  ``Supersymmetries and their Representations,''
  Nucl.\ Phys.\ B {\bf 135} (1978) 149.
  %%CITATION = NUPHA,B135,149;%%

%\cite{Claus:1997cq}
\bibitem{Claus:1997cq}
  P.~Claus, R.~Kallosh and A.~Van Proeyen,
  ``M five-brane and superconformal (0,2) tensor multiplet in six-dimensions,''
  Nucl.\ Phys.\ B {\bf 518} (1998) 117
  [hep-th/9711161].
  %%CITATION = HEP-TH/9711161;%%

%\cite{Bergshoeff:1999db}
\bibitem{Bergshoeff:1999db}
  E.~Bergshoeff, E.~Sezgin and A.~Van Proeyen,
  ``(2,0) tensor multiplets and conformal supergravity in D = 6,''
  Class.\ Quant.\ Grav.\  {\bf 16} (1999) 3193
  [hep-th/9904085].
  %%CITATION = HEP-TH/9904085;%%


\bibitem{Dobrev:2002dt} V.~K.~Dobrev, ``Positive energy unitary irreducible representations of D = 6 conformal supersymmetry,'' J.\ Phys.\ A {\bf 35} (2002) 7079 [hep-th/0201076].
  %%CITATION = HEP-TH/0201076;%%

%%%%%%%%%%%%%%%%%
%% Non-ab gerbes
%%%%%%%%%%%%%%%%%


\bibitem{Breen:2001ie}
  L.~Breen and W.~Messing,
  ``Differential geometry of GERBES,''
  Adv.\ Math.\  {\bf 198} (2005) 732
  [math/0106083 [math-ag]].
  %%CITATION = MATH/0106083;%%

\bibitem{Baez:2010ya}
  J.~C.~Baez and J.~Huerta,
  ``An Invitation to Higher Gauge Theory,''
  arXiv:1003.4485 [hep-th].
  %%CITATION = ARXIV:1003.4485;%%

\bibitem{Saemann:2011nb} C.~Saemann and M.~Wolf, 
``On Twistors and Conformal Field Theories from Six Dimensions,'' arXiv:1111.2539 [hep-th];
%%CITATION = ARXIV:1111.2539;%%

C.~Saemann and M.~Wolf, ``Non-Abelian Tensor Multiplet Equations from Twistor Space,'' arXiv:1205.3108 [hep-th].
%%CITATION = ARXIV:1205.3108;%%

%%%%%%%%%%%%%%%%%%%%%%%
%% aux fields in N=2 and N=4
%%%%%%%%%%%%%%%%%%%%%%


\bibitem{Howe:1981ev}
  P.~S.~Howe,
  ``Off-shell N=2 And N=4 Supergravity In Five-dimensions,''
  CERN-TH-3181.
  %%CITATION = CERN-TH-3181;%%

%%%%%%%%%%%%%%%%%%%%%%%%%

%%%%
% N=4 KK actions
%%%%%%%

\bibitem{Lee:2000kc}
  K.~-M.~Lee and J.~-H.~Park,
  ``5-D actions for 6-D selfdual tensor field theory,''
  Phys.\ Rev.\ D {\bf 64} (2001) 105006
  [hep-th/0008103].
  %%CITATION = HEP-TH/0008103;%%



%%%%%%%%%%%%
%%  Harmonic N=4
%%%%%%%%%%%%   
   

\bibitem{Antoniadis:2007cw}
  I.~Antoniadis, S.~Hohenegger, K.~S.~Narain and E.~Sokatchev,
  ``Harmonicity in N=4 supersymmetry and its quantum anomaly,''
  Nucl.\ Phys.\ B {\bf 794} (2008) 348
  [arXiv:0708.0482 [hep-th]].
  %%CITATION = ARXIV:0708.0482;%%
  

\bibitem{Antoniadis:2009nv}
  I.~Antoniadis, S.~Hohenegger, K.~S.~Narain and E.~Sokatchev,
  ``A New Class of N=2 Topological Amplitudes,''
  Nucl.\ Phys.\ B {\bf 823} (2009) 448
  [arXiv:0905.3629 [hep-th]];
  %%CITATION = ARXIV:0905.3629;%%

   I.~Antoniadis, S.~Hohenegger, K.~S.~Narain and E.~Sokatchev, 
   ``Generalized N=2 Topological Amplitudes and Holomorphic Anomaly Equation,'' 
   Nucl.\ Phys.\ B {\bf 856} (2012) 360 [arXiv:1107.0303 [hep-th]];
   %%CITATION = ARXIV:1107.0303;%%

   I.~Antoniadis and S.~Hohenegger, 
    ``N=4 Topological Amplitudes and Black Hole Entropy," Nucl.\ Phys.\ B {\bf 837} (2010) 61
    [arXiv:0910.5596 [hep-th]];
    %%CITATION = ARXIV:0910.5596;%%

   I.~Antoniadis and S.~Hohenegger, 
   ``Topological amplitudes and physical couplings in string theory," 
   Nucl.\ Phys.\ Proc.\ Suppl.\  {\bf 171} (2007) 176 [hep-th/0701290 [hep-th]];
   %%CITATION = HEP-TH/0701290;%%

   S.~Hohenegger and S.~Stieberger,
   ``BPS Saturated String Amplitudes: K3 Elliptic Genus and Igusa Cusp Form,''
   Nucl.\ Phys.\ B {\bf 856} (2012) 413
   [arXiv:1108.0323 [hep-th]].
   %%CITATION = ARXIV:1108.0323;%%



%%%%%%%%%%%%%%%%%%%%%%%%%%%%%%%

%%%%%%%%%%%%%%%%%%%%%%

\bibitem{Buchbinder:2008ub}
  I.~L.~Buchbinder, O.~Lechtenfeld and I.~B.~Samsonov,
  ``N=4 superparticle and super Yang-Mills theory in USp(4) harmonic superspace,''
  Nucl.\ Phys.\ B {\bf 802} (2008) 208
  [arXiv:0804.3063 [hep-th]].
  %%CITATION = ARXIV:0804.3063;%%
  
   D.~V.~Belyaev and I.~B.~Samsonov,
  ``Wess-Zumino term in the N=4 SYM effective action revisited,''
  JHEP {\bf 1104} (2011) 112
  [arXiv:1103.5070 [hep-th]].
  %%CITATION = ARXIV:1103.5070;%%


\bibitem{Czech:2011dk} B.~Czech, Y.~-t.~Huang and M.~Rozali, 
  ``Amplitudes for Multiple M5 Branes,'' 
   arXiv:1110.2791 [hep-th].
 %%CITATION = ARXIV:1110.2791;%%



%%%%%%%%%%%%%%%


%%%%%%%%%%%%%%%%%



%%%%%



%%%%%%


% 
% 
% 
% 
% 
% 
% 
% 










\end{thebibliography}
\end{document}